# GW Notes

Sep 2010 to Jan 2011

Notes & News for GW science
Editors:
P. Amaro-Seoane and B. F. Schutz

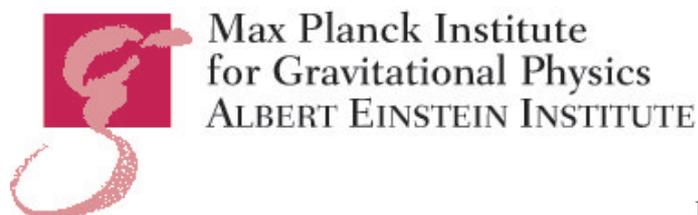

GW Notes was born from the need for a journal where the distinct communities involved in gravitational wave research might gather. While these three communities - Astrophysics, General Relativity and Data Analysis - have made significant collaborative progress over recent years, we believe that it is indispensable to future advancement that they draw closer, and that they speak a common idiom.

> **Editorial**
>
> *GW Notes: a refereed journal*

Since the first issue of GW Notes, back in March 2009, the quality of the articles published made us think that we probably could do much more than what we first envisoned: a short note by a researcher in a field related to any of the interests of LISA. However, we decided to wait and see how things developed. To our surprise, the quality has not only stayed high, but the articles published have grown longer and longer, as you can check by yourself by skimming thought the last releases, and the number of visits to the site has increased too (see the figures with the statistics). The number of download is about 300 every time we release GW Notes.

All of this led us to think that GW Notes was probably in the position of providing its readers with something of even better quality, by refereeing the highlight articles where they are technical enough to warrant it; the criteria for the refereeing is to meet the aims of notes, not of a review journal. We are happy to announce that we accept the challenge that authors have (indirectly) imposed upon us!

From now on we will continue to release GW Notes quarterly, but whether it contains a new highlight article or not will depend on whether there is a refereeing process, which will vary in each individual case.

In this first issue of the new "era" of GW Notes, we present the work of Jonathan Thornburg, who has been fully-refereed, on the Capra research programme for EMRIs.

**brownbag.lisascience.org statistics**

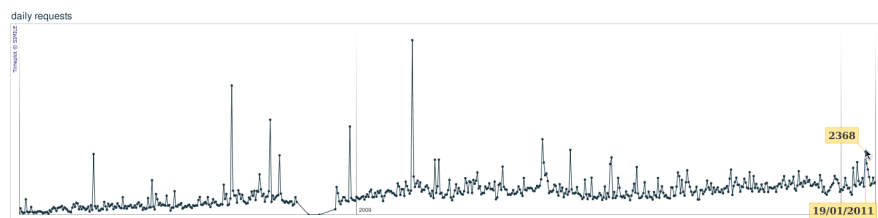

Pau Amaro-Seoane & Bernard F. Schutz, editors

> ### GW Notes highlight article
>
> *The Capra Research Programme and EMRIs*

## THE CAPRA RESEARCH PROGRAM FOR MODELLING EXTREME MASS RATIO INSPIRALS

Jonathan Thornburg
Department of Astronomy and Center for Spacetime Symmetries
Indiana University, Bloomington, Indiana, USA
e-mail: jthorn@astro.indiana.edu

### Abstract

Suppose a small compact object (black hole or neutron star) of mass $m$ orbits a large black hole of mass $M \gg m$. This system emits gravitational waves (GWs) that have a radiation-reaction effect on the particle's motion. EMRIs (extreme–mass-ratio inspirals) of this type will be important GW sources for LISA. To fully analyze these GWs, and to detect weaker sources also present in the LISA data stream, will require highly accurate EMRI GW templates.

In this article I outline the "Capra" research program to try to model EMRIs and calculate their GWs *ab initio*, assuming only that $m \ll M$ and that the Einstein equations hold. Because $m \ll M$ the timescale for the particle's orbit to shrink is too long for a practical direct numerical integration of the Einstein equations, and because this orbit may be deep in the large black hole's strong-field region, a post-Newtonian approximation would be inaccurate. Instead, we treat the EMRI spacetime as a perturbation of the large black hole's "background" (Schwarzschild or Kerr) spacetime and use the methods of black-hole perturbation theory, expanding in the small parameter $m/M$.

The particle's motion can be described either as the result of a radiation-reaction "self-force" acting in the background spacetime or as geodesic motion in a perturbed spacetime. Several different lines of reasoning lead to the (same) basic $\varnothing(m/M)$ "MiSa-TaQuWa" equations of motion for the particle. In particular, the MiSaTaQuWa equations can be derived by modelling the particle as either a point particle or a small Schwarzschild black hole. The latter is conceptually elegant, but the former is technically much simpler and (surprisingly for a nonlinear field theory such as general relativity) still yields correct results.

Modelling the small body as a point particle, its own field is singular along the particle worldline, so it's difficult to formulate a meaningful "perturbation" theory or equations of motion there. Detweiler and Whiting found an elegant decomposition of the particle's metric perturbation into a singular part which is spherically symmetric at the particle and a regular part which is smooth (and non-symmetric) at the particle. If we assume that the singular part (being spherically symmetric at the particle) exerts no force on the particle, then the MiSaTaQuWa equations follow immediately.

The MiSaTaQuWa equations involve gradients of a (curved-spacetime) Green function, integrated over the particle's entire past worldline. These expressions aren't amenable to direct use in practical computations. By carefully analysing the singularity structure of each term in a spherical-harmonic expansion of the particle's field, Barack and Ori found that the self-force can be written as an infinite sum of modes, each of which can be calculated by (numerically) solving a set of wave equations in $1+1$ dimensions, summing the gradients of the resulting fields at the particle position, and then subtracting certain analytically-calculable "regularization parameters". This "mode-sum" regularization scheme has been the basis for much further research including explicit numerical calculations of the self-force in a variety of situations, initially for Schwarzschild spacetime and more recently extending to Kerr spacetime.

Recently Barack and Golbourn developed an alternative "$m$-mode" regularization scheme. This regularizes the physical metric perturbation by subtracting from it a suitable "puncture function" approximation to the Detweiler-Whiting singular field. The residual is then decomposed into a Fourier sum over azimuthal ($e^{im\varphi}$) modes, and the resulting equations solved numerically in $2+1$ dimensions. Vega and Detweiler have developed a related scheme that uses the same puncture-function regularization but then solves the regularized perturbation equation numerically in $3+1$ dimensions, avoiding a mode-sum decomposition entirely. A number of research projects are now using these puncture-function regularization schemes, particularly for calculations in Kerr spacetime.

Most Capra research to date has used 1st order perturbation theory, with the particle moving on a fixed (usually geodesic) worldline. Much current research is devoted to generalizing this to allow the particle worldline to be perturbed by the self-force, and to obtain approximation schemes which remain valid over long (EMRI-inspiral) timescales. To obtain the very high accuracies needed to fully exploit LISA's observations of the strongest EMRIs, 2nd order perturbation theory will probably also be needed; both this and long-time approximations remain frontiers for future Capra research.

"This article is dedicated to the memory of Thomas Radke, my late friend and colleague in many computational adventures."

# Contents

# 1 Introduction

An EMRI (extreme–mass-ratio inspiral) is a binary black hole (BH) system (or a binary BH/neutron-star system) with a highly asymmetric mass ratio. That is, an EMRI consists of a small compact object (a stellar-mass BH or neutron star) of mass $\mu M$ orbiting a large BH of mass $M$, with the mass ratio $\mu \ll 1$. If the small body were a test mass ($m = 0$), then it would orbit on a geodesic of the large BH. However, if $m > 0$, then the system emits gravitational waves (GWs), and there is a corresponding radiation-reaction influence on the small body's motion. Calculating this motion and the emitted GWs is a long-standing research question, and is interesting both as an abstract problem in general relativity and as an essential prerequisite for the full success of LISA. LISA is expected to observe GWs from many EMRIs with $M \sim 10^6 M_\odot$ and $m \sim 10 M_\odot$ (so that $\mu \sim 10^{-5}$) (Amaro-Seoane et al., 2007 and Gair, 2009). To most effectively analyze this LISA data – indeed, even to *detect* much weaker signals that may also be present in the LISA data stream – requires accurately modelling the EMRI GWs, particularly the GW phase (Porter, 2009).

The small body's orbit may be highly relativistic, so post-Newtonian methods (see, for example, (Damour, 1987, section 6.10); (Blanchet, 2006, 2010, Futamase and Itoh, 2007 and Schäfer, 2010) and references therein) may not be accurate for this problem. Since the timescale for radiation reaction to shrink an EMRI orbit is very long ($\sim \mu^{-1} M$) while the required resolution near the small body is very high ($\sim \mu M$), full (nonlinear) numerical-relativity methods (see, for example, (Pretorius, 2007, Hannam et al., 2009, Hannam, 2009, Hannam and Hawke, 2010, Hinder, 2010, Campanelli et al., 2010 and Centrella et al., 2010) and references therein) would be both prohibitively expensive and insufficiently accurate for this problem.[1]

Instead, a variety of other approximation schemes are used to model EMRIs and their GWs. In particular, the "Capra" research program,[2] uses the techniques of BH perturbation theory to model the EMRI spacetime *ab initio* as a perturbation of the massive central BH's Schwarzschild or Kerr spacetime, making no approximations other than that the mass ratio $\mu \ll 1$. In particular, the Capra research program doesn't make any slow-motion or weak-field approximations.

---

[1] The most asymmetric mass ratio yet simulated with full (nonlinear) numerical relativity is $100 : 1$, i.e., $\mu = 10^{-2}$ (Lousto and Zlochower, 2011). A number of researchers have attempted to develop special methods to make EMRI numerical-relativity simulations practical, at least for systems with "intermediate" mass ratios $\mu \sim 10^{-3}$. Although promising initial results have been obtained (see, for example, (Bishop et al., 2003, 2005, Sopuerta et al., 2006, Sopuerta and Laguna, 2006 and Lousto et al., 2010)), it has not (yet) been possible to perform accurate EMRI numerical evolutions lasting for radiation-reaction time scales.

[2] The Capra research program, and the yearly Capra meetings on radiation reaction in general relativity, are named after the late American film director Frank Capra, famous for such films as *It's a Wonderful Life* and *Mr. Smith Goes to Washington* as well as the World War II propaganda series *Why We Fight*. He owned a ranch near San Diego and upon his death donated part of this to Caltech. The first Capra meeting was held there in 1998.

In this article I give a relatively non-technical overview of some of the highlights of the Capra research program, focusing on those aspects most relevant to explicitly calculating radiation-reaction effects in various physical systems. My goal is to give the reader some sense of the "flavor" of Capra research. The reader should have a reasonable background in general relativity and, for some parts of **Sections 2.1** (**page 11**), **2.2** (**page 15**), and **2.4** (**page 20**), be familiar with Green-function methods[3] for solving linear partial differential equations (PDEs). The sections of this article are relatively independent and, with a few exceptions (which should be obvious from cross-references), can be read in any order. In **Sections 2.2** (**page 15**) and **2.4** (**page 20**) I have marked certain passages as somewhat more technical (analogous to the "Track 2" of Misner, Thorne, and Wheeler (Misner et al., 1973)); this material may be skipped if the reader so desires.

This is emphatically *not* a comprehensive review – there are major areas of the Capra program that I only briefly mention, and others which I omit entirely.[4] Except for some of the accuracy arguments in **Section 4** (**page 40**), there's no original research in this article. For more detailed and complete information about the Capra research program, the reader should consult any of a number of excellent review articles, notably those by Poisson (Poisson, 2004, 2005, 2010),[5] Detweiler (Detweiler, 2005), and Barack (Barack, 2009). The websites of recent Capra meetings (Thornburg, 2009 and Lehner et al., 2010) also include archives of meeting presentations.

A key long term goal of the Capra research program is the modelling (and explicit calculation) of highly accurate orbital dynamics and GW templates for generic EMRIs. As discussed in **Section 4** (**page 40**), the highest-accuracy GW templates for LISA will probably require carrying BH perturbation theory to at least 2nd order in the mass ratio $\mu$, and also using special "long-time" approximation schemes. These are ambitious goals, which are still far from being met: most Capra research to date has been devoted to the lesser – but still challenging – problem of trying to model strong-field EMRI radiation-reaction effects using 1st order perturbation theory and, to the best of my knowledge, no Capra GW templates have yet been published. I return to 2nd-order calculations in **Sections 4** (**page 40**) and **5** (**page 43**), but for the rest of this article I consider only 1st-order calculations.

---

[3] We say "Bessel function", not "Bessel's function", so logically the reader should be familiar with "Green-function methods", not "Green's-function methods".

[4] I apologise to the reader for any mistakes there may be in this article, and I particularly apologise to anyone whose work I've slighted or mischaracterized. I welcome corrections for a future revision of this article.

[5] In particular, Poisson's GR17 plenary lecture (Poisson, 2005) contains a short and relatively non-technical review of a large part of the theoretical background underlying the Capra research program. I highly recommend this article to the reader seeking somewhat more detail than I provide in **Section 2** (**page 9**). Poisson's lectures (Poisson, 2010) from the 2008 "Mass and Motion" summer school and 11th Capra meeting provides a somewhat more detailed presentation of this material, and his *Living Reviews in Relativity* article (Poisson, 2004) gives a lengthy and detailed technical account.

In almost all Capra calculations to date, the small body is taken to move on a fixed geodesic worldline of the background (Schwarzschild or Kerr) spacetime, with radiation-reaction effects being manifest as an $O(\mu^2)$ "self-force" acting on the small body.[6] Alternatively, we can view the small body as moving on a geodesic of a $O(\mu)$-perturbed spacetime. These two perspectives can be shown to be fully equivalent (Sago et al., 2008) and are, in some ways, analogous to Eulerian versus Lagrangian approaches to fluid dynamics; we can use whichever is more convenient for any given calculation.

Another important choice in self-force analyses is whether to model the small body as a point particle or as a nonzero-sized small compact body. Modelling it as a nonzero-sized body is conceptually elegant but technically difficult. In contrast, point-particle models are technically simpler but pose difficult conceptual and foundational problems. Indeed, in a nonlinear field theory such as general relativity, the very notion of a "point particle" is difficult to formulate in a self-consistent manner (Geroch and Traschen, 1987). Remarkably, it turns out that these difficulties can be overcome and, in fact, point-particle models have been used for the bulk of Capra research to date. I discuss this point further in **Section 2** (**page 9**).

Starting from the Einstein equations, one can derive the generic $O(\mu)$ "MiSa-TaQuWa" equations of motion for the small body in an arbitrary (strong-field) curved spacetime. These equations give the self-force in terms of a formal Green-function integral over the particle's entire past motion and have now been obtained in several different ways, using both point-particle and nonzero-sized models of the small body.

It's usually not possible to explicitly calculate the Green function appearing in the MiSaTaQuWa equations. Instead, practical computational schemes are usually based on regularizing the (singular) metric-perturbation equations for a point particle; several different ways are now known to do this. The regularized equations can then be solved (usually numerically) to actually compute the self-force for a given physical system. Because these calculations are in many cases both conceptually difficult and computationally demanding, new techniques are often first developed on simpler electromagnetic or scalar-field "model" systems. These retain many of the basic conceptual features of the gravitational case while greatly simplifying the gauge choice[7] and the resulting computations.

---

[6] The small body's mass is $O(\mu)$, so if it were not constrained to moving on a fixed geodesic worldline, the $O(\mu^2)$ self-force would give rise to an $O(\mu)$ "self-acceleration" of the small body away from a geodesic trajectory.

[7] As discussed by Barack and Ori (Barack and Ori, 2001), the self-force is highly gauge-dependent in a somewhat unobvious non-tensorial manner. (For example, there exists a gauge in which the self-force vanishes. Essentially, the gauge transformation follows the small body as it spirals in to the massive BH.) There are thus considerable benefits to computing gauge-invariant effects, an approach particularly championed by Detweiler.

We can categorize Capra self-force calculations along several dimensions of complexity:

- The background spacetime may be either Schwarzschild or Kerr.

- The field equations may be for the scalar-field, electromagnetic, or the full gravitational case.

- The small body may be stationary, in an equatorial circular orbit, in a generic (non-circular) equatorial orbit, or in a fully generic (inclined non-circular) orbit in Kerr spacetime.

The outline of the remainder of this article is as follows: In **Section 2** (**page 9**) I discuss some of the key theoretical foundations of the Capra program including the Barack-Ori mode-sum regularization, the Detweiler-Whiting decomposition of a point particle's metric perturbation, several different derivations of the basic 1st-order "MiSaTaQuWa" equations of motion for a small compact body moving in a curved spacetime, the Barack-Golbourn and Vega-Detweiler puncture-function regularizations and the self-force computational schemes derived from them, and the decomposition of self-force effects into conservative and dissipative parts. In **Section 3** (**page 32**) I summarize a recent self-force calculation of Barack and Sago (Barack and Sago, 2010), which provides an almost complete solution of the 1st-order self-force problem for a particle moving on a fixed geodesic orbit in Schwarzschild spacetime. In **Section 4** (**page 40**) I roughly estimate LISA's accuracy requirements for GW templates, and outline some of the issues in trying to model EMRI orbital dynamics for long (orbital-decay) times to construct such templates. Finally, in **Section 5** (**page 43**) I summarize the progress of the Capra program to date and discuss some of its likely future prospects.

Throughout this article I use $c = G = 1$ units and a $(-, +, +, +)$ metric signature. I use the Penrose abstract-index notation, with $abcde$ as spacetime indices. $\delta(\cdot)$ is the Dirac $\delta$-function, $\tau$ denotes proper time along the small body's worldline, and a subscript $p$ denotes evaluation at the small body (particle)'s current position. $\mu \ll 1$ is the EMRI system's mass ratio and $M$ the central BH's mass. Apart from these, the notation in this article varies somewhat from section to section; it's always described at the start of each section.

## 2 Theoretical Background

In this section I discuss some of the main theoretical background and formalisms which underlie the Capra research program.[8]

---

[8] My exposition in parts of this section draws heavily on that of Poisson's GR17 plenary lecture (Poisson, 2005).

A key early result of Capra research was the derivation in several different ways of the basic 1st-order equations of motion for a small compact body moving in a strong-field curved spacetime. These equations were first derived in 1997 by Mino, Sasaki, and Tanaka (Mino et al., 1997) and Quinn and Wald (Quinn and Wald, 1997), and (abbreviating the authors' names) are now known as the "MiSaTaQuWa" equations.

The MiSaTaQuWa equations involve gradients of a curved-spacetime Green function, integrated over the particle's entire past worldline. We can rarely calculate the Green function explicitly, so the MiSaTaQuWa equations aren't useful for practical computations. In **Section 2.1** (**page 11**) I discuss the "mode-sum regularization" computational scheme due originally to Barack and Ori (Barack and Ori, 2000). This scheme regularizes each mode of a spherical-harmonic decomposition of the (singular) scalar-field or metric perturbation, then solves numerically for each regularized mode in 1+1 dimensions. This scheme has been the basis for much further research, including many practical self-force calculations.

In 2003 Detweiler and Whiting (Detweiler and Whiting, 2003) found a Green-function decomposition – and a corresponding decomposition of the metric perturbation due to a small particle – into singular and radiative fields, which greatly aids understanding the self-force and related phenomena. In **Section 2.2** (**page 15**) I discuss this decomposition and the related Detweiler-Whiting "postulate" concerning the physical significance of the singular and radiative fields.

In **Section 2.3** (**page 19**) I outline how the Detweiler-Whiting postulate allows the MiSaTaQuWa equations to be derived via modelling the small body as a point particle. This derivation of the MiSaTaQuWa equations is quite simple, but it does involve the introduction of point particles and the assumption of the Detweiler-Whiting postulate.

As an alternative, in **Section 2.4** (**page 20**) I outline a different derivation of the MiSaTaQuWa equations, this time modelling the small body as a small nonrotating (Schwarzschild) BH. This derivation is technically more difficult than the point-particle derivation, but it avoids both the introduction of point particles and the assumption of the Detweiler-Whiting postulate.

Recently researchers have developed several new analyses of the self-force problem, leading to much more satisfactory derivations of the MiSaTaQuWa (and analogous) equations. These new analyses are fully rigorous and resolve a number of past conceptual difficulties as well as opening promising avenues for further research. I (very) briefly outline these analyses in **Section 2.5** (**page 23**).

Recently two groups have developed alternate "puncture-function" regularization schemes for self-force computations. Both schemes first subtract from the physical field a suitable "puncture function" approximation to the Detweiler-Whiting singular field, leaving a regular remainder field. Barack, Golbourn, and their coauthors (Barack and Golbourn, 2007, Barack et al., 2007 and Dolan and Barack, 2011) developed an "$m$-mode" regularization scheme which then decomposes the perturbation equation into a Fourier sum over azimuthal ($e^{im\varphi}$) modes, and finally solves numerically for each mode in 2+1 dimensions. Vega, Detweiler, and their coauthors (Vega and Detweiler, 2008 and Vega et al., 2009) have developed a different puncture-function regularization scheme that numerically solves the regularized equation directly in 3+1 dimensions, avoiding a mode-sum decomposition entirely. I describe both of these schemes in **Section 2.6** (**page 24**).

The self force can be decomposed into conservative (time-symmetric) and dissipative (time-antisymmetric) parts. In **Section 2.7** (**page 31**) I discuss this decomposition and its physical significance.

### 2.1 The Barack-Ori Mode-Sum Regularization

The MiSaTaQuWa equations (discussed further in **Sections 2.2** (**page 15**)–**2.5** (**page 23**)) give the self-force in terms of the gradient of a curved-spacetime Green function, integrated over the entire past history of the small body. (The integral must be cut off infinitesimally before the small body's current position.) For most physically-interesting systems we can't explicitly calculate the Green function, so the MiSaTaQuWa equations aren't useful for practical calculations. Instead, almost all practical self-force calculations use other regularized reformulations of the (singular) scalar-field, electromagnetic, or metric-perturbation equations.

Building on earlier suggestions of Ori (Ori, 1995, 1997), in 2000 Barack and Ori (Barack and Ori, 2000) proposed a practical "mode-sum" regularization of the field equations, initially for the model problem of a scalar particle moving in Schwarzschild spacetime. Barack and various coauthors (Barack, 2000, 2001, Barack et al., 2002, Barack and Ori, 2002, 2003a and Barack and Lousto, 2002) soon extended this to include electromagnetic and gravitational particles in a somewhat wider class of spherically symmetric black hole spacetimes, as well as scalar-field, electromagnetic, and gravitational particles in Kerr spacetime (Barack and Ori, 2003b). The mode-sum regularization has been the basis for much further research including many practical self-force calculations.

To explain the mode-sum regularization, I consider the scalar-field case – this contains the essential ideas, but is technically much simpler than the full gravitational case. Thus, consider a point particle of scalar charge $q$, moving along a timelike worldline $\Gamma = \Gamma(\tau)$ in a background Schwarzschild spacetime with metric $g_{ab}$ and covariant derivative operator $\nabla_a$. In this section we raise and lower all indices with the background Schwarzschild metric $g_{ab}$, and we take $(t, r, \theta, \varphi)$ to be the usual Schwarzschild coordinates.

We take the scalar field $\phi$ to satisfy the usual scalar wave equation

$$\Box\phi = -4\pi q \int_{-\infty}^{+\infty} \frac{\delta^4 (x^a - \Gamma^a(\tau'))}{\sqrt{-g}} \, d\tau' =: S \ , \tag{1}$$

where the integral extends over the entire worldline of the particle, and we define $S$ to be the source term (right hand side). We assume that if the scalar field were regular (non-singular) at the particle, it would exert a force

$$F_a = q \, (\nabla_a \phi)_p \tag{2}$$

on the particle.

The scalar wave equation (1) can be formally solved by means of a retarded Green function $\mathcal{G}(x, x')$,

$$\phi(x) = \int_{-\infty}^{+\infty} \mathcal{G}\,(x, \Gamma(\tau')) \, d\tau' \tag{3}$$

where the Green function $\mathcal{G}(x, x')$ satisfies

$$\Box \mathcal{G}(x, x') = -\frac{4\pi}{\sqrt{-g}} \, \delta^4 (x^a - x'^a) \tag{4}$$

and incorporates the appropriate causality relationships.

Ignoring some terms which aren't relevant here, the self-force on the particle at the worldline event $x = (t, r, \theta, \varphi)$ can then be shown to be given by

$$F_a(x) = q^2 \int_{-\infty}^{x^-} \nabla_a \mathcal{G}\,(x; \Gamma(\tau')) \, d\tau' \ , \tag{5}$$

where the upper limit $x^-$ means that the integral extends over the entire past worldline of the particle prior to (but not including) the event $x$.

Now consider a spherical-harmonic decomposition of the scalar field $\phi$ and the self-force $F_a$,

$$\begin{aligned}
\phi(x) &= \sum_{\ell=0}^{\infty} \sum_{m=-\ell}^{+\ell} Y_{\ell m}(\theta, \varphi) \phi^{\ell m}(t, r) \, , \\
F_a(x) &= \sum_{\ell=0}^{\infty} \sum_{m=-\ell}^{+\ell} Y_{\ell m}(\theta, \varphi) F_a^{\ell m}(t, r) \, ,
\end{aligned} \tag{6}$$

and sum over the azimuthal mode number $m$ by defining

$$F_a^\ell(t, r, \theta, \varphi) = \sum_{m=-\ell}^{+\ell} Y_{\ell m}(\theta, \varphi) F_a^{\ell m}(t, r) \ , \tag{7}$$

so that the self-force is given by

$$F_a = \sum_{\ell=0}^{\infty} F_a^\ell \ .$$ (8)

It turns out that each individual spherical-harmonic mode $F_a^\ell$ is finite, but the sum over $\ell$ of these modes in (8) diverges. However, by carefully analysing the divergence of the scalar field and its Green function near the particle, Barack and Ori showed that the related sum

$$\sum_{\ell=0}^{\infty} \left( F_a^\ell - [A_a L + B_a + C_a/L] \right) \ ,$$ (9)

(where $L = \ell + \frac{1}{2}$, and where the quantities $A_a$, $B_a$, and $C_a$ are described below) *does* in fact converge and moreover, that the self-force is given by

$$F_a = \sum_{\ell=0}^{\infty} \left( F_a^\ell - [A_a L + B_a + C_a/L] \right) - Z_a \ ,$$ (10)

where the "regularization parameters" $A_a$, $B_a$, $C_a$, and $Z_a$ are independent of $\ell$ and can be calculated semi-analytically as elliptic integrals depending on the worldline $\Gamma$ and the worldline event $x$.

[More generally, additional even-power terms $D_a^{(2)}/L^2$, $D_a^{(4)}/L^4$, $D_a^{(6)}/L^6$, ..., can be added to the $A_a L + B_a + C_a/L$ inner sum in the mode-sum self-force formula (10) (with corresponding adjustments to the definition of $Z_a$). As discussed by Detweiler, Messaritaki, and Whiting (Detweiler et al., 2003), if the additional regularization parameters $D_a^{(k)}$ can be explicitly calculated (analytically or semi-analytically), then adding these extra terms can greatly accelerate the convergence of the sum over $\ell$. I discuss the numerical treatment of this infinite sum below.]

By virtue of the spherical symmetry of the Schwarzschild background, the spherical-harmonic decomposition (6) separates the scalar wave equation (1). That is, each individual $\phi^{\ell m}(t, r)$ now satisfies a linear wave equation in 1+1 dimensions on the Schwarzschild background,

$$\Box \phi^{\ell m} + V_\ell(r) \phi^{\ell m} = S_{\ell m}(t) \, \delta \left( r - r_p(t) \right) \ ,$$ (11)

where the potential $V_\ell(r)$ and source term $S_{\ell m}(t)$ are known analytically, and where $r_p(t)$ is the particle's (known) Schwarzschild radial coordinate (which is time-dependent if the worldline is anything other than a circular geodesic orbit). The wave equation (11) can then be solved numerically in the time domain to find the field $\phi^{\ell m}$.

Finally, each individual self-force mode $F_a^{\ell m}$ can be calculated from the $\ell m$ component of the basic force law (2),

$$F_a^{\ell m} = q \left( \nabla_a \phi^{\ell m} \right)_p \ ,$$ (12)

and then $F_a^\ell$ can be calculated from (7).

Alternatively, we can take a frequency-domain approach, augmenting the spherical-harmonic expansions (6) with a Fourier transform in time, i.e., we can replace those expansions with

$$\phi(x) = \int_\omega \sum_{\ell=0}^\infty \sum_{m=-\ell}^{+\ell} Y_{\ell m}(\theta, \varphi) e^{i\omega_m t} \phi^{\omega\ell m}(r) \, d\omega \ ,$$

$$F_a(x) = \int_\omega \sum_{\ell=0}^\infty \sum_{m=-\ell}^{+\ell} Y_{\ell m}(\theta, \varphi) e^{i\omega_m t} F_a^{\omega\ell m}(r) \, d\omega \ ,$$

(13)

and correspondingly replace the $m$ summation (7) with

$$F_a^\ell(t, r, \theta, \varphi) = \int_\omega \sum_{m=-\ell}^{+\ell} Y_{\ell m}(\theta, \varphi) e^{i\omega_m t} F_a^{\omega\ell m}(r) \, d\omega \ .$$

(14)

We also similarly Fourier-transform and spherical-harmonic–expand the source term (right hand side) $S$ in the scalar wave equation (1),

$$S(x) = \int_\omega \sum_{\ell=0}^\infty \sum_{m=-\ell}^{+\ell} Y_{\ell m}(\theta, \varphi) e^{i\omega_m t} S_{\omega\ell m}(r) \, d\omega \ .$$

(15)

The rest of the mode-sum regularization then goes through unchanged for each frequency $\omega$, and each individual scalar-field mode $\phi^{\omega\ell m}(r)$ now satisfies the *ordinary* differential equation

$$\frac{d^2\phi^{\omega\ell m}}{dr^2} + H(r)\frac{d\phi^{\ell m\omega}}{dr} + V_{\omega\ell}(r)\phi^{\omega\ell m} = S_{\omega\ell m}(r) \ ,$$

(16)

where the coefficients $H(r)$, $V_{\omega\ell}(r)$, and $S_{\omega\ell m}(r)$ are again all known analytically.

The frequency-domain approach involves much simpler computations (ODEs) than the time-domain approach's PDEs. If the particle orbit is circular, then only a single frequency is needed, and the integrals over $\omega$ are trivial. On the other hand, if the particle orbit is noncircular, many frequencies $\omega$ may be needed to adequately approximate the integrals over $\omega$.

In any case, once the individual $F^{\ell m}$ or $F^{\omega\ell m}$ are known, there remains the problem that the overall $\ell$ sum (8) is an infinite sum. For computational purposes a finite expression is needed. The solution to this lies in the known large-$\ell$ behavior (Detweiler et al., 2003) of $F_{a,\mathrm{reg}}^\ell := F_a^\ell - [A_a L + B_a + C_a/L]$,

$$F_{a,\mathrm{reg}}^\ell = \frac{D_a^{(2)}}{L^2} + \frac{D_a^{(4)}}{L^4} + \frac{D_a^{(6)}}{L^6} + \cdots \qquad \text{for large } L,$$

(17)

where the coefficients $D_a^{(k)}$ are independent of $L$. [Of course, if a term $D_a^{(k)}/L^k$ has been added to the $A_a L + B_a + C_a/L$ inner sum in the mode-sum self-force formula (10), then that term is absent from the large-$L$ series (17).] We partition the overall $\ell$ sum (8) into a finite "numerical" part and an infinite "tail" part,[9]

$$F_a = \sum_{\ell=0}^{\ell_{max}} F_{a,\text{reg}}^\ell + \sum_{\ell=\ell_{max}+1}^{\infty} F_{a,\text{reg}}^\ell \; , \tag{18}$$

where $\ell_{max} \sim 15$–$30$ is a numerical parameter. Then, once all the $F_{a,\text{reg}}^\ell$ in the numerical part of the sum are known, we can fit some number (typically 2 or 3) of terms in the large-$\ell$ series (17) to the numerically-computed $F_{a,\text{reg}}^\ell$ values, and use the fitted coefficients $\{D_a^{(k)}\}$ to estimate the tail term.

### 2.2 The Detweiler-Whiting Decomposition

Suppose we model the small body as a point particle. (In this section we ignore the point-particle foundational issues mentioned in footnote 14 (**page 19**).) Because the particle's own field is singular along the particle worldline, it's not obvious how to write a meaningful "perturbation" theory or formulate equations of motion there. The solution (within the framework of modelling the small body as a point particle) is to somehow regularize the field so that we have finite quantities to manipulate.

The regularization of the (retarded) field of a point-particle source near that source is a long-standing problem in mathematical physics. Dirac (Dirac, 1938) studied this problem for the electromagnetic field of a point charge in flat spacetime and found a decomposition of the field into a singular part which is spherically symmetric about the charge, and a "radiative" part which is regular at the charge. [We might then assume (postulate) that despite being singular, the spherically-symmetric part exerts no force on the charge. I discuss the curved-spacetime generalization of this assumption (postulate) below.] Dirac's analysis was generalized to curved spacetime by DeWitt and Brehme (DeWitt and Brehme, 1960) (with a correction by Hobbs (Hobbs, 1968)).

More recently, Detweiler and Whiting (Detweiler and Whiting, 2003) found a fully satisfactory curved-spacetime decomposition of a point particle's field (whether scalar, electromagnetic, or gravitational) into singular and regular parts. Here I describe this decomposition for the gravitational case. Suppose we have a point mass $m$, moving (unaffected by external forces apart from self-force effects) with 4-velocity $u^a$ along a timelike worldline $\Gamma = \Gamma(\tau)$ in a "background" vacuum spacetime whose typical radius of curvature in a neighborhood of $\Gamma$ is $\mathcal{R} \gg m$. [For a LISA EMRI we might have $m \sim 10 M_\odot$ while $\mathcal{R} \sim 10^6 M_\odot$ (the massive BH's mass).]

---

[9] This terminology is perhaps unfortunate: this usage of "tail" has no connection at all to the usage of "tail term" in **Section 2.2** (**page 15**) and elsewhere in this article.

Suppose $^{(0)}g_{ab}$ is the (vacuum) background metric, i.e., the spacetime metric in the absence of the particle, and $\nabla_a$ is the corresponding covariant derivative.[10] Throughout this section we raise and lower indices with the background metric $^{(0)}g_{ab}$. Let $h_{ab}$ be the metric perturbation produced by the particle, so that the physical spacetime metric is $g_{ab} = {}^{(0)}g_{ab} + h_{ab}$. We introduce the standard trace-reversed metric perturbation $\bar{h}_{ab} = h_{ab} - \frac{1}{2}{}^{(0)}g_{ab}h_c{}^c$ and impose the Lorenz gauge condition $\nabla_a \bar{h}^{ab} = 0$ on the metric perturbation.

[The material from this point up to and including the paragraph containing equation (23) is somewhat more technical than the rest of this article and can be skipped without creating confusion.]

Up to $O(m/\mathcal{R})$ accuracy, the metric perturbation satisfies the linear (wave) equation

$$\nabla^c\nabla_c\bar{h}_{ab} + 2R^c{}_a{}^d{}_b\bar{h}_{cd} = T_{ab} \quad , \tag{19}$$

where $T_{ab}$ is the particle's ($\delta$-function) stress-energy tensor.

The (linear) perturbation equation (19) can be formally solved by introducing a suitable retarded Green function, which for events within a normal convex neighborhood of the particle (i.e., events $x'$ which are linked to the particle event $x$ by a *unique* geodesic) can be written in the Hadamard form

$$\mathcal{G}^{ab}{}_{c'd'}(x,x') = \mathcal{U}^{ab}{}_{c'd'}(x,x')\,\delta\left(\sigma(x,x')\right) + \mathcal{V}^{ab}{}_{c'd'}(x,x')\,\theta\left(-\sigma(x,x')\right) \quad , \tag{20}$$

where $\theta(\cdot)$ is the Heaviside step function, Synge's world function $\sigma(x,x')$ is half the squared geodesic distance between the events $x$ and $x'$,[11] and where $\mathcal{U}^{ab}{}_{c'd'}(x,x')$ and $\mathcal{V}^{ab}{}_{c'd'}(x,x')$ are *smooth* bitensors.[12] The first term in this decomposition is a singular "light-cone part" supported only for $\sigma = 0$, i.e., only when $x'$ lies on the past light cone of $x$. The second term is a smooth "tail part" supported only for $\sigma < 0$, i.e., only when $x'$ lies within (but not on) the past light cone of $x$. Notice that the tail part has support throughout the entire past history of the particle. Heuristically, this is

---

[10] The notation (the presence of the prefix $^{(0)}$ on the background metric, but not on the corresponding covariant derivative operator) is admittedly somewhat inconsistent here.

[11] As explained very clearly in section 2.1.1 of Poisson's Living Reviews in Relativity article (Poisson, 2004), Synge's world function $\sigma(x,x')$ has the following properties:

- $\sigma(x,x') < 0$ if and only if the geodesic connecting $x'$ to $x$ is timelike, i.e., if and only if $x'$ lies within (but not on) $x$'s past light cone.

- $\sigma(x,x') = 0$ if and only if the geodesic connecting $x'$ to $x$ is null, i.e., if and only if $x'$ lies *on* $x$'s past light cone.

- $\sigma(x,x') > 0$ if and only if the geodesic connecting $x'$ to $x$ is spacelike, i.e., if and only if $x'$ lies outside $x$'s past light cone.

[12] For the reader not familiar with bitensors, they are built out of the derivatives $\partial\sigma(x,x')/\partial x$ and $\partial\sigma(x,x')/\partial x'$ somewhat analogously to the way that "ordinary" tensors are built out of vectors and one-forms. Section 2.1.2 of Poisson's Living Reviews in Relativity article (Poisson, 2004) gives a brief and very lucid introduction to bitensors.

because metric perturbations from any time in the particle's past history can scatter off the background spacetime curvature and then return to the event $x$.

The resulting formal solution of the perturbation equation (19) is

$$\bar{h}^{ab}(x) = \frac{4m}{r}\,\mathcal{U}^{ab}{}_{c'd'}(x,x')u^{c'}u^{d'} + \bar{h}^{ab}_{\text{tail}}(x) \quad , \tag{21}$$

where $x'$ is the intersection of $x$'s past light cone with the worldline $\Gamma$, $u^{a'}$ is the small BH's 4-velocity at this retarded point $x'$, and the tail term is given explicitly by

$$\bar{h}^{ab}_{\text{tail}}(x) = 4m \int_{-\infty}^{x'} \mathcal{V}^{ab}{}_{c^\varsigma d^\varsigma}\left(x, \Gamma(\tau^\varsigma)\right) u^{c^\varsigma} u^{d^\varsigma}\, d\tau^\varsigma \tag{22}$$

where the $\varsigma$ accent (superscript) marks the dummy variable of integration (integrating over the worldline), and where the integral is over the particle's entire past history prior to the retarded point $x'$. Fig.1 illustrates the causal relationships between $\Gamma$, $x$, $x'$, and $x^\varsigma$. Physically, the tail term (22) models the effect of "null then scattering then null" paths from $x^\varsigma$ to $x$ (shown in blue in Fig.1, where metric perturbations from the event $x^\varsigma$ scatter off the background spacetime curvature and then return to the event $x$.

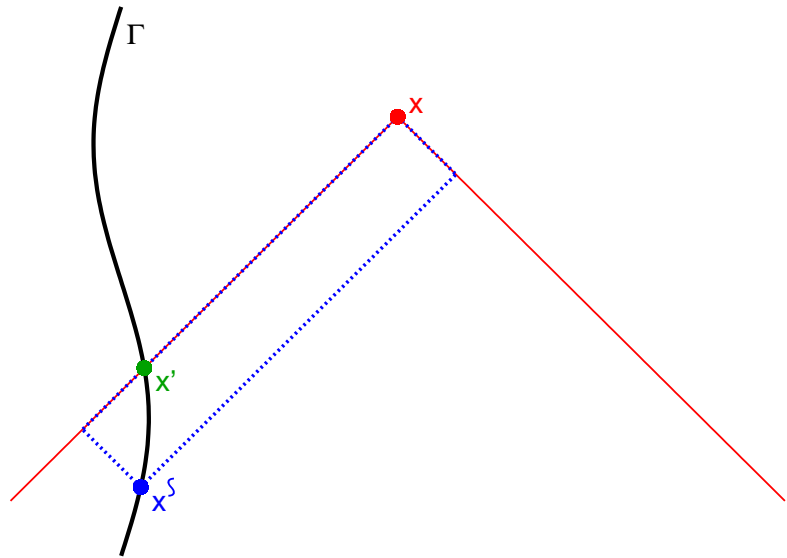

**Fig. 1**  This figure shows (in black) the particle's timelike geodesic worldline $\Gamma$, (in red) a nearby "field" event $x$ and its past lightcone, (in green) the event $x'$ where this lightcone intersects the worldline $\Gamma$, (in blue) an event $x^\varsigma$ on $\Gamma$ in the past of $x'$, and (with blue dashed lines) two possible "null then scattering then null" paths from $x^\varsigma$ to $x$ (these are modelled by the tail term (22)), where (heuristically) metric perturbations from the event $x^\varsigma$ scatter off the background curvature before returning to event $x$.

The Detweiler-Whiting "singular" part of the metric perturbation is then (defined to be)

$$\bar{h}_S^{ab}(x) = \frac{2m}{r}\,\mathcal{U}^{ab}{}_{c'd'}(x,x')u^{c'}u^{d'} + \frac{2m}{r}\,\mathcal{U}^{ab}{}_{c''d''}(x,x'')u^{c''}u^{d''} - 2m\int_{x'}^{x''}\mathcal{V}^{ab}{}_{c^S d^S}\left(x,\Gamma(\tau^S)\right)u^{c^S}u^{d^S}\,d\tau^S \quad,$$

(23)

where (as shown in Fig.2) the events $x'$ and $x''$ are respectively the intersections of the past and future light cones of the event $x$ with the particle worldline $\Gamma$.

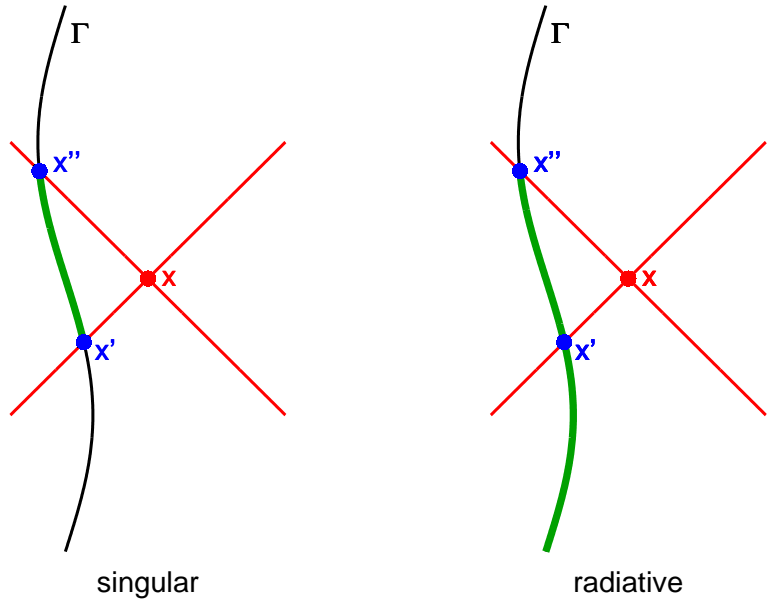

**Fig. 2** This figure shows the causal properties of the Detweiler-Whiting singular and radiative fields. Each subfigure shows (in black) the particle's timelike geodesic worldline $\Gamma$, (in red) a nearby "field" event $x$ and its past and future lightcones, (in blue) the events $x'$ (past) and $x''$ (future) where this lightcone intersects the worldline $\Gamma$, and (in green) the region of the worldline which affects the Detweiler-Whiting singular or radiative fields.

Detweiler and Whiting showed that the (singular) field $\bar{h}_S^{ab}$ defined in this manner satisfies the same metric-perturbation equation (**19**) as the physical (retarded) perturbation $\bar{h}^{ab}$, and furthermore that $\bar{h}_S^{ab}$ is "just as singular" as $\bar{h}^{ab}$ on the worldline $\Gamma$. That is, they showed that the "radiative" field

$$\bar{h}_R^{ab}(x) = \bar{h}^{ab}(x) - \bar{h}_S^{ab}(x)$$

(24)

is in fact *smooth* on the worldline $\Gamma$ (as well as everywhere else). Notice also that the radiative field satisfies the homogeneous form of the metric-perturbation equation (**19**).

The Detweiler-Whiting singular and radiative fields have unusual causal properties, illustrated in Fig. **2**: the singular field depends on the particle's history only between the events $x'$ and $x''$, while the radiative field depends on the particle's entire past history up to the advanced event $x''$. However, in the limit that the event $x$ approaches the worldline $\Gamma$, the radiative field then depends only on the particle's past history.

The Detweiler-Whiting singular field $\bar{h}_S^{ab}(x)$ is spherically symmetric at the particle. That is, it can be shown[13] that if we average the gradient of this field over a 2-sphere of radius $\epsilon$ centered on the particle (as seen in the particle's instantaneous rest frame), then take the limit $\epsilon \to 0$, this average vanishes. This motivates the Detweiler-Whiting postulate: *the singular field exerts no force on the particle; the self-force arises entirely from the action of the (regular) radiative field*. This postulate gives valuable conceptual insight into how the self force "works". This postulate is also closely linked to the MiSaTaQuWa equations (I discuss this in the next paragraph and in **Section 2.3** (**page 19**)) and to puncture-function regularizations and computational schemes for the self force (I discuss these in **Section 2.6** (**page 24**)).

Unfortunately, because the field is singular at the particle, the simple averaging argument described in the previous paragraph doesn't constitute a rigorous proof of the Detweiler-Whiting postulate. However, the Detweiler-Whiting postulate is closely linked to the MiSaTaQuWa equations: if we assume the Detweiler-Whiting postulate, then the MiSaTaQuWa equations follow almost immediately via the argument outlined in **Section 2.3** (**page 19**). Because of this close linkage, we can reverse the direction of logical implication and argue that the validity of the MiSaTaQuWa equations (which are now well-established via the rigorous derivations discussed in **Sections 2.4** (**page 20**) and **2.5** (**page 23**)) supports the correctness of the Detweiler-Whiting postulate. That is, we can argue that the Detweiler-Whiting postulate must be valid, since it is central to a derivation (the one outlined in **Section 2.3** (**page 19**)) which leads to a correct result (the MiSaTaQuWa equations). While not a completely rigorous proof, this argument strongly supports the validity of the Detweiler-Whiting postulate.

Harte (Harte, 2006, 2008, 2009a, 2009b) and Pound (Pound, 2010a, 2010b, 2010c) have recently given rigorous proofs of the Detweiler-Whiting postulate (somewhat generalized in some cases).

### 2.3 Deriving the MiSaTaQuWa Equations via Modelling the Small Body as a Point Particle

If we ignore the foundational issues of point particles,[14] then the Detweiler-Whiting postulate provides a relatively easy route to the MiSaTaQuWa equations: the (Detweiler-Whiting) statement that the self-force arises solely from the action of the (regular) radiative field is equivalent to the statement that the particle moves on a geodesic of the background metric perturbed by this radiative field, i.e., ${}^{(0)}g_{ab} + h_{ab}^R$. The particle's 4-acceleration is thus given by

$$a^a = \left({}^{(0)}g^{ab} + u^a u^b\right)\left(\tfrac{1}{2}\nabla_b h_{cd}^R - \nabla_d h_{bc}^R\right)u^c u^d \quad . \tag{25}$$

---

[13] See Poisson's *Living Reviews in Relativity* article (Poisson, 2004) for details.

[14] Geroch and Traschen (Geroch and Traschen, 1987) have shown that point particles can *not* consistently be described by metrics with $\delta$-function curvature tensors. More general Colombeau-algebra methods

It's now easy to show that on the particle's worldline $\Gamma$,

$$\nabla_c h_{ab}^R = -4m\left(u_{(a} R_{b)dce} + R_{adbe} u_c\right)u^d u^e + \nabla_c h_{ab}^{\text{tail}} \quad . \tag{26}$$

Substituting (26) into (25) then gives the MiSaTaQuWa equations

$$a^a = \left({}^{(0)}g^{ab} + u^a u^b\right)\left(\tfrac{1}{2}h_{cdb}^{\text{tail}} - h_{bcd}^{\text{tail}}\right)u^c u^d \quad . \tag{27}$$

where we define $h_{abc}^{\text{tail}} = \nabla_c h_{ab}^{\text{tail}}$. The corresponding dynamical equations of motion for the particle are

$$u^b \nabla_b u^a = a^a \quad . \tag{28}$$

Notice that the metric ${}^{(0)}g_{ab} + h_{ab}^R$ is smooth on the particle worldline $\Gamma$. Moreover, because ${}^{(0)}g_{ab}$ is a vacuum solution of the Einstein equations everywhere and $h_{ab}^R$ satisfies the homogeneous form of the perturbation equation (19), the metric ${}^{(0)}g_{ab} + h_{ab}^R$ is also a vacuum solution of the Einstein equations everywhere. This gives what Poisson (Poisson, 2005) describes as "a compelling interpretation" to the condition (25): the particle moves on a geodesic of the vacuum spacetime with metric ${}^{(0)}g_{ab} + h_{ab}^R$. Unfortunately, this metric doesn't coincide with the actual physical spacetime metric ${}^{(0)}g_{ab} + h_{ab}$.

## 2.4 Deriving the MiSaTaQuWa Equations via Modelling the Small Body as a Black Hole

This point-particle derivation of the MiSaTaQuWa equations is concise, but depends crucially on the Detweiler-Whiting postulate. In this section I outline a different derivation, based on modelling the small body as a BH. This derivation doesn't require the assumption of the Detweiler-Whiting postulate but it (this derivation) is technically much more involved than the point-particle derivation. This derivation is originally due to Mino, Sasaki, and Tanaka (Mino et al., 1997); my presentation here is based on that of Poisson's GR17 plenary lecture (Poisson, 2005).

In this section I adopt the same notation as in the point-particle derivation (Section 2.3 (page 19)) except that the small body is no longer modelled as a point particle. Because the small body is (locally) free-falling, its motion is actually independent of its internal structure (ignoring spin and tidal effects). This "effacement

---

may be able to resolve this problem (Steinbauer and Vickers, 2006), but the precise meaning of the phrase "point particle" in general relativity remains a very delicate question.

of internal structure" is a fundamental property of general relativity (*not* shared by most other relativistic gravity theories) and is discussed in detail in Damour's fascinating review article in the *Three Hundred Years of Gravitation* volume (Damour, 1987). In view of this property, we are free to choose the small body's internal structure for maximum convenience in our analysis; here we choose it to be a nonrotating (Schwarzschild) BH.

Our analysis will be based on matched asymptotic expansions of the spacetime metric: Sufficiently far from the small BH (the "far zone"), the metric is that of the background spacetime, perturbed by the presence of the small BH,

$$g_{ab} = g_{ab}^{\text{background}} + O(m/r) \tag{29}$$

Sufficiently near to the small BH (the "near zone"), the metric is that of the small (Schwarzschild) BH perturbed by the tidal field of the background spacetime,

$$g_{ab} = g_{ab}^{\text{Schwarzschild}} + O\left((r/\mathcal{R})^2\right) \ . \tag{30}$$

Since $m \ll \mathcal{R}$, there exists an intermediate "matching zone" where $m/r \ll 1$ and $r/\mathcal{R} \ll 1$, and hence both the expansions (29) and (30) are simultaneously valid. In that region these expansions must represent the *same* vacuum solution of the Einstein equations (modulo gauge choice). The small BH's motion is then determined by the matching conditions.

[The material from this point up to the start of **Section 2.4.3** (**page 22**) is somewhat more technical than the rest of this article and can be skipped without creating confusion.]

We begin by introducing suitable retarded coordinates centered on the worldline $\Gamma$: $v$ is a backwards null coordinate constant on each ingoing null cone centered on $\Gamma$ (and is equal to proper time on $\Gamma$), and $r$ is an affine parameter on the cone's null generators. In this section I use $ijk$ as Penrose abstract indices ranging over the spatial (non-$v$) coordinates only. The angular coordinates $\Omega^i = x^i/r$ are constant on each generator (one can think of $\Omega^i$ as spatial coordinates on a 2-sphere centered on the small BH).

### 2.4.1 The Near-Zone Metric

For present purposes it suffices to approximate the spacetime metric sufficiently near the small BH (the "near zone") by that of a Schwarzschild BH subject to a quadrupole perturbation.[15] In a suitable perturbation $(\bar{v}, \bar{x} = \bar{r}\bar{\Omega}^i)$ of the coordinates $(v, x^i = r\Omega^i)$, the null-null component of the near-zone perturbed spacetime metric is

$$g_{\bar{v}\bar{v}}^{\text{near}} = -\left(1 - \frac{2m}{\bar{r}}\right),$$
$$= -\bar{r}^2\left(1 - \frac{2m}{\bar{r}}\right)^2 \mathcal{E}_{ij}\bar{\Omega}^i\bar{\Omega}^j + O\left((\bar{r}/\mathcal{R})^3\right), \tag{31}$$

where $\mathcal{E}_{ij} = C_{vivj} = O(1/\mathcal{R}^2)$ are the electric components of the Weyl tensor $C_{abcd}$; these components measure the tidal distortion induced by the background spacetime.

---

[15] This quadrupole term is in general just the leading order in a multipolar expansion.

### 2.4.2 The Far-Zone Metric

Sufficiently far from the small BH (i.e., in the "far zone"), the null-null component of the background metric is

$${}^{(0)}g_{vv} = -\left(1 + 2ra_i\Omega^i + r^2\mathcal{E}_{ij}\Omega^i\Omega^j\right) + O\left((r/\mathcal{R})^3\right), \tag{32}$$

where $\mathcal{E}_{ij}$ are once again the electric components of the Weyl tensor, evaluated on $\Gamma$.

The metric perturbation $\bar{h}_{ab}$ produced by the small BH satisfies the linear perturbation equation (19), with $T_{ab} = 0$ in the far zone. Assuming that the far zone lies within a normal convex neighborhood of the small BH, it can be shown that the null-null component of the far-zone perturbed spacetime metric is

$$g_{vv}^{\text{far}} = -\left(1 - \frac{2m}{r}\right),$$
$$+ h_{uu}^{\text{tail}}$$
$$+ r\left(2m\mathcal{E}_{ij}\Omega^i\Omega^j - 2a_i\Omega^i + h_{uuu}^{\text{tail}} + h_{uui}^{\text{tail}}\Omega^i\right)$$
$$- r^2\mathcal{E}_{ij}\Omega^i\Omega^j$$
$$+ O\left((m/\mathcal{R})(r/\mathcal{R})^2\right) + O\left((r/\mathcal{R})^3\right), \tag{33}$$

Differentiating the tail term (22), we find that in terms of the original (physical) retarded Green function $\mathcal{G}^{ab}{}_{c'd'}(x,x'')$, $h_{abc}^{\text{tail}}$ is given by

$$h_{abc}^{\text{tail}} = 4m\int_{-\infty}^{x'^-}\nabla_c\left(\mathcal{G}_{aba'b'} - \tfrac{1}{2}{}^{(0)}g_{ab}\mathcal{G}^d{}_{da'b'}\right)(x,x')u^{a'}u^{b'}\,d\tau', \tag{34}$$

where the upper integration limit $x'^-$ means that the integral extends over the entire past worldline of the small BH prior to (but not including) the event $x'$. By cutting off the integration infinitesimally before $x'$ we include the (regular) tail part of the Green function and exclude the (singular) light-cone part. As a result, $h_{abc}^{\text{tail}}$ is finite (although generally only $C^0$, i.e., continuous but not differentiable) on the worldline.

### 2.4.3 Matching

The coordinate transformation between the near-zone coordinates $(\bar{v}, \bar{x}^i = \bar{r}\bar{\Omega}^i)$ and the far-zone coordinates $(v, x^i = r\Omega^i)$ can be computed explicitly (up to sufficient orders in the small-in-the-matching-zone quantities $m/\mathcal{R}$ and $r/\mathcal{R}$) in terms of $h_{ab}^{\text{tail}}$, its integrals and gradients, and $\mathcal{E}_{ij}$. Using this to transform the far-zone metric into the near-zone coordinates gives the null-null metric component

$$
\begin{aligned}
g_{\tilde{v}\tilde{v}}^{\text{far}} = & -\left(1 - \frac{2m}{\tilde{r}}\right) \\
& - 2\tilde{r}\left(a_i - \tfrac{1}{2}h_{uui}^{\text{tail}} + h_{uui}^{\text{tail}}\right)\tilde{\Omega}^i \\
& + (4m\tilde{r} - \tilde{r}^2)\mathcal{E}_{ij}\tilde{\Omega}^i\tilde{\Omega}^j \\
& + O\left((m/\mathcal{R})(\tilde{r}/\mathcal{R})^2\right) + O\left((\tilde{r}/\mathcal{R})^3\right),
\end{aligned}
\tag{35}
$$

Requiring that this match the same near-zone metric component (**31**) up to $O(m/\mathcal{R})$ now gives the 3-acceleration of the small BH's worldline as

$$
a_i = \tfrac{1}{2}h_{uui}^{\text{tail}} - h_{uiu}^{\text{tail}} ,
\tag{36}
$$

from which the MiSaTaQuWa equations (**27**) follow directly.

Although the full derivation (including all the steps I've omitted in this brief synopsis) is somewhat lengthy, it can be made quite rigorous, requiring no unproven assumptions about the physical system.

## 2.5 Other Derivations of the MiSaTaQuWa Equations

In this section I briefly mention a number of other derivations of the MiSaTaQuWa equations. In the interests of keeping this review both short and broadly accessible, I won't describe any of these derivations in detail.

As well as the matched-asymptotic-expansions derivation outlined in **Section 2.4** (**page 20**), Mino, Sasaki, and Tanaka (Mino et al., 1997) also gave another derivation based on an extension of the electromagnetic radiation-reaction analysis of DeWitt and Brehme (DeWitt and Brehme, 1960).

Quinn and Wald (Quinn and Wald, 1997) took an axiomatic approach, showing that the electromagnetic self force can be derived by (i) using a "comparison axiom" which relates the electromagnetic force acting on charged particles with the same charge and 4-acceleration in two possibly-different spacetimes, and in addition (ii) assuming that in Minkowski spacetime the half-advanced, half-retarded electromagnetic field exerts no force on a uniformly accelerating charged particle. Quinn and Wald also derived the gravitational self force (the MiSaTaQuWa equations) using a similar set of axioms.

Building on earlier work by Harte (Harte, 2006, 2008, 2009a, 2009b), Gralla, Harte, and Wald (Gralla et al., 2009) have recently provided a rigorous rederivation of the classical (DeWitt-Brehme) electromagnetic self-force based on taking the limit of a 1-parameter family of spacetimes corresponding to the small body being "scaled down" in charge and mass simultaneously. (Harte's analysis also includes a rigorous proof of a generalized form of the Detweiler-Whiting postulate for the scalar-field and electromagnetic cases, and for the linearized Einstein equations.) Gralla and Wald (Gralla and Wald, 2008) have rederived the gravitational self-force (the MiSaTaQuWa equations) based on a similar technique; here the small body is "scaled down" in size and mass simultaneously. Both of these derivations are mathematically rigorous and make no assumptions beyond the existence and appropriate smoothness and limit properties of the 1-parameter families of spacetimes.

Pound (Pound, 2010a, 2010b, 2010c) has reviewed various derivations of the MiSaTaQuWa equations (including both the ones I've outlined here, and others) and developed several new mathematical techniques for analyzing the self-force problem. Using these, he has rederived the MiSaTaQuWa equation in a highly rigorous manner.[16] His analysis includes a rigorous proof of the (gravitational) Detweiler-Whiting postulate and also provides many valuable insights into future directions for the Capra research program; I outline some of these "future directions" in **Section 5.2 (page 45)**.

### 2.6 Puncture-Function Regularizations

In this section I describe two recently-developed alternate regularization schemes for self-force computations. Both schemes begin by considering a "residual field", defined as the difference between the particle's physical field and a suitably chosen "puncture function" which approximates the particle's Detweiler-Whiting singular field near the particle. By construction, the residual field is finite (although of limited differentiability) at the particle position, and it yields the correct self-force in the force law (**2**). The residual field satisfies a scalar wave equation similar to the usual one (**1**), except that by construction the right hand side is now a nonsingular "effective source" that can be calculated analytically. I describe the puncture function, the effective source, and the basic outline of how they can be used to regularize the (singular) field equation in **Section 2.6.1 (page 25)**.

The puncture function and effective source are constructed to have certain specified properties near the particle. Their behavior far from the particle can equivalently be described as either (i) they are undefined far from the particle but the computational scheme is formulated so as not to make use of them there, or (ii) they are defined everywhere but vanish (or are negligibly small) far from the particle. Following Wardell and his colleagues (Wardell, 2009, Ottewill and Wardell, 2008, 2009a, 2009b, Wardell and Vega, 2011 and Vega et al., 2011), I use the terminology (i); note that some other authors use the terminology (ii).

Given the puncture function and effective source, Barack and Golbourn's "$m$-mode" scheme (Barack and Golbourn, 2007, Barack et al., 2007 and Dolan and Barack, 2011) does a Fourier decomposition of the resulting equation into azimuthal ($e^{im\varphi}$) modes, and uses a "world tube" technique to remove any dependence on the puncture function or effective source far from the particle. The authors then solve numerically for each $m$-mode of the residual field (using a time-domain numerical evolution in 2+1 dimensions for each mode), and compute the final self-force by summing over all the modes' contributions. I discuss the Barack-Golbourn $m$-mode scheme further in **Section 2.6.2** (**page 27**).

---

[16] At the conclusion of his main analysis, Pound writes: "This concludes what might seem to be the most egregiously lengthy derivation of the self-force yet performed.".

Vega and Detweiler (Vega and Detweiler, 2008 and Vega et al., 2009) take a different approach: Given the puncture function and effective source, they introduce a smooth "window function" to remove any dependence on the puncture function or effective source far from the particle, then numerically solve the resulting equation directly in 3+1 dimensions. I discuss the Vega-Detweiler scheme further in **Section 2.6.3** (**page 29**).

For either scheme, there are actually many possible choices for the puncture function and effective source. These differ in their tradeoffs between the difficulty of analytically calculating the puncture function and effective source, and how accurately the puncture function approximates the particle's Detweiler-Whiting singular field (and correspondingly, how small the effective source is and how smooth the puncture function and effective source are at the particle). I discuss this further in **Section 2.6.4** (**page 30**).

Throughout this section we consider a point particle of scalar charge $q$, moving along a timelike worldline $\Gamma$ in (say) Kerr spacetime, whose typical radius of curvature in a neighborhood of $\Gamma$ is $\mathcal{R}$.

### 2.6.1 The Basic Puncture-Function Scheme

In this section I describe the basic puncture-function regularization in its simplest form. This is directly applicable to the Barack-Golbourn $m$-mode scheme discussed in **Section 2.6.2** (**page 27**) but is slightly modified for the Vega-Detweiler scheme discussed in **Section 2.6.3** (**page 29**).

We take the scalar field $\phi$ to satisfy the usual scalar wave equation (**1**). In general the Detweiler-Whiting singular field $\phi^S$ isn't known analytically but, by careful analysis of the scalar field's singularity structure near the particle, we can construct approximations to the singular field. Thus, we define an "$n$th order puncture function" $\phi^{S(n)}$ as a specific approximation – one that *is* known analytically – to the Detweiler-Whiting singular field $\phi^S$ near the particle, which satisfies

$$\phi^{S(n)} - \phi^S = O\left(|\lambda|^{n-1}\right) \tag{37}$$

near the particle, where $\lambda$ is (roughly) the geodesic distance from the particle (see (Dolan and Barack, 2011) for a precise definition). Notice that at this point, the puncture function need only be defined near the particle; in practice, it's usually only defined within at most a normal convex neighborhood of the particle worldline. I discuss the construction of the puncture function further in **Section 2.6.4** (**page 30**).

We define the "residual" field

$$\phi^{R(n)} = \phi - \phi^{S(n)} \tag{38}$$

near the particle. The residual field is $C^{n-2}$ at the particle (and $C^{\infty}$ elsewhere near the particle) and satisfies the wave equation

$$\Box \phi^{R(n)} = S^{\text{eff}(n)} \quad , \tag{39}$$

with the "effective source" $S^{\text{eff}(n)}$ given by

$$S^{\text{eff}(n)} = -\Box \phi^{S(n)} - 4\pi q \int_{-\infty}^{+\infty} \frac{\delta^4 \left(x^a - \Gamma^a(\tau')\right)}{\sqrt{-g}} \, d\tau' \quad , \tag{40}$$

where as in **Section 2.1** (**page 11**), the integral extends over the entire worldline of the particle. In general the effective source is $O\left(\lambda^{n-3}\right)$ at the particle (and $C^{\infty}$ elsewhere near the particle).

The subtraction in the effective-source definition (**40**) can't be evaluated numerically (both terms are singular at the particle), but it can be evaluated analytically using a (lengthy) series-expansion analysis of the field's singularity structure. I discuss this further in **Section 2.6.4** (**page 30**).

If $\phi^{S(n)}$ is a sufficiently good approximation to the true Detweiler-Whiting singular field $\phi^S$ near the particle (i.e., if the order $n$ is large enough), and the Detweiler-Whiting postulate holds (i.e., the singular field exerts no force on the particle), then it's easy to see that $q\nabla\phi^{R(n)}$ at the particle position gives precisely the desired self-force acting on the particle. Thus (apart from the difficulties outlined in the next two paragraphs) the self-force can be calculated by analytically calculating the effective source, then numerically solving the wave equations (**39**), and then finally taking the gradient of $\phi^{R(n)}$ at the particle position.

Accurately solving the wave equation (**39**) is made more difficult by the limited differentiability of the effective source and residual field at the particle. With standard finite differencing methods, this limited differentiability limits the order of finite-differencing convergence attainable very near the particle. Current research is exploring a variety of techniques to alleviate this problem including ignoring it (i.e., simply accepting the lower order of convergence),[17] modifying the finite differencing scheme near the particle, and using finite-element or domain-decomposition pseudospectral methods that naturally accommodate well-localized non-differentiability in the fields (Sopuerta et al., 2006, Sopuerta and Laguna, 2006, Cañizares and Sopuerta, 2009a, 2009b and Vega et al., 2009).

---

[17] It's not clear to me how much of the overall numerical error occurs within a finite-difference-molecule radius of the particle. If this fraction is small at practical grid resolutions, then a lower order of convergence at those few grid points might have only a minor impact on the overall numerical accuracy.

Another difficulty with puncture-function regularization schemes is that in general the puncture function and effective source are only defined within at most a normal convex neighborhood of the particle whereas the physically appropriate boundary conditions for the wave equation (39) are applied (to the physical field $\phi$) at infinity. The Barack-Golbourn $m$-mode scheme and the Vega-Detweiler scheme take very different approaches to resolving this difficulty; I discuss these in (respectively) **Sections 2.6.2 (page 27)** and **2.6.3 (page 29)** below.

### 2.6.2 The Barack-Golbourn $m$-mode Scheme

The Barack-Golbourn $m$-mode scheme for self-force computation (Barack and Golbourn, 2007, Barack et al., 2007 and Dolan and Barack, 2011) begins by defining the puncture function $\phi^{S(n)}$ and effective source $S^{\text{eff}(n)}$ exactly as just described (**Section 2.6.1 (page 25)**). The authors then decompose the residual field $\phi^{R(n)}$ and effective source $S^{\text{eff}(n)}$ into Fourier series in the azimuthal ($\varphi$) direction,

$$
\begin{aligned}
\phi^{R(n)}(x) &= \sum_{m=-\infty}^{\infty} \phi_m^{R(n)}(t, r, \theta) e^{im\varphi} \,, \\
S^{\text{eff}(n)}(x) &= \sum_{m=-\infty}^{\infty} S_m^{\text{eff}(n)}(t, r, \theta) e^{im\varphi} \,,
\end{aligned}
\tag{41}
$$

Away from the particle, the physical scalar field $\phi$ is smooth and can be similarly decomposed,

$$
\phi(x) = \sum_{m=-\infty}^{\infty} \phi_m(t, r, \theta) e^{im\varphi}.
$$

Since we're working on a Kerr background, the wave equation (39) now separates, so that each (complex) residual-field $m$-mode $\phi_m^{R(n)}(t, r, \theta)$ satisfies a modified wave equation

$$
\Box_m \phi_m^{R(n)} = S_m^{\text{eff}(n)}
\tag{42}
$$

in 2+1 dimensions, where the operator $\Box_m$ is easily derived analytically and where the effective-source $m$ modes are given explicitly by

$$
S_m^{\text{eff}(n)}(t, r, \theta) = \frac{1}{2\pi} \int_{-\pi}^{\pi} S^{\text{eff}(n)}(t, r, \theta, \varphi') e^{-im\varphi'} \, d\varphi' \quad .
\tag{43}
$$

This integral can be done analytically in some cases, but otherwise must be done numerically.

There still remains the difficulty that the puncture function and effective source are only defined near the particle, while the physical field $\phi$ has has outgoing-wave boundary conditions at infinity. To resolve this problem, the authors introduce a worldtube (whose size is a numerical parameter, and shouldn't be "too large" in a sense described below) whose interior contains the particle worldline $\Gamma$. The authors then define a new "numerical" field

$$\phi_m^{N(n)} = \begin{cases} \phi_m^{R(n)} & \text{inside the worldtube} \\ \phi_m & \text{outside the worldtube} \end{cases}$$

and solve numerically for this. The numerical field evidently satisfies the equations

$$\Box_m \phi_m^{N(n)} = \begin{cases} S_m^{\text{eff}(n)} & \text{inside the worldtube} \\ 0 & \text{outside the worldtube} \end{cases}$$

Equivalently, one could say that the authors numerically solve the equations

$$\begin{aligned} \Box_m \phi_m^{R(n)} &= S_m^{\text{eff}(n)} \quad \text{inside the worldtube} \\ \Box_m \phi_m &= 0 \quad \text{outside the worldtube} \\ \phi_m^{R(n)} &= \phi_m - \phi_m^{S(n)} \quad \text{on the worldtube boundary} \end{aligned} \qquad (44)$$

The authors solve the piecewise modified wave equation (42) numerically for each $m$ using a standard time-domain finite-difference numerical evolution code in 2+1 dimensions.[18] The authors use arbitrary initial data on a large domain, in the same manner discussed in **Section 3.4 (page 36)** below.

[In a finite-difference numerical code, the piecewise aspect of the equations (44) is trivial to implement (Barack and Golbourn, 2007 and Dolan and Barack, 2011): the code stores the grid function $\phi_m^{N(n)}$, and for each finite differencing operation, checks if the finite difference molecule crosses the worldtube boundary. If so, then the code "adjusts" the grid function values being finite differenced (which in this case might well be a temporary copy of a molecule-sized region of the actual grid function $\phi_m^{N(n)}$) as appropriate using (44).]

With this scheme neither the puncture function nor the effective source are ever needed more than a short distance (the maximum finite-difference molecule size) outside the worldtube. Hence, so long as the worldtube isn't too large, it's not a problem that the puncture function and effective source aren't defined far from the particle. Outside the worldtube, the piecewise equations (44) reduce to $\Box_m \phi_m = 0$, so it's easy to impose the appropriate outgoing-radiation outer boundary conditions.

Finally, the authors show that the self-force $F_{\text{self}}^a$ is given by

$$F_{\text{self}}^a(\tau) = q \sum_{m=0}^{\infty} \left( \nabla^a \tilde{\phi}_m^{R(n)} \right) \Big|_{\Gamma(\tau)} , \qquad (45)$$

where the gradient is evaluated at the particle, and where the real fields $\tilde{\phi}_m^{R(n)}$ are defined by

$$\tilde{\phi}_m^{R(n)} = \begin{cases} \phi_m^{R(n)} & \text{if } m = 0 \\ 2\text{Re}\left( \phi_m^{R(n)} e^{im\varphi} \right) & \text{if } m > 0 \end{cases}$$

---

[18] Other numerical methods are of course also possible.

The infinite sum over $m$ in the self-force law (**45**) can be approximated with a finite computation using a tail-fitting procedure analogous to that described in **Section 2.1** (**page 11**).

The Barack-Golbourn $m$-mode scheme provides a practical and efficient route to self-force computations for a variety of physical systems. It is currently the basis for a number of such calculations. Where the original mode-sum scheme described in **Section 2.1** (**page 11**) reduced the self-force problem to the numerical solution of a 2-dimensional set of PDEs in 1+1 dimensions,[19] the $m$-mode scheme reduces the self-force problem to the solution of a 1-dimensional set of PDEs in 2+1 dimensions. Both schemes have the major advantage that the problem-domain size, grid resolution, and/or other numerical parameters can be varied from one PDE to another. This greatly increases the efficiency of the numerical solutions.

### 2.6.3 The Vega-Detweiler Scheme

The Vega-Detweiler scheme for self-force computation (Vega and Detweiler, 2008 and Vega et al., 2009) takes a somewhat different approach: it begins by defining the puncture function $\phi^{S(n)}$ and effective source $S^{\text{eff}(n)}$ exactly as described in **Section 2.6.1** (**page 25**). The authors then introduce a real $C^\infty$ "window function" $W$ chosen (in a manner described further below) such that

$$W = 1 + O\left((\lambda/\mathcal{R})^4\right) \tag{46}$$

near the particle, and $W \to 0$ "sufficiently fast" (i.e., $W$ is either exactly zero or has decayed to a negligible value) far from the particle, including in the wave zone and at any event horizon(s) in the spacetime.

The authors then define the residual field in a manner slightly different from the definition (**38**) of **Section 2.6.1** (**page 25**): using a subscript $W$ to denote "windowed" quantities, the authors define

$$\phi_W^{R(n)} = \phi - W\phi^{S(n)} \ . \tag{47}$$

so that the windowed residual field satisfies the wave equation

$$\Box \phi_W^{R(n)} = S_W^{\text{eff}(n)} \ , \tag{48}$$

with the windowed effective source given by

$$S_W^{\text{eff}(n)} = -\Box(W\phi^{S(n)}) - 4\pi q \int_{-\infty}^{+\infty} \frac{\delta^4\left(x^a - \Gamma^a(\tau')\right)}{\sqrt{-g}} \, d\tau' \ , \tag{49}$$

where once again the integral extends over the entire worldline of the particle. By construction, the residual field and effective source so defined have the same continuity properties at the particle as described in **Section 2.6.1** (**page 25**).

---

[19] I'm describing the time-domain case here; somewhat similar arguments would also apply to a frequency-domain solution.

In the same manner as in **Section 2.6.1** (**page 25**), if the puncture function $\phi^{S(n)}$ is of sufficiently high order and the Detweiler-Whiting postulate holds, then it's easy to see that the windowed residual-field gradient $\nabla \phi_W^{R(n)}$ at the particle position gives precisely the desired self-force acting on the particle. Thus the self-force can be calculated by analytically calculating the effective source, then numerically solving the wave equation (**48**) in 3+1 dimensions with the effective source (**49**), then finally taking the gradient of the windowed residual field $\phi_W^{R(n)}$ at the particle position.

Since the window function is chosen to approach zero "sufficiently fast" far from the particle, it's not a problem for this scheme that the puncture function and effective source aren't defined far from the particle. That is, far from the particle we have (either exactly or to an excellent approximation) $W = 0$ and hence $\phi_W^{R(n)} = \phi$ and $S_W^{\text{eff}(n)} = 0$, so the wave equation (**48**) becomes simply $\Box \phi = 0$. This also makes it easy to impose the appropriate outgoing-radiation outer boundary conditions on $\phi$.

Like the Barack-Golbourn $m$-mode scheme, the Vega-Detweiler scheme provides a practical and efficient route to self-force computations for a variety of physical systems. The Vega-Detweiler scheme is designed to reduce the self-force problem to the numerical solution of a (single) wave equation in 3+1 dimensions. This type of problem is quite similar to that solved by many existing 3+1 numerical relativity codes, so the Vega-Detweiler scheme can often reuse existing numerical-relativity codes and/or infrastructure.

### 2.6.4 Constructing the Puncture Function and Effective Source

The key to the success of puncture-function schemes (in either the Barack-Golbourn or Vega-Detweiler variants) is the construction of the puncture function $\phi^{S(n)}$. This essentially requires a careful local analysis of the field's singularity structure near the particle. This can be done exactly only in very simple cases (for example, for a static particle in Schwarzschild spacetime). In more general cases, such an analysis uses lengthy series expansions and (particularly for higher orders $n$) is usually done using a symbolic algebra system. Once the puncture function is known, the effective source can then be computed (again symbolically) directly from the definition (**40**). The resulting algebraic expressions are very lengthy, so usually the symbolic algebra system is also used to directly generate C or Fortran for inclusion in a numerical code.

The difficulty (complexity of the expressions) in computing the puncture function and effective source in this way rises very rapidly with the puncture function's order $n$. In practice, 4th order seems to be both practical and a good compromise between the difficulty of computing the puncture function and effective source, the expense of evaluating the resulting (machine-generated C or Fortran) expressions in a numerical code, and the differentiability (and hence order of accuracy) of the numerical solution.

Wardell and his colleagues (Wardell, 2009, Ottewill and Wardell, 2008, 2009a, 2009b, Wardell and Vega, 2011 and Vega et al., 2011) have developed efficient software for computing puncture functions and their corresponding effective sources at (in theory) any order, and these are now being used in a number of self-force research projects. In the interests of brevity I won't try to describe the details of how these puncture functions are calculated, but Wardell and Vega (Wardell and Vega, 2011) give a very clear description of this.

## 2.7 Conservative versus Dissipative Effects

The self force can be decomposed into conservative (time-symmetric) and dissipative (time-antisymmetric) parts. This decomposition is an important conceptual tool for understanding the physical meaning of the self-force. This decomposition is also important for practical computations, for reasons described below.

To actually compute the conservative and dissipative parts of the self force, consider that thus far, we have used solely the *retarded* scalar field $\phi^{\mathrm{ret}} := \phi$ (or metric perturbation $h_{ab}^{\mathrm{ret}} := h_{ab}$), and our goal has been to compute the corresponding *retarded* self-force $F_{\mathrm{ret}}^a := F_{\mathrm{self}}^a$. If we introduce an *advanced* scalar field $\phi^{\mathrm{adv}}$ (or metric perturbation $h_{ab}^{\mathrm{adv}}$) and the corresponding advanced self-force $F_{\mathrm{adv}}^a$ (both computed in a manner that's the time-reversal of that for the corresponding retarded quantities), then as described in more detail by Dolan and Barack (Dolan and Barack, 2011), the conservative part of the self-force $F_{\mathrm{cons}}^a$ and the dissipative part $F_{\mathrm{diss}}^a$ can easily be computed via

$$F_{\mathrm{cons}}^a = \tfrac{1}{2}(F_{\mathrm{ret}}^a + F_{\mathrm{adv}}^a) \tag{50}$$

$$F_{\mathrm{diss}}^a = \tfrac{1}{2}(F_{\mathrm{ret}}^a - F_{\mathrm{adv}}^a) \tag{51}$$

This decomposition can also be performed mode-by-mode in a mode-sum or $m$-mode calculation. In some cases there are also ways of computing this decomposition without needing to explicitly compute the advanced self-force; I describe one such scheme in **Section 3.6 (page 38)**.

The Detweiler-Whiting singular field is time-symmetric, so it cancels out in the subtraction (**51**) and hence doesn't affect the dissipative part of the self-force. This means that the dissipative part can be computed without regularizing the singular field, i.e., given a suitable computational scheme, the dissipative part can be computed much more easily than the conservative part. In mode-sum and puncture-function regularization schemes, the dissipative part of the mode sums also converges much faster (exponentially instead of polynomially) than the conservative part.

The dissipative part of the self-force directly causes secular drifts in the small body's orbital energy, angular momentum, and (for non-equatorial orbits in Kerr space-time) Carter constant. Mino (Mino, 2003, 2005b, 2005a, 2006, 2008) has argued that dissipative self-force alone can be used to calculate the correct long-term (secular) orbital evolution of an EMRI system: the conservative part of the self-force appears to cause only quasi-periodic oscillations in the orbital parameters, not long-term secular drifts. This "adiabatic approximation" is very useful and can provide a route to EMRI orbital evolution that's much simpler and more efficient than the full Capra calculations that are the main subject of this article.

However, Drasco and Hughes (Drasco and Hughes, 2006) and Pound and Poisson (Pound et al., 2005 and Pound and Poisson, 2008a) have found that the adiabatic approximation isn't as accurate as had previously been thought. In particular, they have found that conservative effects also lead to long-term secular changes in the orbital motion. Huerta and Gair (Huerta and Gair, 2009) have recently estimated the magnitude of these latter effects for a quasicircular EMRI inspiral. In their approximate model of a LISA EMRI whose GWs accumulate $\sim 10^6$ radians of phase in the last year of inspiral, conservative effects contribute $\sim 20$ radians during this time interval. Conservative effects are likely to be much larger for eccentric EMRI inspirals.[20] While a small fraction of the total phase, even the quasicircular-inspiral conservative effects are still large enough to be easily measurable by LISA (in **Section 4.1 (page 40)** I estimate that LISA will be able to detect GW phase differences as small as $\sim 10^{-2}$ radians). Thus conservative effects must be included to model EMRIs sufficiently accurately for LISA.

Pound and Poisson (Pound et al., 2005 and Pound and Poisson, 2008a) have also drawn useful distinctions between "adiabatic", "secular", and "radiative" approximation schemes, which have often been confused in the past.

## 3 Self-Force via the Barack-Ori Mode-Sum Regularization

In this section I summarize the recent work of Barack and Sago (Barack and Sago, 2010) in which they calculate the $O(\mu)$ gravitational self-force on a particle in an arbitrary (fixed) bound geodesic orbit in Schwarzschild spacetime.

The calculation is done in the Lorenz gauge, decomposing the metric perturbation due to the particle into tensor spherical harmonics, solving for the $\ell = 0$ and $\ell = 1$ harmonics via a frequency-domain method, for the $\ell \geq 2$ harmonics by numerically evolving a 1+1-dimensional wave equation in the time domain, and then computing the final self-force using the mode-sum regularization described in **Section 2.1 (page 11)**.

---

[20] Building on their self-force calculation for arbitrary bound geodesic orbits in Schwarzschild spacetime (Barack and Sago, 2010) (discussed in detail in **Section 3 (page 32)**), Barack and Sago (Barack and Sago, 2011) have recently studied conservative self-force effects for eccentric orbits in Schwarzschild spacetime.

This calculation marks a major milestone in the Capra research program and uses techniques typical of many other self-force calculations using time-domain integration of the mode-sum–regularized perturbation equations. This calculation also illustrates something of the (high) level of complexity involved in self-force calculations for astrophysically "interesting" physical systems – the authors report that (even after many years of preparatory research) it took over 2 years to develop and debug the techniques and computer code for this calculation.

In this section I use $\ell m$ as spherical-harmonic indices, and $ijk$ as "tensor component" indices ranging from 1 to 10 (these indices are always enclosed in parentheses, and index the individual coordinate components of symmetric rank-2 4-tensors).

### 3.1 Particle Orbit

I take the Schwarzschild line element to be

$$
\begin{aligned}
ds^2 &= -f\,dt^2 + f^{-1}\,dr^2 + r^2(d\theta^2 + \sin^2\theta\,d\varphi^2)\,, \\
&= -f\,du\,dv + r^2(d\theta^2 + \sin^2\theta\,d\varphi^2)\,,
\end{aligned}
\tag{52}
$$

where $M$ is the mass of the Schwarzschild spacetime, $f = 1 - 2M/r$, $(t, r, \theta, \varphi)$ are the usual Schwarzschild coordinates, and $(v, u)$ are null coordinates defined by $v = t + r_*$ and $u = t - r_*$, where

$$
r_* = r + 2M \log\left|\frac{r}{2M} - 1\right|
\tag{53}
$$

is the "tortise" radial coordinate.

Without loss of generality the authors take the particle orbit to lie in the equatorial plane $\theta_p = \frac{\pi}{2}$. The orbit may be parameterized by its (dimensionless) semi-latus rectum $p$ and eccentricity $e$, defined by

$$
\begin{aligned}
p &= \frac{2r_{\min}r_{\max}}{M(r_{\min} + r_{\max})}\,, \\
e &= \frac{r_{\max} - r_{\min}}{r_{\max} + r_{\min}}\,,
\end{aligned}
\tag{54}
$$

where $r_{\min}$ and $r_{\max}$ are the minimum and maximum $r$ coordinate along the orbit. (For a circular geodesic orbit, $p = r/M$ and $e = 0$.) Fig.3 shows the $(p, e)$ corresponding to unstable, marginally stable, and stable orbits.

We normalize $\tau$ (proper time along the particle worldline) to be zero at a (i.e., at some arbitrary) periastron passage $r = r_{\min}$. The particle's geodesic motion $x^a = x^a(\tau; p, e)$ can then be computed by integrating an appropriate set of ODEs (Sago, 2009), or semi-analytically in terms of elliptic integrals.

### 3.2 Mode-Sum Regularization

Using the mode-sum regularization discussed in **Section 2.1 (page 11)**, Barack and Ori (Barack and Ori, 2000), Barack (Barack, 2001), and Barack *et al* (Barack et al., 2002), have shown that the (Lorenz-gauge) 4-vector gravitational self-force $F^a$ at any event along the particle's worldline is given by

$$F^a = \sum_{\ell=0}^{\infty} F_{\text{reg}}^{a\ell} \ , \tag{55}$$

where the "regularized self-force mode" $F_{\text{reg}}^{a\ell}$ is given by

$$F_{\text{reg}}^{a\ell} = F_{\text{full},\pm}^{a\ell} - \left( A_{\pm}^a (\ell + \tfrac{1}{2}) + B^a \right) \ , \tag{56}$$

where the "full self-force mode" $F_{\text{full},\pm}^{a\ell}$ is computed for each $\ell$ as described in **Section 3.3** (**page 36**), the $\pm$ refers to two different ways of doing this computation (taking one-sided radial derivatives of the metric perturbation at the particle from either the outside or inside), and $A_{\pm}^a$ and $B^a$ are "regularization parameters" given semi-analytically in terms of certain elliptic integrals of the orbit parameters. Each full self-force mode $F_{\text{full},\pm}^{a(\ell)}$ is itself finite at the particle, but their sum diverges whereas the regularized sum (**55**) converges.

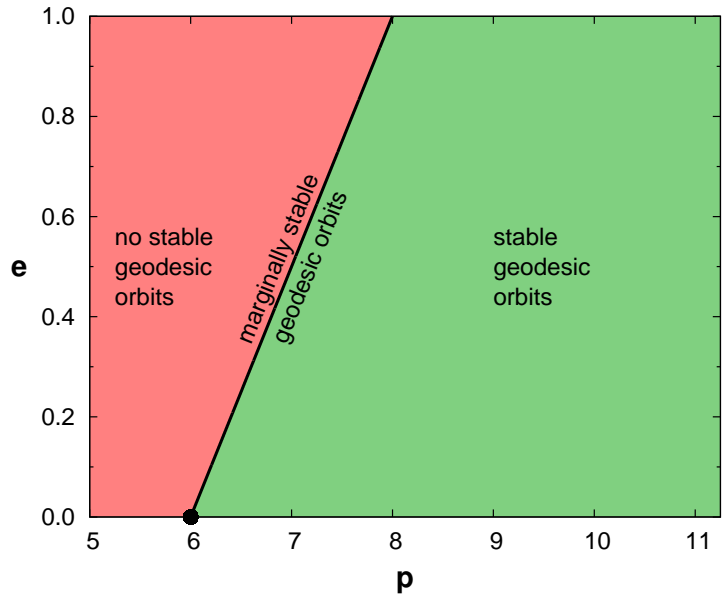

**Fig. 3** This figure shows the $(p, e)$ parameter space for bound geodesic orbits in Schwarzschild spacetime. The region $p > 6 + 2e$ where stable geodesic orbits are possible is shown in green. In the region shown in red ($p < 6 + 2e$) there are no stable geodesic orbits, only unstable "plunge" ones. The point at $p=6$, $e=0$ marks the innermost stable circular orbit (ISCO). The line $p = 6 + 2e$ marks the locus of marginally stable orbits, while zoom-whirl orbits are those just to the right of this line.

The computation of $F_{\text{full},\pm}^{a(\ell)}$ is based on a tensor-spherical-harmonic decomposition of the Lorenz-gauge metric perturbation induced by the particle. Let $g_{ab}$ be the background Schwarzschild metric (used to raise and lower all indices in this section), $g$ be the determinant of $g_{ab}$, $\nabla$ be the corresponding (background) covariant derivative operator, $h_{ab}$ be the physical (retarded) metric perturbation due to the particle, $h = h_\mu{}^\mu$ be the trace of $h_{ab}$, and $\bar{h}_{ab} = h_{ab} - \frac{1}{2} g_{ab} h$ be the trace-reversed metric perturbation. We take $h_{ab}$ to satisfy the Lorenz gauge condition

$$\nabla^a \bar{h}_{ab} = 0 \quad . \tag{57}$$

Let $u^a$ be the particle's 4-velocity.

To first order in $h_{ab}$, the linearized Einstein equations are then

$$\nabla^c \nabla_c \bar{h}_{ab} + 2 R_a{}^c{}_b{}^d \bar{h}_{cd} = -16\pi T_{ab} \tag{58}$$

where

$$T_{ab} = \mu \int_{-\infty}^{\infty} \frac{u_a u_b \, \delta^{(4)}(x^c - x_p^c)}{\sqrt{-g}} \, d\tau \tag{59}$$

is the particle's ($\delta$-function) stress-energy tensor.

Due to the symmetry of the Schwarzschild background, the linearized Einstein equations (58) (together with the added constraint-damping terms discussed in **Section 3.4 (page 36)**) are separable into tensorial spherical harmonics via the *ansatz*

$$\bar{h}_{ab} = \frac{\mu}{r} \sum_{\ell=0}^{\infty} \sum_{m=-\ell}^{+\ell} \sum_{i=1}^{10} a_\ell^{(i)} \, \bar{h}^{(i)\ell m}(r,t) \, Y_{ab}^{(i)\ell m}(\theta, \varphi; r) \quad , \tag{60}$$

and similarly for $T_{ab}$.

The resulting separated equations take the form of the coupled linear wave equations

$$\Box \bar{h}^{(i)\ell m} + \sum_{j,a} \mathcal{N}_{(j)a}^{(i)\ell} \partial_a \bar{h}^{(j)\ell m} + \sum_j \mathcal{M}_{(j)}^{(i)\ell} \bar{h}^{(j)\ell m} = S^{(i)\ell m} \delta(r - r_p) \quad , \tag{61}$$

where $\Box$ is the 2-D scalar wave operator on the Schwarzschild background, and where $\mathcal{N}_{(j)a}^{(i)\ell}$, $\mathcal{M}_{(j)}^{(i)\ell}$, and $S^{(i)\ell m}$ are given analytically as known functions of the indices ($i$) and/or ($j$), position, $u^a$, and $\ell$ and $m$.[21]

The authors solve the wave equations (61) numerically to obtain the Lorenz-gauge metric perturbation modes $\bar{h}^{(i)\ell m}$ and their gradients along the particle worldline. I describe this numerical solution in **Section 3.4 (page 36)**.

---

[21] The reader is warned that my notation here differs from that of Barack and Sago: I make all derivatives explicit in the wave equations (61), so that $N$ and $M$ are algebraic *coefficients*, whereas Barack and Sago use $\mathcal{M}$ to denote a single set of 1st-order *differential operators* which contains both the 1st derivative and 0th derivative terms in the wave equations (61).

### 3.3 The Full Force Modes

Given the Lorenz-gauge metric perturbation modes $\bar{h}^{(i)\ell m}$ in a neighborhood of the particle worldline, the authors next compute a set of coefficients $f^{a\ell m}_{(k)\pm}$ defined along the particle worldline in terms of (analytically-known) linear combinations of $\bar{h}^{(i)\ell m}$, $\partial_{r\pm}\bar{h}^{(i)\ell m}$, and $\partial_{t\pm}\bar{h}^{(i)\ell m}$, where the $\pm$ corresponds to taking one-sided derivatives from the outside or inside of the particle orbit respectively. (Due to the $\delta$-function source term in the wave equation (61), $\bar{h}^{(i)\ell m}$ is typically $C^0$ near the particle, i.e., $\bar{h}^{(i)\ell m}$ is continuous at the particle worldline but its 1st derivatives have a jump discontinuity there.)

Taking into account the *tensor* spherical harmonic expansion of $\bar{h}_{ab}$ and $T_{ab}$ (as compared to the *scalar* spherical harmonic expansion implicit in the definition of the $f^{a\ell m}_{(k)\pm}$), the authors then compute

$$F^{a(\ell)}_{\text{full},\pm} = \frac{\mu^2}{r_p^2} \sum_{m=-\ell}^{+\ell} \left( \sum_{p=-3}^{+3} \mathcal{F}^{a(\ell+p)m}_{p\pm} \right) Y^{\ell m}(\theta_p, \varphi_p) \ , \tag{62}$$

where each $\mathcal{F}^{a\ell m}_{p\pm}$ is a certain (analytically-known) linear combination of the $f^{a\ell m}_{(k)\pm}$ with the same $\ell$ and $m$. Because of the definition (62) – and more generally because of the decomposition of tensor spherical harmonics into scalar spherical harmonics – a given full force mode $F^{a(\ell)}_{\text{full},\pm}$ depends on the Lorenz-gauge metric perturbation modes $\bar{h}^{(i)\ell' m}$ for $\ell-3 \leq \ell' \leq \ell+3$.

### 3.4 Numerical Solution of the Wave Equations (61)

For each $(\ell, m)$ the authors solve the 10 coupled wave equations (61) numerically for the 10 $\bar{h}^{(i)\ell m}$ fields, using 4th order finite differencing on a uniform characteristic (double-null) $(v, u)$ grid.

The authors' "diamond integral" finite differencing scheme is adapted from those of (Lousto, 2005 and Haas, 2007). Since the $\delta$-function source term in the wave equation (61) is nonzero only on the particle's worldline, grid cells away from the worldline have no source-term contribution, allowing a relatively straightforward finite differencing scheme. The handling of the source term for those grid cells which are intersected by the particle worldline – or where the finite difference molecule is intersected by the particle worldline – is much more complicated, particularly since (for a non-circular orbit) the particle generally crosses grid cells obliquely, with no particular symmetry.

Several other aspects of the numerical solution are worth of note here:

- The authors found that a direct numerical solution of the wave equations (61) was unstable, with rapidly growing violations of the Lorenz gauge constraint (57). Following Barack and Lousto (Barack and Lousto, 2005), the authors added constraint-damping terms to the evolution equations so as to dynamically damp these gauge violations.

- The correct initial data for the wave equations (**61**) isn't known. Instead, the authors use zero initial data. This results in the evolution initially being dominated by spurious radiation induced by the imperfect initial data. Fortunately, this spurious radiation dies out (radiates away) within a few orbital periods, so in a sufficiently long evolution its influence eventually becomes negligible.[22]

- As always for mode-sum schemes, the numerical calculations are done independently for each $\ell$ and $m$. This makes the calculation trivial to parallelize.

- The length of evolution required (or equivalently, given the authors' characteristic grid setup, the size of the grid) isn't known *a priori*. Rather, the evolution must be long enough (the grid must be large enough) for the initial-data spurious radiation to have decayed to a sufficiently small level and, more generally, for the $\bar{h}^{(i)\ell m}$ field configuration to have reached an equilibrium. In practice, the authors monitor $\bar{h}^{(i)\ell m}$ and its gradient along the particle worldline, and stop the evolution once these become periodic (with the particle-orbit period) to within a numerical error threshold of $\sim 10^{-4}$. If the fields don't meet this criterion before the evolution ends, the authors increase the size of the grid and rerun the evolution.

### 3.5  Monopole and Dipole Modes

The authors were unable to obtain stable numerical evolutions of the wave equations (**61**) for $\ell = 0$ or $\ell = 1$. Instead, they (Barack, Ori, and Sago (Barack et al., 2008)) used a frequency-domain method to solve for $\bar{h}^{(i)\ell m}$ in these cases.

Because $\bar{h}^{(i)\ell m}$ is only $C^0$ at the particle worldline ($\bar{h}^{(i)\ell m}$ is continuous at the particle worldline but its 1st derivatives have a jump discontinuity there), a naive frequency-domain method would have very poor convergence due to Gibbs-phenomenon oscillations. The authors (Barack, Ori, and Sago (Barack et al., 2008)) have developed an elegant solution to this problem, using the *homogeneous* modes of the wave equations (**61**) as a basis for the numerical solution. They report that this "method of extended homogeneous solutions" works very well, with the resulting frequency-domain Fourier sums converging exponentially fast to the desired $\bar{h}^{(i)\ell m}$.

---

[22] Recently Field, Hesthaven, and Lau (Field et al., 2010) suggested that some effects of the spurious radiation would in fact *not* die out even after long evolutions. However, Jaramillo, Sopuerta, and Cañizares (Jaramillo et al., 2011) argue that such "Jost junk solutions" are an artifact of a particular (inconsistent) implementation of the $\delta$-function source term. In practice, almost all time-domain mode-sum self-force calculations – including the Barack-Sago one being presented here – ignore this issue with no apparent ill effect. Thornburg (Thornburg, 2010a, 2010b) calculated the self-force to $\lesssim 1$ part per million relative accuracy using a time-domain mode-sum code which ignored the possibility of Jost (junk) solutions, suggesting that the Jost-solution errors, if present, are very small.

### 3.6 Conservative and Dissipative Parts of the Self-Force

In general, the technique described in **Section 2.7 (page 31)** for decomposing the self-force into conservative and dissipative parts requires explicitly (numerically) computing the advanced metric perturbation $h_{ab}^{\mathrm{adv}}$ as well as the usual retarded metric perturbation $h_{ab}^{\mathrm{ret}}$. This essentially doubles the overall computational effort.

As an alternative, the authors describe another way of computing the conservative and dissipative parts of the self-force using only the usual (retarded) self-force, assuming only that the particle orbit is periodic with a single intrinsic frequency. (This is the case for the authors' system of an arbitrary bound geodesic particle orbit in Schwarzschild spacetime, as well as for some types of orbits in Kerr spacetime.) Given this condition, the authors extend an argument of Hinderer and Flanagan (Hinderer and Flanagan, 2008) to infer that

$$
\begin{aligned}
F_{\mathrm{adv}}^{t}(\tau) &= -F_{\mathrm{ret}}^{t}(-\tau) , \\
F_{\mathrm{adv}}^{r}(\tau) &= +F_{\mathrm{ret}}^{r}(-\tau) , \\
F_{\mathrm{adv}}^{\theta}(\tau) &= +F_{\mathrm{ret}}^{\theta}(-\tau) , \\
F_{\mathrm{adv}}^{\varphi}(\tau) &= -F_{\mathrm{ret}}^{\varphi}(-\tau) .
\end{aligned}
\tag{63}
$$

Assuming that the usual $F_{\mathrm{ret}}^{a}$ is known for an entire particle orbit, the conservative and dissipative parts of the self-force can then be calculated via (**50**). As noted in **Section 2.7 (page 31)**, this decomposition can also be performed mode-by-mode.

### 3.7 The $\ell$ Sum

The mode sum (**55**) is an *infinite* sum. For computational purposes a finite expression is required. To this end, the authors partition the mode sum, rewriting (**55**) as

$$
F^{a} = \sum_{\ell=0}^{\ell_{\max}} F_{\mathrm{reg}}^{a(\ell)} + \sum_{\ell=\ell_{\max}+1}^{\infty} F_{\mathrm{reg}}^{a(\ell)} ,
\tag{64}
$$

where $\ell_{\max} \sim 15$ is a numerical parameter, and compute the first term numerically. (This is the main computation; recall that it requires numerically solving the wave equations (**61**) for $0 \le \ell \le \ell_{\max}+3$.)

To estimate the second term in the partitioned mode sum (**64**), the authors consider the conservative and dissipative parts of the self-force separately. For the conservative part, the authors make use of the known large-$\ell$ asymptotic series

$$
F_{\mathrm{cons,reg}}^{a(\ell)} = \frac{D_{2}^{a}}{(\ell + \frac{1}{2})^{2}} + \frac{D_{4}^{a}}{(\ell + \frac{1}{2})^{4}} + \frac{D_{6}^{a}}{(\ell + \frac{1}{2})^{6}} + \cdots ,
\tag{65}
$$

where the coefficients $D_{k}^{a}$ don't depend on $\ell$. The authors least-squares fit the first two terms in this series to the numerically computed values of $F_{\mathrm{cons,reg}}^{a(\ell)}$ for $\ell_{\min} \le \ell \le \ell_{\max}$, where $\ell_{\min} \sim 10$ is another numerical parameter. Given the fitted coefficients $\{D^a_2, D^a_4\}$, the second term of the partitioned mode sum (64) can then be estimated in terms of polygamma functions.

The dissipative part of the mode sum (64) converges much faster (in fact, exponentially fast) and is thus much easier to handle numerically: in practice, $F^{a(\ell)}_{\text{diss,reg}}$ falls below the numerical error even before $\ell = \ell_{\text{max}}$, so the second term in the partitioned mode sum (64) is negligible.

### 3.8 Results and Discussion

The basic result of the authors' computations is the 4-vector Lorenz-gauge gravitational self-force $F^a$ as a function of time along (around) the particle orbit. Fig. 4 shows an example of these results.

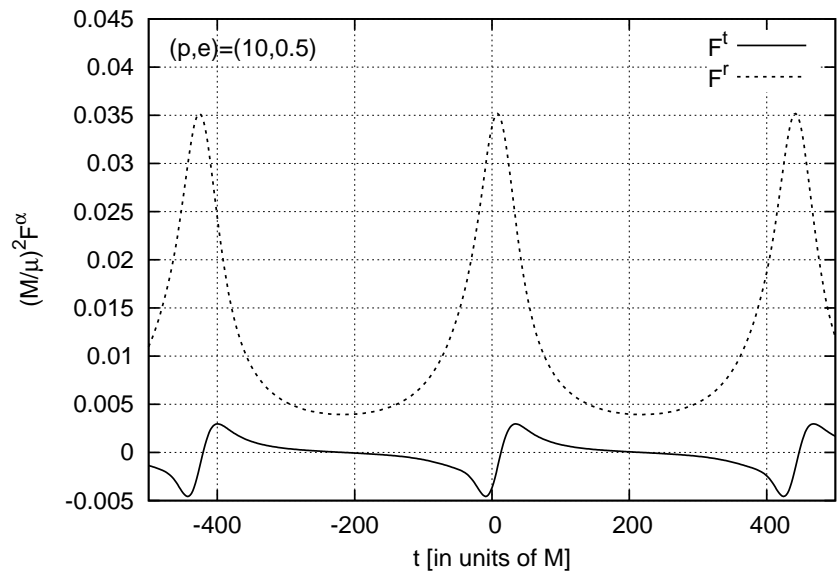

**Fig. 4** This figure shows the self-force components $F^r$ (dashed line) and $F^t$ (solid line) as a function of scaled Schwarzschild time $t/M$, for a particle orbit with $p = 10$ and $e = 0.5$ (so that $r_{\text{min}} = 6\frac{2}{3}M$ and $r_{\text{max}} = 20M$). The orbital period is $434M$. Notice the slight asymmetry of the self-force with respect to the orbit (for example, $F^r$ peaks slightly *after* the particle's periastron passage at $t/M = 0$). This is a genuine physical effect, not a numerical artifact.

The authors also studied zoom-whirl orbits, finding and analyzing interesting behavior of the self-force during the whirl phase of the orbit. In the interests of brevity, I won't discuss this phenomenon here.

The authors have also used their code to make the first calculation of the $O(\mu)$ self-force corrections to the location and angular frequency of the ISCO (Barack and Sago, 2009). This calculation is numerically quite delicate, since the ISCO is a singular point in the $(p, e)$ space of particle orbits (Fig. 3). The authors used two different techniques to make the calculation, and also made a number of other tests to validate the accuracy of their results. They found that self-force effects shift the ISCO inwards, from $r = 6M$ to $(6 - 3.269 \, \mu)M$, and slightly raise its frequency, from $\Omega = 1/(6^{3/2}M)$ to $(1 + 0.4869 \, \mu/M)\big/(6^{3/2}M)$.

These results are of great interest, both as a benchmark of the current state of the Capra research program and for comparison with other approaches to modelling EMRI dynamics. Notably, they can usefully be compared with post-Newtonian (PN) and effective one-body (EOB) calculations (Damour, 2010 and Barack et al., 2010). As well as providing valuable tests of each formalism, this can help to "calibrate" various undetermined coefficients in the PN and EOB expansions.

## 4 Accuracy

In trying to model EMRI orbital dynamics and calculate EMRI GW templates, it's essential to know what accuracy (in the GW phase) is needed, and what accuracy is achievable with various approximation schemes. In this section I briefly discuss these issues.

### 4.1 The Accuracy Needed by LISA

Matched filtering of the entire years-long LISA data stream would be impractically expensive for *detecting* EMRIs with hitherto-unknown parameters (Gair et al., 2004, section 3). However, once EMRIs have been detected by more economical search algorithms (Amaro-Seoane et al., 2007 and Porter, 2009), precision modelling and matched filtering of the full LISA data set[23] become practical and even essential to allow detecting and characterizing weaker sources in the presence of strong EMRIs.

The strongest LISA EMRIs may have signal/noise ratios of up to $\rho \sim 100$ after matched filtering (Amaro-Seoane et al., 2007 and Porter, 2009), so phase differences on the order of $1/\rho$ radians should be just detectable in matched filtering. If we want the maximum possible science return from the LISA mission, i.e., if we wish to avoid having this science return limited by the finite accuracy of our GW templates, then these templates should have GW phase errors of somewhat less than 10 milliradians. Any phase errors larger than this run the risk of significantly increasing the overall parameter-estimation error budget. Indeed, if we are lucky and LISA detects a very strong EMRI with (say) $\rho \sim 300$, the allowable GW phase errors may be even smaller, perhaps $\sim 1$–2 milliradians.

---

[23] Flanagan and Hinderer (Éanna É. Flanagan and Hinderer, 2010) have recently found that many LISA EMRI inspirals will include several strong transient resonance crossings. The EMRI osculating-geodesic orbital state vector exiting such a resonance crossing is a very sensitive function of the orbital state vector entering the resonance crossing, with the Jacobian $\partial(\text{post} - \text{resonance state})\big/\partial(\text{pre} - \text{resonance state}) \sim \mu^{-1/2} \sim 300$ for a canonical $10^6 : 10 M_\odot$ EMRI. This may limit precision modelling and matched filtering to the intervals between strong resonance crossings, i.e., to perhaps $O(1/3)$ of the full EMRI-inspiral data set.

As a rough approximation, a typical LISA EMRI accumulates $\sim 10^6$ radians of orbital phase during its last year of inspiral (Amaro-Seoane et al., 2007), so maintaining a phase error of somewhat less than 10 milliradians implies a fractional error of somewhat less than 10 parts per billion[24] in the instantaneous GW frequency (whose integral gives the cumulative GW phase), and hence also in the instantaneous EMRI orbital frequency.

It's non-trivial to translate "required accuracy in the instantaneous EMRI orbital frequency" into "required accuracy in a self-force calculation", but we can make a crude estimate using the recent analyses of Huerta and Gair (Huerta and Gair, 2009, table 1). As discussed in **Section 2.7 (page 31)**, for a quasicircular inspiral they estimated that $O(\mu^2)$ self-force effects (an $O(\mu) \sim 10^{-5}$ fraction of the overall self-force) contribute $\sim 20$ radians to the cumulative GW phase of our typical LISA EMRI. This suggests that a GW phase error tolerance of $\lesssim 10$ milliradians corresponds to a fractional accuracy of roughly

$$\frac{10 \text{ milliradians}}{20 \text{ radians}} \times 10^{-5} = 5 \times 10^{-9} \tag{66}$$

in the overall self-force. This is only a crude estimate, but it does suggest the general order of magnitude of self-force computation accuracy needed to match LISA's data quality for strong EMRIs.

### 4.2 High-Accuracy Capra Computations

How accurate is (will) (can) a Capra-based GW template be? This obviously depends on many factors. Astrophysically, there are a variety of possible perturbations to the vacuum–Einstein-equations (Kerr + compact-object) model used in most Capra calculations to date: magnetic fields, accretion disks (Giampieri, 1993, Chakrabarti, 1996, Narayan, 2000, Levin, 2007 and Barausse and Rezzolla, 2008), or even another supermassive BH within a few tenths of a parsec (Yunes et al., 2011). It remains an open research problem to incorporate such perturbations within Capra models.

Within vacuum–Einstein-equations Capra models, there are two major sources of error in our models of EMRI dynamics and GW emission/propagation:

- Our Capra computational schemes are based on approximations to the Einstein equations. For example, the use of 1st order BH perturbation theory implies fractional errors in self-force effects of at least $O(\mu) \sim 10^{-5}$ due to neglected 2nd order effects; 2nd order BH perturbation theory should bring these errors down to $O(\mu^2) \sim 10^{-10}$, subject to the long-time-approximation issues discussed in **Section 4.3** (**page 42**) below. No practical 2nd-order Capra computational schemes exist yet; I discuss prospects for their construction in the future in **Section 5.2** (**page 45**).

---

[24] I use the North American definition that billion = $10^9$.

- In practice, we numerically solve our equations using finite-precision arithmetic, and using discrete approximations to the ordinary or partial differential equations (ODEs or PDEs). As discussed in **Section 2.1** (**page 11**), frequency-domain methods only require integrating ODEs; these are potentially very accurate. For example, Detweiler, Messaritaki, and Whiting (Detweiler et al., 2003) achieved fractional numerical errors $\lesssim 1.5 \times 10^{-8}$ in a computation of scalar-field self-force acting on a scalar particle in a circular orbit in Schwarzschild spacetime, and Blanchet *et al* (Blanchet et al., 2010a) achieved fractional numerical errors $\lesssim 10^{-13}$ in a computation of the gravitational self-force for a point mass in a circular orbit in Schwarzschild spacetime. However, time-domain methods require numerically solving PDEs, and have typical fractional numerical accuracies of at best $\lesssim 10^{-4}$, although Thornburg (Thornburg, 2010a, 2010b) achieved fractional accuracies $\lesssim 10^{-6}$ in a time-domain code by combining adaptive mesh refinement with extended-precision floating-point arithmetic.

For very high accuracy *all* error sources need to be small, i.e., the EMRI's astrophysical environment must be accurately modelled *and* the Capra computational scheme must accurately approximate the Einstein equations (probably using 2nd order BH perturbation theory and a long-time approximation of the type described below) *and* the numerical computations must be very accurate.

### 4.3 Long-Time Approximation Schemes

Most Capra research to date has used 1st order BH perturbation theory *and* taken the small body to move on a fixed (timelike) geodesic worldline in the background Schwarzschild or Kerr spacetime. Assuming some computational scheme for the self-force, an obvious improvement is to use the computed self-force to perturb the particle's worldline away from being a geodesic, updating a "deviation vector" with the $O(\mu)$ equations of motion (**28**). Unfortunately, the worldline is fixed by the 1st-order Bianchi identity, so it's not obvious that such a scheme can be self-consistent. As very clearly described by Pound (Pound, 2010b), a "gauge relaxation" technique can be used to allow the particle worldline to vary, but such a scheme can still be valid for at most a short time: since the particle's orbit gradually shrinks due to GW emission, the particle's orbital phase differs from that of a reference geodesic by $\sim 1$ radian after the "dephasing time", which is quite short – $O(\mu^{-1/2})$. At this point the deviation vector is large, and the whole approximation scheme breaks down.

A much more sophisticated orbital-evolution scheme is needed to avoid this problem, i.e., to remain accurate for the (long) orbital-decay timescales $O(\mu^{-1})$. Hinderer and Flanagan (Hinderer and Flanagan, 2008) and Pound (Pound, 2010b) discuss various aspects of how such schemes might be constructed. The detailed definition and implementation of such schemes remains a topic for future research.

# 5 Summary and Future Prospects

## 5.1 Past Light Cone

A major area of Capra research has long been the effort to analyze the singularity structure of the scalar-field, electromagnetic, and gravitational perturbations induced by point charges/masses. Much of our current understanding of this structure is based on a decomposition due to Detweiler and Whiting (Detweiler and Whiting, 2003). As discussed in **Section 2.2** (**page 15**), the Detweiler-Whiting decomposition splits the perturbation into a singular part (which is, in a suitable sense, spherically symmetric at the particle) and a "radiative" part which is finite at the particle. In this context it's common to assume the Detwiler-Whiting postulate, which asserts that by virtue of its symmetry the singular field exerts no force on the particle; self-force effects arise solely from the particle's interaction with the radiative field. This postulate has recently been rigorously proved by Harte (Harte, 2006, 2008, 2009a, 2009b) and Pound (Pound, 2010a, 2010b, 2010c).

A variety of lines of reasoning lead to the (same) "MiSaTaQuWa" equations of motion for the 1st-order-perturbation-theory self-force acting on a small body moving in an external field. The original derivations of these equations are due to Mino, Sasaki, and Tanaka (Mino et al., 1997) and Quinn and Wald (Quinn and Wald, 1997) (thus the name "MiSaTaQuWa"). More recently, the rigorous derivations of Gralla and Wald (Gralla and Wald, 2008), Gralla, Harte, and Wald (Gralla et al., 2009), and Pound (Pound, 2010a, 2010b, 2010c) have helped to put the MiSaTaQuWa equations on a solid mathematical foundation.

The notion of "point particle" has serious foundational difficulties in a nonlinear field theory such as general relativity (Geroch and Traschen, 1987) (see also footnote **14** (**page 19**)). However, at least in 1st-order perturbation theory these difficulties seem to be surmountable. As Poisson writes (Poisson, 2004, section 5.5.4),

> "The introduction of a point mass in a nonlinear theory of gravitation would appear at first sight to be severely misguided. The lesson learned here is that *one can in fact get away with it.* The derivation of the MiSaTaQuWa equations of motion based on the method of matched asymptotic expansions does indeed show that results obtained on the basis of a point-particle description can be reliable, in spite of all their questionable aspects. This is a remarkable observation, and one that carries a lot of convenience: It is much easier to implement the point-mass description than to perform the matching of two metrics in two coordinate systems."

The MiSaTaQuWa equations involve a curved-spacetime Green function which can only rarely be explicitly calculated. Instead, almost all practical calculations of self-force effects have returned to the scalar-field, Maxwell, or Einstein equations (as appropriate), and regularized the point-particle singularity.

Barack and Ori (Barack and Ori, 2000) (see also (Barack, 2000, 2001, Barack et al., 2002, Barack and Ori, 2002, 2003a and Barack and Lousto, 2002)) developed the "mode-sum" regularization, which provides a practical route to self-force computations. As discussed in **Section 2.1** (**page 11**), this scheme first decomposes the field perturbation into spherical harmonics. Each individual spherical-harmonic mode can be calculated by numerically solving a linear wave equation in 1+1 dimensions, or by solving an ODE if a frequency-domain approach is used. The self-force is then obtained by subtracting certain analytically-calculable regularization parameters from the gradient of each mode's field at the particle, and finally summing over all modes.

The Barack-Ori mode-sum scheme has been the basis for much further research as well as having been used for a large number of self-force calculations in various physical systems. In **Section 3** (**page 32**), I summarize a noteworthy recent self-force calculation using this scheme, due to Barack and Sago (Barack and Sago, 2010). They calculate the 1st-order-perturbation-theory gravitational self-force acting on a particle in an arbitrary bound geodesic orbit in Schwarzschild spacetime, and find and analyze a variety of interesting physical effects such as the $O(\mu)$ self-force corrections to the ISCO position and orbital frequency. This marks a major milestone in the Capra research program.

Barack and Golbourn (Barack and Golbourn, 2007) (see also (Barack et al., 2007 and Dolan and Barack, 2011)) and Vega and Detweiler (Vega and Detweiler, 2008) (see also (Vega et al., 2009)) have recently developed "puncture-function" regularization schemes. As discussed in **Section 2.6** (**page 24**), these schemes first subtract a puncture function (a suitable analytically-calculable approximation to the Detweiler-Whiting singular field) from the physical field near the particle, leaving a finite "residual" field. The Barack-Golbourn "$m$-mode" scheme decomposes the residual field into a Fourier series in the azimuthal ($\varphi$) direction, and calculates each azimuthal mode by numerically solving a linear wave equation in 2+1 dimensions on the Kerr background. The self-force is then obtained by summing the field gradient at the particle over all modes. The Vega-Detweiler scheme numerically solves the regularized field equation (a linear wave equation) for the residual field in 3+1 dimensions, bypassing any mode-sum decomposition. Both the Barack-Golbourn $m$-mode scheme and the Vega-Detweiler scheme are now in use for a variety of self-force calculations.

Because of the complexity of self-force calculations, many techniques have first been developed for scalar-field or electromagnetic particles, and then extended to the gravitational case. Historically, self-force calculations first considered particles moving in Schwarzschild spacetime, but in recent years a growing number of researchers have considered particles in Kerr spacetime.

The self force and its effects can usefully be decomposed into conservative (time-symmetric) and dissipative (time-asymmetric) parts; the latter are often much easier to calculate. It was once thought that accurate EMRI orbital evolutions could be obtained ignoring the conservative part, but Drasco and Hughes (Drasco and Hughes, 2006) and Pound and Poisson (Pound et al., 2005 and Pound and Poisson, 2008a) have found that this isn't the case. Huerta and Gair (Huerta and Gair, 2009) have estimated the magnitude of conservative effects as $\sim 20$ radians of GW phase (out of a total accumulated GW phase of $\sim 10^6$ radians) during the final year of a typical LISA quasicircular EMRI's inspiral; conservative effects are likely to be much larger for eccentric inspirals.

### 5.2 Future Light Cone

Many of the topics discussed in this article remain active areas of research. Some of the areas where I expect to see major developments are:

- Further study of the transient resonance crossings recently identified by Flanagan and Hinderer (Éanna É. Flanagan and Hinderer, 2010) (described in footnote **23** (**page 40**)). These could have a major impact on LISA EMRI data analysis.

- Many more self-force calculations for particles orbiting in Kerr spacetime. To this end, several research groups are actively working on implementing and extending the puncture-function computational schemes described in **Section 2.6** (**page 24**). Warburton and Barack (Warburton and Barack, 2010) have also used the "classic" Barack-Ori mode-sum scheme (discussed in **Section 2.1** (**page 11**)) for Kerr calculations.

- The quantitative comparison of different Capra calculations. Sago, Barack, and Detweiler (Sago et al., 2008) made a very important comparison of this type, showing the consistency of a frequency-domain calculation by Detweiler and a time-domain calculation by Sago and Barack, despite these using different gauges. Such comparisons serve as valuable checks on all the methods and codes involved.

- The use of Capra calculations to help determine free parameters in post-Newtonian approximation schemes, such as the recent work of Blanchet *et al* (Blanchet et al., 2010a, 2010b). These comparisons also serve as valuable checks on both the Capra and post-Newtonian schemes and computations.

- The development of improved long-time approximation schemes, initially using simple "orbit perturbation" ideas as described in **Section 4.2** (**page 41**), and later possibly along the lines suggested by Hinderer and Flanagan (Hinderer and Flanagan, 2008) and/or Pound (Pound, 2010b) (I briefly discuss the need for these in **Section 4.3** (**page 42**)).

- The development of analyses, and eventually practical computational schemes, based on 2nd-order BH perturbation theory.[25,26] Pound (Pound, 2010b) has recently reviewed this problem, and has suggested at least one possible route to the construction of a practical 2nd-order scheme. The computations required are likely to be very complicated (both analytically and numerically) but seem to be possible.

- The computation of Capra GW templates, and later the incorporation of many of the other developments mentioned above into the computation of these templates.

The result of these and many other developments will (I hope) eventually be the calculation of highly accurate EMRI orbital dynamics and GW templates. The numerical calculations involved in doing this will almost certainly be very expensive, perhaps comparable or even larger in magnitude to those for a full numerical-relativity simulation of a comparable-mass binary BH inspiral/coalescence/ringdown. Thus it won't be practical to calculate in this way the huge numbers of GW templates that will be needed for LISA data-analysis template banks. Rather, moderate numbers of Capra GW templates will be used to calibrate other (cheaper) approximation schemes (perhaps $n$th-generation descendents of the "kludge" waveforms used today,[27] or perhaps new schemes like the effective one-body (EOB) ones described by Yunes (Yunes, 2009 and Yunes et al., 2010)). These cheaper schemes will then be used to generate the actual template banks.

Given the very talented people working on these problems, I predict that Capra EMRI GW templates meeting the accuracy goal described in **Section 4.2** (**page 41**) ($\lesssim 10$ milliradians of phase error over a full million-radian inspiral) will be published within 10 years of this article's appearance. I hope many readers of this article will participate in this effort and that within most of our lifetimes we will see actual LISA data being filtered with these templates. There is much work to do.

# 6 Acknowledgements

I thank Leor Barack for introducing me to the self-force problem, and Leor Barack, Norichika Sago, Darren Golbourn, Sam Dolan, and Barry Wardell for many useful conversations. Leor Barack, Barry Wardell, and Virginia J. Vitzthum provided many valuable comments on various drafts of this article. I thank a referee for many valuable comments on an earlier version of this article.

[25] Rosenthal (Rosenthal, 2005a, 2005b, 2006a, 2006b) has obtained a formal expression for the 2nd-order self-force, but unfortunately this is in a gauge which is very inconvenient for practical calculations (in this gauge the 1st-order self-force vanishes).

[26] As well as their use for EMRIs, 2nd-order schemes would be of great value in modelling *intermediate* mass ratio inspirals.

[27] For an introduction to kludge waveforms see, for example, Barack and Cutler (Barack and Cutler, 2004) or Babak *et al* (Babak et al., 2007, 2008).

# References in the highlight article

---

*Selected abstracts*

*Sep 2010 to Jan 2011*

---

# Double white dwarfs and LISA

**Authors:** Marsh, T. R.

**Abstract:** Close pairs of white dwarfs are potential progenitors of Type~ Ia supernovae and they are common, with of order 100 – 300 million in the Galaxy. As such they will be significant, probably dominant, sources of the gravitational waves detectable by LISA. In the context of LISA's goals for fundamental physics, double white dwarfs are a source of noise, but from an astrophysical perspective, they are of considerable interest in their own right. In this paper I discuss our current knowledge of double white dwarfs and their close relatives (and possible descendants) the AM~ CVn stars. LISA will add to our knowledge of these systems by providing the following unique constraints: (i) an almost direct measurement of the Galactic merger rate of DWDs from the detection of short period systems and their period evolution, (ii) an accurate and precise normalisation of binary evolution models at the shortest periods, (iii) a determination of the evolutionary pathways to the formation of AM~ CVn stars, (iv) measurements of the influence of tidal coupling in white dwarfs and its significance for stabilising mass transfer, and (v) discovery of numerous examples of eclipsing white dwarfs with the potential for optical follow-up to test models of white dwarfs.

# Probing the size of extra dimension with gravitational wave astronomy

**Authors:** Yagi, Kent; Tanahashi, Norihiro; Tanaka, Takahiro

**Abstract:** In Randall-Sundrum II (RS-II) braneworld model, it has been conjectured according to the AdS/CFT correspondence that brane-localized black hole (BH) larger than the bulk AdS curvature scale $\ell$ cannot be static, and it is dual to a four dimensional BH emitting the Hawking radiation through some quantum fields. In this scenario, the number of the quantum field species is so large that this radiation changes the orbital evolution of a BH binary. We derived the correction to the gravitational waveform phase due to this effect and estimated the upper bounds on $\ell$ by performing Fisher analyses. We found that DECIGO/BBO can put a stronger constraint than the current table-top result by detecting gravitational waves from small mass BH/BH and BH/neutron star (NS) binaries. Furthermore, DECIGO/BBO is expected to detect $10^5$ BH/NS binaries per year. Taking this advantage, we found that DECIGO/BBO can actually measure $\ell$ down to $\ell = 0.33\mu m$ for 5 year observation if we know that binaries are circular a priori. This is about 40 times smaller than the upper bound obtained from the table-top experiment. On the other hand, when we take eccentricities into binary parameters, the detection limit weakens to $\ell = 1.5\mu m$ due to strong degeneracies between $\ell$ and eccentricities. We also derived the upper bound on $\ell$ from the expected detection number of extreme mass ratio inspirals (EMRIs) with LISA and BH/NS binaries with DECIGO/BBO, extending the discussion made recently by McWilliams. We found that these less robust constraints are weaker than the ones from phase differences.

## Supermassive black holes do not correlate with dark matter halos of galaxies

**Authors:** Kormendy, John; Bender, Ralf

**Abstract:** Supermassive black holes have been detected in all galaxies that contain bulge components when the galaxies observed were close enough so that the searches were feasible. Together with the observation that bigger black holes live in bigger bulges, this has led to the belief that black hole growth and bulge formation regulate each other. That is, black holes and bulges "coevolve". Therefore, reports of a similar correlation between black holes and the dark matter halos in which visible galaxies are embedded have profound implications. Dark matter is likely to be non-baryonic, so these reports suggest that unknown, exotic physics controls black hole growth. Here we show - based in part on recent measurements of bulgeless galaxies - that there is almost no correlation between dark matter and parameters that measure black holes unless the galaxy also contains a bulge. We conclude that black holes do not correlate directly with dark matter. They do not correlate with galaxy disks, either. Therefore black holes coevolve only with bulges. This simplifies the puzzle of their coevolution by focusing attention on purely baryonic processes in the galaxy mergers that make bulges.

## Mergers of Supermassive Black Holes in Astrophysical Environments

**Authors:** Bode, Tanja; Bogdanovic, Tamara; Haas, Roland; Healy, James; Laguna, Pablo; Shoemaker, Deirdre

**Abstract:** Modeling the late inspiral and merger of supermassive black holes is central to understanding accretion processes and the conditions under which electromagnetic emission accompanies gravitational waves. We use fully general relativistic, hydrodynamics simulations to investigate how electromagnetic signatures correlate with black hole spins, mass ratios, and the gaseous environment in this final phase of binary evolution. In all scenarios, we find some form of characteristic electromagnetic variability whose pattern depends on the spins and binary mass ratios. Binaries in hot accretion flows exhibit a flare followed by a sudden drop in luminosity associated with the plunge and merger, as well as quasi-periodic oscillations correlated with the gravitational waves during the inspiral. Conversely, circumbinary disk systems are characterized by a low luminosity of variable emission, suggesting challenging prospects for their detection.

## Cosmic Lighthouses : Unveiling the nature of high-redshift galaxies

**Authors:** Dayal, Pratika

**Abstract:** We are in the golden age for the search for high-redshift galaxies, made possible by a combination of new instruments and innovative search techniques. One of the major aims of such searches is to constrain the epoch of reionization (EoR), which marks the second major change in the ionization state of the Universe. Understanding the EoR is difficult since whilst it is galaxy evolution which drives reionization, reionization itself influences galaxy evolution through feedback effects. Unraveling the interplay of reionization and galaxy evolution is further complicated by of a lack of understanding of the metal enrichment and dust distribution in high redshift galaxies. To this end, a class of galaxies called Lyman Alpha Emitters (LAEs) have been gaining enormous popularity as probes of all these three processes. In this thesis, we couple state of the art cosmological SPH simulations (GADGET-2) with a physically motivated, self-consistent model for LAEs, so as to be able to understand the importance of the intergalactic medium (IGM) ionization state, dust and peculiar velocities in shaping their observed properties. By doing so, the aim is to gain insight on the nature of LAEs, put precious constraints on their elusive physical properties and make predictions for future instruments such as the Atacama Large Millimeter Array (ALMA). Using our LAE model in conjunction with a code that builds the MW merger tree (GAMETE), we build a bridge between the high-redshift and the local Universe. We also use SPH simulations (GADGET-2) to study the nature of the earliest galaxies that have been detected as of yet, place constraints on their contribution to reionization, and predict their detectability using the next generation of instruments, such as the James Web Space Telescope (JWST).

## Neural network interpolation of the magnetic field for the LISA Pathfinder Diagnostics Subsystem

**Authors:** Diaz-Aguilo, Marc; Lobo, Alberto; García-Berro, Enrique

**Abstract:** LISA Pathfinder is a science and technology demonstrator of the European Space Agency within the framework of its LISA mission, which aims to be the first space-borne gravitational wave observatory. The payload of LISA Pathfinder is the so-called LISA Technology Package, which is designed to measure relative accelerations between two test masses in nominal free fall. Its disturbances are monitored and dealt by the diagnostics subsystem. This subsystem consists of several modules, and one of these is the magnetic diagnostics system, which includes a set of of four tri-axial fluxgate magnetometers, intended to measure with high precision the magnetic field at the positions of the test masses. However, since the magnetometers are located far from the positions of the test masses, the magnetic field at their positions must be interpolated. It has been recently shown that because there are not enough magnetic channels, classical interpolation methods fail to derive reliable measurements at the positions of the test masses, while neural network interpolation can provide the required measurements at the desired accuracy. In this paper we expand these studies and we assess the reliability and robustness of the neural network interpolation scheme for variations of the locations and possible offsets of the magnetometers, as well as for changes in environmental conditions.

We find that neural networks are robust enough to derive accurate measurements of the magnetic field at the positions of the test masses in most circumstances.

## Transition from adiabatic inspiral to plunge into a spinning black hole

**Authors:** Kesden, Michael

**Eprint:** http://arxiv.org/abs/1101.3749

**Keywords:** astro-ph.CO; EMRI; general relativity; gr-qc; spin

**Abstract:** A test particle of mass $\mu$ on a bound geodesic of a Kerr black hole of mass $M \gg \mu$ will slowly inspiral as gravitational radiation extracts energy and angular momentum from its orbit. This inspiral can be considered adiabatic when the orbital period is much shorter than the timescale on which energy is radiated, and quasi-circular when the radial velocity is much less than the azimuthal velocity. Although the inspiral always remains adiabatic provided $\mu \ll M$, the quasi-circular approximation breaks down as the particle approaches the innermost stable circular orbit (ISCO). In this paper, we relax the quasi-circular approximation and solve the radial equation of motion explicitly near the ISCO. We use the requirement that the test particle's 4-velocity remain properly normalized to calculate a new contribution to the difference between its energy and angular momentum. This difference determines how a black hole's spin changes following a test-particle merger, and can be extrapolated to help predict the mass and spin of the final black hole produced in finite-mass-ratio black-hole mergers. Our new contribution is particularly important for nearly maximally spinning black holes, as it can affect whether a merger produces a naked singularity.

## Measuring parameters of massive black hole binaries with partially aligned spins

**Authors:** Lang, Ryan N.; Hughes, Scott A.; Cornish, Neil J.

**Eprint:** http://arxiv.org/abs/1101.3591

**Keywords:** astro-ph.CO; gr-qc; massive binaries of black holes; parameter estimation; spin

**Abstract:** The future space-based gravitational wave detector LISA will be able to measure parameters of coalescing massive black hole binaries, often to extremely high accuracy. Previous work has demonstrated that the black hole spins can have a strong impact on the accuracy of parameter measurement. Relativistic spin-induced precession modulates the waveform in a manner which can break degeneracies between parameters, in principle significantly improving how well they are measured. Recent studies have indicated, however, that spin precession may be weak for an important subset of astrophysical binary black holes: those in which the spins are aligned due to interactions with gas. In this paper, we examine how well a binary's parameters can be measured when its spins are partially aligned and compare results using waveforms that include higher post-Newtonian harmonics to those that are truncated at leading quadrupole order. We find that the weakened precession can substantially degrade parameter estimation, particularly for the "extrinsic" parameters sky position and distance. Absent higher harmonics, LISA typically localizes the sky position of a nearly aligned binary about an order of magnitude less accurately than one for which the spin orientations are random. Our knowledge of a source's sky position will thus be worst for the gas-rich systems which are most likely to produce electromagnetic counterparts. Fortunately, higher harmonics of the waveform can make up for this degradation. By including harmonics beyond the quadrupole in our waveform model, we find that the accuracy with which most of the binary's parameters are measured can be substantially improved. In some cases, the improvement is such that they are measured almost as well as when the binary spins are randomly aligned.

## Beyond the geodesic approximation: conservative effects of the gravitational self-force in eccentric orbits around a Schwarzschild black hole

**Authors:** Barack, Leor; Sago, Norichika

**Abstract:** We study conservative finite-mass corrections to the motion of a particle in a bound (eccentric) strong-field orbit around a Schwarzschild black hole. We assume the particle's mass $\mu$ is much smaller than the black hole mass $M$, and explore post-geodesic corrections of $O(\mu/M)$. Our analysis uses numerical data from a recently developed code that outputs the Lorenz-gauge gravitational self-force (GSF) acting on the particle along the eccentric geodesic. First, we calculate the $O(\mu/M)$ conservative correction to the periastron advance of the orbit, as a function of the (gauge dependent) semi-latus rectum and eccentricity. A gauge-invariant description of the GSF precession effect is made possible in the circular-orbit limit, where we express the correction to the periastron advance as a function of the invariant azimuthal frequency. We compare this relation with results from fully nonlinear numerical-relativistic simulations. In order to obtain a gauge-invariant measure of the GSF effect for fully eccentric orbits, we introduce a suitable generalization of Detweiler's circular-orbit "red shift" invariant. We compute the $O(\mu/M)$ conservative correction to this invariant, expressed as a function of the two invariant frequencies that parametrize the orbit. Our results are in good agreement with results from post-Newtonian calculations in the weak field regime, as we shall report elsewhere. The results of our study can inform the development of analytical models for the dynamics of strongly-gravitating binaries. They also provide an accurate benchmark for future numerical-relativistic simulations.

## Tuning Time-Domain Pseudospectral Computations of the Self-Force on a Charged Scalar Particle

**Authors:** Canizares, Priscilla; Sopuerta, Carlos F.

**Eprint:** http://arxiv.org/abs/1101.2526

**Keywords:** astro-ph.HE; EMRI; general relativity; geodesic motion; gr-qc; math-ph; math.MP; numerical methods

**Abstract:** The computation of the self-force constitutes one of the main challenges for the construction of precise theoretical waveform templates in order to detect and analyze extreme-mass-ratio inspirals with the future space-based gravitational-wave observatory LISA. Since the number of templates required is quite high, it is important to develop fast algorithms both for the computation of the self-force and the production of waveforms. In this article we show how to tune a recent time-domain technique for the computation of the self-force, what we call the Particle without Particle scheme, in order to make it very precise and at the same time very efficient. We also extend this technique in order to allow for highly eccentric orbits.

## Resolution requirements for Smoothed Particle Hydrodynamics simulations of self-gravitating accretion discs

**Authors:** Lodato, Giuseppe; Clarke, Cathie C.

**Eprint:** http://arxiv.org/abs/1101.2448

**Keywords:** accretion discs; astro-ph.CO; astro-ph.EP; astro-ph.SR; astrophysics; numerical methods; supermassive black holes

**Abstract:** Stimulated by recent results by Meru and Bate (2010a,b), we revisit the issue of resolution requirements for simulating self-gravitating accretion discs with Smoothed Particle Hydrodynamics (SPH). We show that the results by Meru and Bate (2010a) are consistent with those of Meru and Bate (2010b) if they are both interpreted as driven by resolution effects, therefore implying that the resolution criterion for cooling gaseous discs is a function of the imposed cooling rate. We discuss two possible numerical origins of such dependence, which are both consistent with the limited number of available data. Our results tentatively indicate that convergence for current simulations is being reached for a number of SPH particles approaching 10 millions (for a disc mass of order 10 per cent of the central object mass), which would set the critical cooling time for fragmentation at about $15\Omega^{-1}$, roughly a factor two larger than previously thought. More in general, we discuss the extent to which the large number of recent numerical results are reliable or not. We argue that those results that pertain to the dynamics associated with gravitational instabilities (such as the locality of angular momentum transport, and the relationship between density perturbation and induced stress) are robust, while those pertaining to the thermodynamics of the system (such as the determination of the critical cooling time for fragmentation) can be affected by poor resolution.

## The Black-Hole Mass in M87 from Gemini/NIFS Adaptive Optics Observations

**Authors:** Gebhardt, Karl; Adams, Joshua; Richstone, Douglas; Lauer, Tod R.; Faber, S. M.; Gultekin, Kayhan; Murphy, Jeremy; Tremaine, Scott

**Abstract:** We present the stellar kinematics in the central 2" of the luminous elliptical galaxy M87 (NGC 4486), using laser adaptive optics to feed the Gemini telescope integral-field spectrograph, NIFS. The velocity dispersion rises to 480 km/s at 0.2". We combine these data with extensive stellar kinematics out to large radii to derive a black-hole mass equal to $(6.6 + -0.4)x10^9 M_\odot$, using orbit-based axisymmetric models and including only the NIFS data in the central region. Including previously-reported ground-based data in the central region drops the uncertainty to $0.25x10^9 M_\odot$ with no change in the best-fit mass; however, we rely on the values derived from the NIFS-only data in the central region in order to limit systematic differences. The best-fit model shows a significant increase in the tangential velocity anisotropy of stars orbiting in the central region with decreasing radius; similar to that seen in the centers of other core galaxies. The black-hole mass is insensitive to the inclusion of a dark halo in the models — the high angular-resolution provided by the adaptive optics breaks the degeneracy between black-hole mass and stellar mass-to-light ratio. The present black-hole mass is in excellent agreement with the Gebhardt & Thomas value, implying that the dark halo must be included when the kinematic influence of the black hole is poorly resolved. This degeneracy implies that the black-hole masses of luminous core galaxies, where this effect is important, may need to be re-evaluated. The present value exceeds the prediction of the black hole-dispersion and black hole-luminosity relations, both of which predict about $1 \times 10^9 M_\odot$ for M87, by close to twice the intrinsic scatter in the relations. The high-end of the black hole correlations may be poorly determined at present.

## Relation Between Globular Clusters and Supermassive Black Holes in Ellipticals as a Manifestation of the Black Hole Fundamental Plane

**Authors:** Snyder, Gregory F.; Hopkins, Philip F.; Hernquist, Lars

**Eprint:** http://arxiv.org/abs/1101.1299

**Keywords:** astro-ph.CO; astrophysics; globular clusters; observations; supermassive black holes

**Abstract:** We analyze the relation between the mass of the central supermassive black hole (Mbh) and the number of globular clusters (Ngc) in elliptical galaxies and bulges as a ramification of the black hole fundamental plane, the theoretically predicted and observed multi-variable correlation between Mbh and bulge binding energy. Although the tightness of the Mbh-Ngc correlation suggests an unlikely causal link between supermassive black holes and globular clusters, such a correspondence can exhibit small scatter even if the physical relationship is indirect. We show that the relatively small scatter of the Mbh-Ngc relation owes to the mutual residual correlation of Mbh and Ngc with stellar mass when the velocity dispersion is held fixed. Thus, present observations lend evidence for feedback-regulated models in which the bulge binding energy is most important; they do not necessarily imply any 'special' connection between globular clusters and Mbh. This raises the question of why Ngc traces the formation of ellipticals and bulges sufficiently well to be correlated with binding energy.

## HST WFC3/IR Observations of Active Galactic Nucleus Host Galaxies at z 2: Supermassive Black Holes Grow in Disk Galaxies

**Authors:** Schawinski, Kevin; Treister, Ezequiel; Urry, C. Megan; Cardamone, Carolin N.; Simmons, Brooke; Yi, Sukyoung K.

Abstract: We present the rest-frame optical morphologies of active galactic nucleus (AGN) host galaxies at $1.5 < z < 3$, using near-infrared imaging from the Hubble Space Telescope Wide Field Camera 3, the first such study of AGN host galaxies at these redshifts. The AGN are X-ray selected from the Chandra Deep Field South and have typical luminosities of $10^{42} < L_X < 10^{44}$ erg/s. Accreting black holes in this luminosity and redshift range account for a substantial fraction of the total space density and black hole mass growth over cosmic time; they thus represent an important mode of black hole growth in the universe. We find that the majority ($\sim$ 80%) of the host galaxies of these AGN have low Sersic indices indicative of disk-dominated light profiles, suggesting that secular processes govern a significant fraction of the cosmic growth of black holes. That is, many black holes in the present-day universe grew much of their mass in disk-dominated galaxies and not in early-type galaxies or major mergers. The properties of the AGN host galaxies are furthermore indistinguishable from their parent galaxy population and we find no strong evolution in either effective radii or morphological mix between $z \sim 2$ and $z \sim 0.05$.

## Dynamical Black Hole Masses of BL Lac Objects from the Sloan Digital Sky Survey

Authors: Plotkin, Richard M.; Markoff, Sera; Trager, Scott C.; Anderson, Scott F.

Abstract: We measure black hole masses for 71 BL Lac objects from the Sloan Digital Sky Survey with redshifts out to $z \sim 0.4$. We perform spectral decompositions of their nuclei from their host galaxies and measure their stellar velocity dispersions. Black hole masses are then derived from the black hole mass - stellar velocity dispersion relation. We find BL Lac objects host black holes of similar masses, $\sim 10^{8.5} M_\odot$, with a dispersion of 0.4 dex, similar to the uncertainties on each black hole measurement. Therefore, all BL Lac objects in our sample have the same indistinguishable black hole mass. These 71 BL Lac objects follow the black hole mass - bulge luminosity relation, and their narrow range of host galaxy luminosities confirm previous claims that BL Lac host galaxies can be treated as standard candles. We conclude that the observed diversity in the shapes of BL Lac object spectral energy distributions is not strongly driven by black hole mass or host galaxy properties.

## A New Catalog of Globular Clusters in the Milky Way

**Authors:** Harris, William E.

**Abstract:**

A new revision of the McMaster catalog of Milky Way globular clusters is available. This is the first update since 2003 and the biggest single revision since the original version of the catalog published in 1996. The list now contains a total of 157 objects classified as globular clusters. Major upgrades have been made especially to the cluster coordinates, metallicities, and structural profile parameters, and the list of parameters now also includes central velocity dispersion.

NB: This paper is a stand-alone publication available only on the astro-ph archive; it will not be published separately in a journal.

## Tidal stellar disruptions by massive black hole pairs: II. Decaying binaries

**Authors:** Chen, Xian; Sesana, Alberto; Madau, Piero; Liu, Fukun

**Abstract:** Tidal stellar disruptions have traditionally been discussed as a probe of the single, massive black holes (MBHs) that are dormant in the nuclei of galaxies. In Chen et al. (2009), we used numerical scattering experiments to show that three-body interactions between bound stars in a stellar cusp and a non-evolving "hard" MBH binary will also produce a burst of tidal disruptions, caused by a combination of the secular "Kozai effect" and by close resonant encounters with the secondary hole. Here we derive basic analytical scalings of the stellar disruption rates with the system parameters, assess the relative importance of the Kozai and resonant encounter mechanisms as a function of time, discuss the impact of general relativistic (GR) and extended stellar cusp effects, and develop a hybrid model to self-consistently follow the shrinking of an MBH binary in a stellar background, including slingshot ejections and tidal disruptions. In the case of a fiducial binary with primary hole mass $M_1 = 10^7 M_\odot$ and mass ratio $q = M_2/M_1 = 1/81$, embedded in an isothermal cusp, we derive a stellar disruption rate $\dot{N}_* \sim 0.2/\text{yr}$ lasting $\sim 3 \times 10^5$ yr. This rate is 3 orders of magnitude larger than the corresponding value for a single MBH fed by two-body relaxation, confirming our previous findings. For q<$\sim$ 10% of the tidal-disruption events may originate in MBH binaries.

## Black-hole binaries go to eleven orbits

**Authors:** Sperhake, Ulrich; Bruegmann, Bernd; Mueller, Doreen; Sopuerta, Carlos F.

**Abstract:** We analyse an eleven-orbit inspiral of a non-spinning black-hole binary with mass ratio q=M1/M2=4. The numerically obtained gravitational waveforms are compared with post-Newtonian (PN) predictions including several subdominant multipoles up to multipolar indices (l=5,m=5). We find that (i) numerical and post-Newtonian predictions of the phase of the (2,2) mode accumulate a phase difference of about 0.35 rad at the PN cut off frequency 0.1 for the Taylor T1 approximant; (ii) in contrast to previous studies of equal-mass and specific spinning binaries, we find the Taylor T4 approximant to agree less well with numerical results, provided the latter are extrapolated to infinite extraction radius; (iii) extrapolation of gravitational waveforms to infinite extraction radius is particularly important for subdominant multipoles with l unequal m; (iv) 3PN terms in post-Newtonian multipole expansions significantly improve the agreement with numerical predictions for sub-dominant multipoles.

## The Status of Black-Hole Binary Merger Simulations with Numerical Relativity

**Authors:** McWilliams, Sean T.

**Abstract:** The advent of long-term stability in numerical relativity has yielded a windfall of answers to long-standing questions regarding the dynamics of space-time, matter, and electromagnetic fields in the strong-field regime of black-hole binary mergers. In this review, we will briefly summarize the methodology currently applied to these problems, emphasizing the most recent advancements. We will discuss recent results of astrophysical relevance, and present some novel interpretation. Though we primarily present a review, we also present a simple analytical model for the time-dependent Poynting flux from two orbiting black holes immersed in a magnetic field, which compares favorably with recent numerical results. Finally, we will discuss recent advancements in our theoretical understanding of merger dynamics and gravitational waveforms that have resulted from interpreting the ever-growing body of numerical relativity results.

# Binary black hole coalescence in the extreme-mass-ratio limit: testing and improving the effective-one-body multipolar waveform

**Authors:** Bernuzzi, Sebastiano; Nagar, Alessandro; Zenginoglu, Anil

**Abstract:** We discuss the properties of the effective-one-body (EOB) multipolar gravitational waveform emitted by nonspinning black-hole binaries of masses $\mu$ and $M$ in the extreme-mass-ratio limit, $\mu/M = \nu \ll 1$. We focus on the transition from quasicircular inspiral to plunge, merger and ringdown.We compare the EOB waveform to a Regge-Wheeler-Zerilli (RWZ) waveform computed using the hyperboloidal layer method and extracted at null infinity. Because the EOB waveform keeps track analytically of most phase differences in the early inspiral, we do not allow for any arbitrary time or phase shift between the waveforms. The dynamics of the particle, common to both wave-generation formalisms, is driven by leading-order $O(\nu)$ analytically–resummed radiation reaction. The EOB and the RWZ waveforms have an initial dephasing of about $5 \times 10^{-4}$ rad and maintain then a remarkably accurate phase coherence during the long inspiral ($\sim 33$ orbits), accumulating only about $-2 \times 10^{-3}$ rad until the last stable orbit, i.e. $\Delta\phi/\phi \sim -5.95 \times 10^{-6}$. We obtain such accuracy without calibrating the analytically-resummed EOB waveform to numerical data, which indicates the aptitude of the EOB waveform for LISA-oriented studies. We then improve the behavior of the EOB waveform around merger by introducing and tuning next-to-quasi-circular corrections both in the gravitational wave amplitude and phase. For each multipole we tune only four next-to-quasi-circular parameters by requiring compatibility between EOB and RWZ waveforms at the light-ring. The resulting phase difference around merger time is as small as ±0.015 rad, with a fractional amplitude agreement of 2.5%. This suggest that next-to-quasi-circular corrections to the phase can be a useful ingredient in comparisons between EOB and numerical relativity waveforms.

## Gravitational radiation from radial infall of a particle into a Schwarzschild black hole. A numerical study of the spectra, quasi-normal modes and power-law tails

**Authors:** Mitsou, Ermis

**Eprint:** http://arxiv.org/abs/1012.2028

**Keywords:** EMRI; general relativity; geodesic motion; gr-qc

**Abstract:** The computation of the gravitational radiation emitted by a particle falling into a Schwarzschild black hole is a classic problem studied already in the 1970s. Here we present a detailed numerical analysis of the case of radial infall starting at infinity with no initial velocity. We compute the radiated waveforms, spectra and energies for multipoles up to $l = 6$, improving significantly on the numerical accuracy of existing results. This is done by integrating the Zerilli equation in the frequency domain using the Green's function method. The resulting wave exhibits a "ring-down" phase whose dominant contribution is a superposition of the quasi-normal modes of the black hole. The numerical accuracy allows us to recover the frequencies of these modes through a fit of that part of the wave. Comparing with direct computations of the quasi-normal modes we reach a $\sim 10^{-4}$ to $\sim 10^{-2}$ accuracy for the first two overtones of each multipole. Our numerical accuracy also allows us to display the power-law tail that the wave develops after the ring-down has been exponentially cut-off. The amplitude of this contribution is $\sim 10^2$ to $\sim 10^3$ times smaller than the typical scale of the wave.

## Sgr A*: The Optimal Testbed of Strong-Field Gravity

**Authors:** Psaltis, Dimitrios; Johannsen, Tim

**Eprint:** http://arxiv.org/abs/1012.1602

**Keywords:** astro-ph.HE; astrophysics; detectors; gr-qc; instruments; supermassive black holes

**Abstract:** The black hole in the center of the Milky Way has been observed and modeled intensely during the last decades. It is also the prime target of a number of new experiments that aim to zoom into the vicinity of its horizon and reveal the inner working of its spacetime. In this review we discuss our current understanding of the gravitational field of Sgr A* and the prospects of testing the Kerr nature of its spacetime via imaging, astrometric, and timing observations.

## Reducing orbital eccentricity in quasi-circular binary black-hole evolutions in presence of spins

**Authors:** Buonanno, Alessandra; Kidder, Lawrence E.; MrouÃĺ, Abdul H.; Pfeiffer, Harald P.; Taracchini, Andrea

**Abstract:** Building initial conditions for generic binary black-hole evolutions without initial spurious eccentricity remains a challenge for numerical-relativity simulations. This problem can be overcome by applying an eccentricity-removal procedure which consists in evolving the binary for a couple of orbits, estimating the eccentricity, and then correcting the initial conditions. The presence of spins can complicate this procedure. As predicted by post-Newtonian theory, spin-spin interactions and precession prevent the binary from moving along an adiabatic sequence of spherical orbits, inducing oscillations in the radial separation and in the orbital frequency. However, spin-induced oscillations occur at approximately twice the orbital frequency, therefore they can be distinguished from the initial spurious eccentricity, which occurs at approximately the orbital frequency. We develop a new removal procedure based on the derivative of the orbital frequency and find that it is successful in reducing the eccentricity measured in the orbital frequency to less than 0.0001 when moderate spins are present. We test this new procedure using numerical-relativity simulations of binary black holes with mass ratios 1.5 and 3, spin magnitude 0.5 and various spin orientations. The numerical simulations exhibit spin-induced oscillations in the dynamics at approximately twice the orbital frequency. Oscillations of similar frequency are also visible in the gravitational-wave phase and frequency of the dominant mode.

## Spinning super-massive objects in galactic nuclei up to $a_* > 1$

**Authors:** Bambi, Cosimo

**Abstract:** Today we believe that a typical galaxy contains about $10^7$ stellar-mass black holes and a single super-massive black hole at its center. According to general relativity, these objects are characterized solely by their mass $M$ and by their spin parameter $a_*$. A fundamental limit for a black hole in general relativity is the Kerr bound $|a_*| \leq 1$, but the accretion process can spin it up to $a_* \approx 0.998$. If a compact object is not a black hole, the Kerr bound does not hold and in this letter I show that the accretion process can spin the body up to $a_* > 1$. While this fact should be negligible for stellar-mass objects, most of the super-massive objects at the center of galaxies may actually be super-spinning bodies exceeding the Kerr bound. Such a possibility can be tested by gravitational wave detectors like LISA or by sub-millimeter very long baseline interferometry facilities.

# From laboratory experiments to LISA Pathfinder: achieving LISA geodesic motion

**Authors:** Antonucci, F; Armano, M; Audley, H; Auger, G; Benedetti, M; Binetruy, P; Boatella, C; Bogenstahl, J; Bortoluzzi, D; Bosetti, P; Brandt, N; Caleno, M; Cavalleri, A; Cesa, M; Chmeissani, M; Ciani, G; Conchillo, A; Congedo, G; Cristofolini, I; Cruise, M; Danzmann, K; De Marchi, F; Diaz-Aguilo, M; Diepholz, I; Dixon, G; Dolesi, R; Dunbar, N; Fauste, J; Ferraioli, L; Fertin, D; Fichter, W; Fitzsimons, E; Freschi, M; Marin, A Garcí■a; Marirrodriga, C Garcí■a; Gerndt, R; Gesa, L; Giardini, D; Gibert, F; Grimani, C; Grynagier, A; Guillaume, B; GuzmÁan, F; Harrison, I; Heinzel, G; Hewitson, M; Hollington, D; Hough, J; Hoyland, D; Hueller, M; Huesler, J; Jeannin, O; Jennrich, O; Jetzer, P; Johlander, B; Killow, C; Llamas, X; Lloro, I; Lobo, A; Maarschalkerweerd, R; Madden, S; Mance, D; Mateos, I; McNamara, P W; MendestÁñ, J; Mitchell, E; Monsky, A; Nicolini, D; Nicolodi, D; Nofrarias, M; Pedersen, F; Perreur-Lloyd, M; Perreca, A; Plagnol, E; Prat, P; Racca, G D; Rais, B; Ramos-Castro, J; Reiche, J; Perez, J A Romera; Robertson, D; Rozemeijer, H; Sanjuan, J; Schleicher, A; Schulte, M; Shaul, D; Stagnaro, L; Strandmoe, S; Steier, F; Sumner, T J; Taylor, A; Texier, D; Trenkel, C; Tombolato, D; Vitale, S; Wanner, G; Ward, H; Waschke, S; Wass, P; Weber, W J; Zweifel, P

**Abstract:** This paper presents a quantitative assessment of the performance of the upcoming LISA Pathfinder geodesic explorer mission. The findings are based on the results of extensive ground testing and simulation campaigns using flight hardware and flight control and operations algorithms. The results show that, for the central experiment of measuring the stray differential acceleration between the LISA test masses, LISA Pathfinder will be able to verify the overall acceleration noise to within a factor two of the LISA requirement at 1 mHz and within a factor 10 at 0.1 mHz. We also discuss the key elements of the physical model of disturbances, coming from LISA Pathfinder and ground measurement, that will guarantee the LISA performance.

## Phenomenological gravitational waveforms from spinning coalescing binaries

**Authors:** Sturani, R.; Fischetti, S.; Cadonati, L.; Guidi, G. M.; Healy, J.; Shoemaker, D.; Vicere', A.

**Abstract:** An accurate knowledge of the coalescing binary gravitational waveform is crucial for match filtering techniques, which are currently used in the observational searches performed by the LIGO-Virgo collaboration. Following an earlier paper by the same authors we expose the construction of analytical phenomenological waveforms describing the signal sourced by generically spinning binary systems. The gap between the initial inspiral part of the waveform, described by spin-Taylor approximants, and its final ring-down part, described by damped exponentials, is bridged by a phenomenological phase calibrated by comparison with the dominant spherical harmonic mode of a set of waveforms including both numerical and phenomenological waveforms of a different type. All waveforms considered describe equal mass systems with dimension-less spin magnitudes equal to 0.6. The noise-weighted overlap integral between numerical and phenomenological waveforms ranges between 0.93 and 0.98 for a wide span of mass values.

## Forced motion near black holes

**Authors:** Gair, Jonathan R.; Flanagan, Eanna E.; Drasco, Steve; Hinderer, Tanja; Babak, Stanislav

**Abstract:** We present two methods for integrating forced geodesic equations in the Kerr spacetime, which can accommodate arbitrary forces. As a test case, we compute inspirals under a simple drag force, mimicking the presence of gas. We verify that both methods give the same results for this simple force. We find that drag generally causes eccentricity to increase throughout the inspiral. This is a relativistic effect qualitatively opposite to what is seen in gravitational-radiation-driven inspirals, and similar to what is observed in hydrodynamic simulations of gaseous binaries. We provide an analytic explanation by deriving the leading order relativistic correction to the Newtonian dynamics. If observed, an increasing eccentricity would provide clear evidence that the inspiral was occurring in a non-vacuum environment. Our two methods are especially useful for evolving orbits in the adiabatic regime. Both use the method of osculating orbits, in which each point on the orbit is characterized by the parameters of the geodesic with the same instantaneous position and velocity. Both methods describe the orbit in terms of the geodesic energy, axial angular momentum, Carter constant, azimuthal phase, and two angular variables that increase monotonically and are relativistic generalizations of the eccentric anomaly. The two methods differ in their treatment of the orbital phases and the representation of the force. In one method the geodesic phase and phase constant are evolved together as a single orbital phase parameter, and the force is expressed in terms of its components on the Kinnersley orthonormal tetrad. In the second method, the phase constants of the geodesic motion are evolved separately and the force is expressed in terms of its Boyer-Lindquist components. This second approach is a generalization of earlier work by Pound and Poisson for planar forces in a Schwarzschild background.

## Multiple Tidal Disruptions as an Indicator of Binary Super-Massive Black Hole Systems

**Authors:** Wegg, Christopher; Bode, J. Nate

**Abstract:** We find that the majority of systems hosting multiple tidal disruptions are likely to contain hard binary SMBH systems, and also show that the rates of these repeated events are high enough to be detected by LSST over its lifetime. Therefore, these multiple tidal disruption events provide a novel method to identify super-massive black hole (SMBH) binary systems with parsec to sub-parsec separations. The rates of tidal disruptions are investigated using simulations of non-interacting stars initially orbiting a primary SMBH and the potential of the model stellar cusp. The stars are then evolved forward in time and perturbed by a secondary SMBH inspiraling from the edge of the cusp to its stalling radius. We find with conservative magnitude estimates that the next generation transient survey LSST should detect multiple tidal disruptions in approximately 3 galaxies over 5 years of observation, though less conservative estimates could increase this rate by an order of magnitude.

# Numerical Parameter Survey of Nonradiative Black Hole Accretion – Flow Structure and Variability of the Rotation Measure

**Authors:** Pang, Bijia; Pen, Ue-Li; Matzner, Christopher D.; Green, Stephen R.; Liebendörfer, Matthias

**Abstract:**

We conduct a survey of numerical simulations to probe the structure and appearance of non-radiative black hole accretion flows like the supermassive black hole at the Galactic centre. We find a generic set of solutions, and make specific predictions for currently feasible rotation measure (RM) observations, which are accessible to current instruments including the EVLA, GMRT and ALMA. The slow time variability of the RM is a key quantitative signature of this accretion flow. The time variability of RM can be used to quantitatively measure the nature of the accretion flow, and to differentiate models. Sensitive measurements of RM can be achieved using RM synthesis or using pulsars.

Our energy conserving ideal magneto-hydrodynamical simulations, which achieve high dynamical range by means of a deformed-mesh algorithm, stretch from several Bondi radii to about one thousandth of that radius, and continue for tens of Bondi times. Magnetized flows which lack outward convection possess density slopes around -1, almost independent of physical parameters, and are more consistent with observational constraints than are strongly convective flows We observe no tendency for the flows to become rotationally supported in their centres, or to develop steady outflow.

We support these conclusions with formulae which encapsulate our findings in terms of physical and numerical parameters. We discuss the relation of these solutions to other approaches. The main potential uncertainties are the validity of ideal MHD and the absence of a fully relativistic inner boundary condition. The RM variability predictions are testable with current and future telescopes.

## Performance of astrometric detection of a hotspot orbiting on the innermost stable circular orbit of the galactic centre black hole

**Authors:** Vincent, F. H.; Paumard, T.; Perrin, G.; Mugnier, L.; Eisenhauer, F.; Gillessen, S.

**Abstract:**

The galactic central black hole Sgr A* exhibits outbursts of radiation in the near infrared (so-called IR flares). One model of these events consists in a hotspot orbiting on the innermost stable circular orbit (ISCO) of the hole. These outbursts can be used as a probe of the central gravitational potential. One main scientific goal of the second generation VLTI instrument GRAVITY is to observe these flares astrometrically. Here, the astrometric precision of GRAVITY is investigated in imaging mode, which consists in analysing the image computed from the interferometric data. The capability of the instrument to put in light the motion of a hotspot orbiting on the ISCO of our central black hole is then discussed.

We find that GRAVITY's astrometric precision for a single star in imaging mode is smaller than the Schwarzschild radius of Sgr $A^*$. The instrument can also demonstrate that a body orbiting on the last stable orbit of the black hole is indeed moving. It yields a typical size of the orbit, if the source is as bright as $m_K = 14$.

These results show that GRAVITY allows one to study the close environment of Sgr $A^*$. Having access to the ISCO of the central massive black hole probably allows constraining general relativity in its strong regime. Moreover, if the hotspot model is appropriate, the black hole spin can be constrained.

## Magnetothermal and magnetorotational instabilities in hot accretion flows

**Authors:** Bu, De-Fu; Yuan, Feng; Stone, James M

**Abstract:** For magnetized accretion flows with very low accretion rates such as that in the supermassive black hole in our Galactic center, $SgrA^*$, the mean free path of electrons is much greater than the Larmor radius and is an appreciable fraction of the size of the system. In this case, the thermal conduction is anisotropic and dynamically important. Provided that the magnetic field is weak, magnetothermal instability (MTI) exists . It can amplify the magnetic field and align the field lines with the temperature gradient (i.e., the radial direction). If the accretion flow is differentially rotating, magnetorotational instability (MRI) also exists as well known. In this paper, we investigate the possible interaction of these two instabilities. We study a hot accretion flow around Bondi radius, where the infall timescale of gas is longer than the MTI and MRI growth timescales, thus MTI and MRI coexist. We focus on the interaction between MTI and MRI by examining the magnetic field amplification induced by the two instabilities. We find that MTI and MRI mainly amplify the radial and toroidal components of the magnetic field, respectively. Most importantly, we find that if MTI alone can amplify the magnetic field by a factor of $F_t$ and MRI alone by a factor of $F_r$, when MTI and MRI coexist, the magnetic field can be amplified by a factor of $F_t F_r$. We therefore conclude that MTI and MRI operate separately. The physical reason for the decouple of MTI and MRI is that they are two intrinsically different physical process. We also find that MTI helps to transfer angular momentum, because MTI can enhance the Maxwell stress (by amplifying the magnetic field) and Reynolds stress. Finally, we find that thermal conduction makes the temperature slope flatter by transporting energy outward. This makes the mass accretion rate smaller.

## Type 2 Active Galactic Nuclei with Double-Peaked [OIII] Lines. II. Resulting More from Narrow-Line Region Kinematics than from Merging Supermassive Black Hole Pairs

**Authors:** Shen, Yue; Liu, Xin; Greene, Jenny; Strauss, Michael

**Abstract:** (Abridged) Approximately 1% of low redshift (z<0.3) optically-selected type 2 AGNs show a double-peaked [OIII] narrow emission line profile in their spatially-integrated spectra. Such features are usually interpreted as due either to kinematics, such as biconical outflows and/or disk rotation of the narrow line region (NLR) around single black holes, or to the relative motion of two distinct NLRs in a merging pair of AGNs. Here we report follow-up near infrared (NIR) imaging and optical slit spectroscopy of 31 double-peaked [OIII] type 2 AGNs drawn from the SDSS parent sample presented in Liu et al (2010). These data reveal a mixture of origins for the double-peaked feature. Roughly 10% of our objects are best explained by binary AGNs at (projected) kpc-scale separations, where two stellar components with spatially coincident NLRs are seen. ∼ 50% of our objects have [OIII] emission offset by a few kpc, corresponding to the two velocity components seen in the SDSS spectra, but there are no corresponding double stellar components seen in the NIR imaging. For those objects with sufficiently high quality slit spectra, we see velocity and/or velocity dispersion gradients in [OIII] emission, suggestive of the kinematic signatures of a single NLR. The remaining ∼ 40% of our objects are ambiguous, and will need higher spatial resolution observations to distinguish between the two scenarios. Our observations therefore favor the kinematics scenario with a single AGN for the majority of these double-peaked [OIII] type 2 AGNs. We emphasize the importance of combining imaging and slit spectroscopy in identifying kpc binary AGNs, i.e., in no cases does one of these alone allow an unambiguous identification. We estimate that ∼ 0.2-1% of the z∼ 150 km/s.

## Strong lensing of gravitational waves as seen by LISA

**Authors:** Sereno, M.; Sesana, A.; Bleuler, A.; Jetzer, Ph.; Volonteri, M.; Begelman, M. C.

**Abstract:** We discuss strong gravitational lensing of gravitational waves from merging of massive black hole binaries in the context of the LISA mission. Detection of multiple events would provide invaluable information on competing theories of gravity, evolution and formation of structures and, with complementary observations, constraints on $H_0$ and other cosmological parameters. Most of the optical depth for lensing is provided by intervening massive galactic halos, for which wave optics effects are negligible. Probabilities to observe multiple events are sizable for a broad range of formation histories. For the most optimistic models, up to 4 multiple events with a signal to noise ratio => 8 are expected in a 5-year mission. Chances are significant even for conservative models with either light (<= 60%) or heavy (<= 40%) seeds. Due to lensing amplification, some intrinsically too faint signals are brought over threshold (<= 2 per year).

## Fractal Geometry of Angular Momentum Evolution in Near-Keplerian Systems

**Authors:** Gürkan, M. Atakan

Abstract: In this paper, we propose a method to study the nature of resonant relaxation in near-Keplerian systems. Our technique is based on measuring the fractal dimension of the angular momentum trails and we use it to analyze the outcome of N-body simulations. With our method, we can reliably determine the timescale for resonant relaxation, as well as the rate of change of angular momentum in this regime. We find that growth of angular momentum is more rapid than random walk, but slower than linear growth. We also determine the presence of long term correlations, arising from the bounds on angular momentum growth. We develop a toy model that reproduces all essential properties of angular momentum evolution.

# Color Behavior Of BL Lacertae Object OJ 287 During Optical Outburst

Authors: Dai, Yan; Wu, Jianghua; Zhu, Zong-Hong; Zhou, Xu; Ma, Jun

Abstract: This paper aims to study the color behavior of the BL Lac object OJ 287 during optical outburst. According to the revisit of the data from the OJ-94 monitoring project and the analysis the data obtained with the 60/90 cm Schmidt Telescope of NAOC, we found a bluer-when-brighter chromatism in this object. The amplitude of variation tends to decrease with the decrease of frequency. These results are consistent with the shock-in-jet model. We made some simulations and confirmed that both amplitude difference and time delay between variations at different wavelengths can result in the phenomenon of bluer-when-brighter. Our observations confirmed that OJ 287 underwent a double-peaked outburst after about 12 years from 1996, which provides further evidence for the binary black hole model in this object.

# Pop III Stellar Masses and IMF

Authors: Norman, Michael L.

**Abstract:** We provide a status report on our current understanding of the mass scales for Pop III.1 and Pop III.2 stars. Since the last review (Norman 2008), substantial progress has been made both numerically and analytically on the late stages of protostellar cloud core collapse, protostar formation and accretion, and stellar evolution taking into account cloud core properties and radiative feedback effects. Based on this, there are growing indications that primordial stars forming from purely cosmological initial conditions (Pop III.1) were substantially more massive than stars forming in preionized gas (Pop III.2) where HD cooling is important. Different stellar endpoints are predicted for these two types of Pop III stars with different chemical enrichment signatures: the former die as pair instability supernovae or intermediate mass black holes, whereas the latter die as iron core-collapse supernovae, leaving behind neutron star and stellar black hole remnants. We review recent simulations which show evidence for binary fragmentation at high densities, and comment on the significance of these results. We then summarize an attempt to directly calculate the Pop III.1 IMF taking into account the latest numerical and analytical models. We conclude with suggestions for the kind of simulations needed next to continue improving our understanding of Pop III star formation, which is a necessary input to understanding high redshift galaxy formation.

## Massive black holes in stellar systems: 'quiescent' accretion and luminosity

**Authors:** Volonteri, Marta; Dotti, Massimo; Campbell, Duncan; Mateo, Mario

**Abstract:** Only a small fraction of local galaxies harbor an accreting black hole, classified as an active galactic nucleus (AGN). However, many stellar systems are plausibly expected to host black holes, from globular clusters to nuclear star clusters, to massive galaxies. The mere presence of stars in the vicinity of a black hole provides a source of fuel via mass loss of evolved stars. In this paper we assess the expected luminosities of black holes embedded in stellar systems of different sizes and properties, spanning a large range of masses. We model the distribution of stars and derive the amount of gas available to a central black hole through a geometrical model. We estimate the luminosity of the black holes under simple, but physically grounded, assumptions on the accretion flow. Finally we discuss the detectability of âĂĂŸquiescentâĂĂŹ black holes in the local Universe.

## The coordinated key role of wet, mixed, and dry major mergers in the buildup of massive early-type galaxies at z<~1

**Authors:** Eliche-Moral, M. Carmen; Prieto, Mercedes; Gallego, Jesus; Zamorano, Jaime

**Abstract:** Hierarchical models predict that massive early-type galaxies (mETGs) derive from the most massive and violent merging sequences occurred in the Universe. However, the role of wet, mixed, and dry major mergers in the assembly of mETGs is questioned by some recent observations. We have developed a semi-analytical model to test the feasibility of the major-merger origin hypothesis for mETGs, just accounting for the effects on galaxy evolution of the major mergers strictly reported by observations. The model proves that it is feasible to reproduce the observed number density evolution of mETGs since z~ 1, just accounting for the coordinated effects of wet/mixed/dry major mergers. It can also reconcile the different assembly redshifts derived by hierarchical models and by mass downsizing data for mETGs, just considering that a mETG observed at a certain redshift is not necessarily in place since then. The model predicts that wet major mergers have controlled the mETGs buildup since z~ 1, although dry and mixed mergers have also played an essential role in it. The bulk of this assembly took place at 0.7<z<1, being nearly frozen at z<~ 0.7 due to the negligible number of major mergers occurred per existing mETG since then. The model suggests that major mergers have been the main driver for the observational migration of mass from the massive end of the blue galaxy cloud to that of the red sequence in the last ~ 8 Gyr.

## Star Formation in Quasar Disk

**Authors:** Jiang, Yanfei; Goodman, Jeremy

**Abstract:** Using a version of the ZEUS code, we carry out two-dimensional simulations of self-gravitating shearing sheets, with application to QSO accretion disks at a few thousand Schwarzschild radii, corresponding to a few hundredths of a parsec for a $10^8$ solar-mass black hole. Radiation pressure and optically thick radiative cooling are implemented via vertical averages. We determine dimensionless versions of the maximum surface density, accretion rate, and effective viscosity that can be sustained by density-wave turbulence without fragmentation. Where fragments do form, we study the final masses that result. The maximum Shakura-Sunyaev viscosity parameter is approximately 0.4. Fragmentation occurs when the cooling time is less than about twice the shearing time, as found by Gammie and others, but can also occur at very long cooling times in sheets that are strongly radiation-pressure dominated. For accretion at the Eddington rate onto a $10^8$ solar-mass black hole, fragmentation occurs beyond four thousand Schwarzschild radii, $r_s$. Near this radius, initial fragment masses are several hundred suns, consistent with estimates from linear stability; final masses after merging increase with the size of the sheet, reaching several thousand suns in our largest simulations. With increasing black-hole mass at a fixed Eddington ratio, self-gravity prevails to smaller multiples of $r_s$, where radiation pressure is more important and the cooling time is longer compared to the dynamical time; nevertheless, fragmentation can occur and produces larger initial fragment masses. We also find energy conservation is likely to be a challenge for all eulerian codes in self-gravitating regimes where radiation pressure dominates.

## Fast variability of gamma-ray emission from supermassive black hole binary OJ 287

**Authors:** Neronov, Andrii; Vovk, Ievgen

**Abstract:** We report the discovery of fast variability of gamma-ray flares from blazar OJ 287. This blazar is known to be powered by binary system of supermassive black holes. The observed variability time scale $T_v ar < 3$-10 hr is much shorter than the light crossing time of more massive ($1.8 \times 10^{10}$ solar masses) black hole and is comparable to the light crossing time of the less massive ($1.3 \times 10^8$ solar masses) black hole. This indicates that gamma-ray emission is produced by relativistic jet ejected by the black hole of smaller mass. Detection of gamma rays s with energies in excess of 10 GeV during the fast variable flares constrains the Doppler factor of the jet to be larger than 4. Possibility of the study of orbital modulation of emission from relativistic jet makes OJ 287 a unique laboratory for the study of the mechanism(s) of formation of jets by black holes, in particular, of the response of the jet parameters to the changes of the parameters of the medium from which the black hole accretes and into which the jet expands.

## The coupling of a young stellar disc with the molecular torus in the Galactic centre

**Authors:** Haas, Jaroslav; Subr, Ladislav; Kroupa, Pavel

**Abstract:** The Galactic centre hosts, according to observations, a number of early-type stars. About one half of those which are orbiting the central supermassive black hole on orbits with projected radii $\gtrsim 0.03$ pc form a coherently rotating disc. Observations further reveal a massive gaseous torus and a significant population of late-type stars. In this paper, we investigate, by means of numerical N-body computations, the orbital evolution of the stellar disc, which we consider to be initially thin. We include the gravitational influence of both the torus and the late-type stars, as well as the self-gravity of the disc. Our results show that, for a significant set of system parameters, the evolution of the disc leads, within the lifetime of the early-type stars, to a configuration compatible with the observations. In particular, the disc naturally reaches a specific - perpendicular - orientation with respect to the torus, which is indeed the configuration observed in the Galactic centre. We, therefore, suggest that all the early-type stars may have been born within a single gaseous disc.

## Perturbative effects of spinning black holes with applications to recoil velocities

**Authors:** Nakano, Hiroyuki; Campanelli, Manuela; Lousto, Carlos O.; Zlochower, Yosef

**Abstract:** Recently, we proposed an enhancement of the Regge-Wheeler-Zerilli formalism for first-order perturbations about a Schwarzschild background that includes first-order corrections due to the background black-hole spin. Using this formalism, we investigate gravitational wave recoil effects from a spinning black-hole binary system analytically. This allows us to better understand the origin of the large recoils observed in full numerical simulation of spinning black hole binaries.

## Evidence for Low Black Hole Spin and Physically Motivated Accretion Models from Millimeter VLBI Observations of Sagittarius A*

**Authors:** Broderick, Avery E.; Fish, Vincent L.; Doeleman, Sheperd S.; Loeb, Abraham

**Abstract:** Millimeter very-long baseline interferometry (mm-VLBI) provides the novel capacity to probe the emission region of a handful of supermassive black holes on sub-horizon scales. For Sagittarius A* (Sgr A*), the supermassive black hole at the center of the Milky Way, this provides access to the region in the immediate vicinity of the horizon. Broderick et al. (2009) have already shown that by leveraging spectral and polarization information as well as accretion theory, it is possible to extract accretion-model parameters (including black hole spin) from mm-VLBI experiments containing only a handful of telescopes. Here we repeat this analysis with the most recent mm-VLBI data, considering a class of aligned, radiatively inefficient accretion flow (RIAF) models. We find that the combined data set rules out symmetric models for Sgr A*'s flux distribution at the 3.9-sigma level, strongly favoring length-to-width ratios of roughly 2.4:1. More importantly, we find that physically motivated accretion flow models provide a significantly better fit to the mm-VLBI observations than phenomenological models, at the 2.9-sigma level. This implies that not only is mm-VLBI presently capable of distinguishing between potential physical models for Sgr A*'s emission, but further that it is sensitive to the strong gravitational lensing associated with the propagation of photons near the black hole. Based upon this analysis we find that the most probable magnitude, viewing angle, and position angle for the black hole spin are a=0.0(+0.64+0.86), theta=68(+5+9)(-20-28) degrees, and xi=-52(+17+33)(-15-24) east of north, where the errors quoted are the 1-sigma and 2-sigma uncertainties.

## A Very Close Binary Black Hole in a Giant Elliptical Galaxy 3C 66B and its Black Hole Merger

**Authors:** Iguchi, Satoru; Okuda, Takeshi; Sudou, Hiroshi

**Abstract:** Recent observational results provide possible evidence that binary black holes (BBHs) exist in the center of giant galaxies and may merge to form a super-massive black hole in the process of their evolution. We first detected a periodic flux variation on a cycle of $93 \pm 1$ days from the 3-mm monitor observations of a giant elliptical galaxy 3C 66B for which an orbital motion with a period of $1.05 \pm 0.03$ years had been already observed. The detected signal period being shorter than the orbital period can be explained by taking into consideration the Doppler-shifted modulation due to the orbital motion of a BBH. Assuming that the BBH has a circular orbit and that the jet axis is parallel to the binary angular momentum, our observational results demonstrate the presence of a very close BBH that has the binary orbit with an orbital period of $1.05 \pm 0.03$ years, an orbital radius of $(3.9 \pm 1.0) \times 10^{-3}$ pc, an orbital separation of $(6.1^{+1.0}_{-0.9}) \times 10^{-3}$ pc, the larger black hole mass of $(1.2^{+0.5}_{-0.2}) \times 10^9$ $M_\odot$, and the smaller black hole mass of $(7.0^{+4.7}_{-6.4}) \times 10^8$ $M_\odot$. The BBH decay time of $(5.1^{+60.5}_{-2.5}) \times 10^2$ years provides evidence for the occurrence of black hole mergers. This Letter will demonstrate the interesting possibility of black hole collisions to form a supermassive black hole in the process of evolution, one of the most spectacular natural phenomena in the universe.

## Retrograde Accretion and Merging Supermassive Black Holes

**Authors:** Nixon, C. J.; Cossins, P. J.; King, A. R.; Pringle, J. E.

**Abstract:** We investigate whether a circumbinary gas disc can coalesce a supermassive black hole binary system in the centre of a galaxy. This is known to be problematic for a prograde disc. We show that in contrast, interaction with a retrograde circumbinary disc is considerably more effective in shrinking the binary because there are no orbital resonances. The binary directly absorbs negative angular momentum from the circumbinary disc by capturing gas into a disc around the secondary black hole, or discs around both holes if the binary mass ratio is close to unity. In many cases the binary orbit becomes eccentric, shortening the pericentre distance as the eccentricity grows. In all cases the binary coalesces once it has absorbed the angular momentum of a gas mass comparable to that of the secondary black hole. Importantly, this conclusion is unaffected even if the gas inflow rate through the disc is formally super–Eddington for either hole. The coalescence timescale is therefore always $\sim M_2/\dot{M}$, where $M_2$ is the secondary black hole mass and $\dot{M}$ the inflow rate through the circumbinary disc.

## Fundamental physics and cosmology with LISA

**Authors:** Babak, Stanislav; Gair, Jonathan R.; Petiteau, Antoine; Sesana, Alberto

**Eprint:** http://arxiv.org/abs/1011.2062

**Keywords:** cosmology; gr-qc; supermassive black holes

**Abstract:** In this article we give a brief review of the fundamental physics that can be done with the future space-based gravitational wave detector LISA. This includes detection of gravitational wave bursts coming from cosmic strings, measuring a stochastic gravitational wave background, mapping spacetime around massive compact objects in galactic nuclei with extreme-mass-ratio inspirals and testing the predictions of General Relativity for the strong dynamical fields of inspiralling binaries. We give particular attention to new results which show the capability of LISA to constrain cosmological parameters using observations of coalescing massive Black Hole binaries.

## The Murmur of The Hidden Monster: Chandra's Decadal View of The Super-massive Black Hole in M31

**Authors:** Li, Zhiyuan; Garcia, Michael R.; Forman, William R.; Jones, Christine; Kraft, Ralph P.; Lal, Dharam V.; Murray, Stephen S.; Wang, Q. Daniel

**Eprint:** http://arxiv.org/abs/1011.1224

**Keywords:** astro-ph.HE; observations; supermassive black holes

**Abstract:** The Andromeda galaxy (M31) hosts a central super-massive black hole (SMBH), known as M31*, which is remarkable for its mass ($\sim 10^8 M_\odot$) and extreme radiative quiescence. Over the past decade, the Chandra X-ray observatory has pointed to the center of M31 nearly 100 times and accumulated a total exposure of $\sim 900$ ks. Based on these observations, we present an X-ray study of the temporal behavior of M31*. We find that M31* remained in a quiescent state from late 1999 to 2005, exhibiting an average 0.5-8 keV luminosity $\sim 10^{36}$erg/s, or only $\sim 10^{-10}$ of its Eddington luminosity. We report the discovery of an outburst that occurred on January 6, 2006, during which M31* radiated at $\sim 4.3 \times 10^{37}$erg/s. After the outburst, M31* apparently entered a more active state that lasts to date, characterized by frequent flux variability around an average luminosity of $\sim 4.8 \times 10^{36}$ erg/s. These strong flux variations are similar to the X-ray flares found in the SMBH of our Galaxy (Sgr A*), which may be explained by an episodic ejection of relativistic plasma inflated by magnetic field reconnection in the accretion disk.

## Modeling maximum astrophysical gravitational recoil velocities

**Authors:** Lousto, Carlos O.; Zlochower, Yosef

**Eprint:** http://arxiv.org/abs/1011.0593

**Keywords:** astro-ph.CO; astro-ph.GA; astro-ph.HE; gr-qc; massive binaries of black holes; numerical relativity

**Abstract:** We measure the recoil velocity as a function of spin for equal-mass, highly-spinning black-hole binaries, with spins in the orbital plane, equal in magnitude and opposite in direction. We confirm that the leading-order effect is linear in the spin and the cosine of angle between the spin direction and the infall direction at merger. We find higher-order corrections that are proportional to the odd powers in both the spin and cosine of this angle. Taking these corrections into account, we predict that the maximum recoil will be 3680+-130 km/s.

## Probing Intermediate Mass Black Holes With Optical Emission Lines from Tidally Disrupted White Dwarfs

**Authors:** Clausen, Drew; Eracleous, Michael

**Eprint:** http://arxiv.org/abs/1010.6087

**Keywords:** astro-ph.CO; EM counterparts; IMRI; intermediate-mass black holes; observations

**Abstract:** We calculate the emission line spectrum produced by the debris released when a white dwarf (WD) is tidally disrupted by an intermediate-mass black hole (IMBH; $M \sim 10^2 - 10^5 M_\odot$) and we explore the possibility of using the emission lines to identify such events and constrain the properties of the IMBH. To this end, we adopt and adapt the techniques developed by Strubbe & Quataert to study the optical emission lines produced when a main sequence (MS) star is tidally disrupted by a supermassive black hole. WDs are tidally disrupted outside of the event horizon of a $< 10^5 M_\odot$ black hole, which makes these tidal disruption events good signposts of IMBHs. We focus on the optical and UV emission lines produced when the accretion flare photoionizes the stream of debris that remains unbound during the disruption. We find that the spectrum is dominated by lines due to ions of C and O, the strongest of which are C species $\lambda 1549$ at early times and [O species] $\lambda 5007$ at later times. Furthermore, we model the profile of the emission lines in the [O species] $\lambda\lambda 4959, 5007$ doublet and find that it is highly asymmetric with velocity widths of up to $\sim 2500$ km s$^{-1}$, depending on the properties of the WD-IMBH system and the orientation of the observer. Finally, we compare the models with observations of X-ray flares and optical emission lines in the cores of globular clusters and propose how future observations can test if these features are due to a WD that has been tidally disrupted by an IMBH.

## The impact of realistic models of mass segregation on the event rate of extreme-mass ratio inspirals and cusp re-growth

**Authors:** Amaro-Seoane, Pau; Preto, Miguel

**Abstract:** One of the most interesting sources of gravitational waves (GWs) for LISA is the inspiral of compact objects on to a massive black hole (MBH), commonly referred to as an "extreme-mass ratio inspiral" (EMRI). The small object, typically a stellar black hole (bh), emits significant amounts of GW along each orbit in the detector bandwidth. The slowly, adiabatic inspiral of these sources will allow us to map space-time around MBHs in detail, as well as to test our current conception of gravitation in the strong regime. The event rate of this kind of source has been addressed many times in the literature and the numbers reported fluctuate by orders of magnitude. On the other hand, recent observations of the Galactic center revealed a dearth of giant stars inside the inner parsec relative to the numbers theoretically expected for a fully relaxed stellar cusp. The possibility of unrelaxed nuclei (or, equivalently, with no or only a very shallow cusp) adds substantial uncertainty to the estimates. Having this timely question in mind, we run a significant number of direct-summation $N-$body simulations with up to half a million particles to calibrate a much faster orbit-averaged Fokker-Planck code. We then investigate the regime of strong mass segregation (SMS) for models with two different stellar mass components. We show that, under quite generic initial conditions, the time required for the growth of a relaxed, mass segregated stellar cusp is shorter than a Hubble time for MBHs with $M_\bullet \lesssim 5\times10^6 M_\odot$ (i.e. nuclei in the range of LISA). SMS has a significant impact boosting the EMRI rates by a factor of $\sim 10$ for our fiducial models of Milky Way type galactic nuclei.

# Black-hole binaries, gravitational waves, and numerical relativity

**Authors:** Centrella, Joan M.; Baker, John G.; Kelly, Bernard J.; van Meter, James R.

**Abstract:** Understanding the predictions of general relativity for the dynamical interactions of two black holes has been a long-standing unsolved problem in theoretical physics. Black-hole mergers are monumental astrophysical events, releasing tremendous amounts of energy in the form of gravitational radiation, and are key sources for both ground- and space-based gravitational wave detectors. The black-hole merger dynamics and the resulting gravitational waveforms can only be calculated through numerical simulations of Einstein's equations of general relativity. For many years, numerical relativists attempting to model these mergers encountered a host of problems, causing their codes to crash after just a fraction of a binary orbit could be simulated. Recently, however, a series of dramatic advances in numerical relativity has, for the first time, allowed stable, robust black hole merger simulations. We chronicle this remarkable progress in the rapidly maturing field of numerical relativity, and the new understanding of black-hole binary dynamics that is emerging. We also discuss important applications of these fundamental physics results to astrophysics, to gravitational-wave astronomy, and in other areas.

# Electromagnetic Counterparts to Black Hole Mergers

**Authors:** Schnittman, Jeremy D.

**Abstract:** During the final moments of a binary black hole (BH) merger, the gravitational wave (GW) luminosity of the system is greater than the combined electromagnetic output of the entire observable universe. However, the extremely weak coupling between GWs and ordinary matter makes these waves very difficult to detect directly. Fortunately, the inspiraling BH system will interact strongly–on a purely Newtonian level–with any surrounding material in the host galaxy, and this matter can in turn produce unique electromagnetic (EM) signals detectable at Earth. By identifying EM counterparts to GW sources, we will be able to study the host environments of the merging BHs, in turn greatly expanding the scientific yield of a mission like LISA.

## Simulating merging binary black holes with nearly extremal spins

**Authors:** Lovelace, Geoffrey; Scheel, Mark. A.; Szilagyi, Bela

**Eprint:** http://arxiv.org/abs/1010.2777

**Keywords:** gr-qc; massive binaries of black holes; numerical relativity; spin

**Abstract:** Astrophysically realistic black holes may have spins that are nearly extremal (i.e., close to 1 in dimensionless units). Numerical simulations of binary black holes—important tools both for calibrating analytical templates for gravitational-wave detection and for exploring the nonlinear dynamics of curved spacetime—are particularly challenging when the holes' spins are nearly extremal. Typical initial data methods cannot yield simulations with nearly extremal spins; e.g., Bowen-York data cannot produce simulations with spins larger than about 0.93. In this paper, we present the first binary black hole inspiral, merger, and ringdown with initial spins larger than the Bowen-York limit. Specifically, using the Spectral Einstein Code (SpEC), we simulate the inspiral (through 12.5 orbits), merger and ringdown of two equal-mass black holes with equal spins of magnitude 0.95 antialigned with the orbital angular momentum.

## Properties of Accretion Flows Around Coalescing Supermassive Black Holes

**Authors:** Bogdanovic, Tamara; Bode, Tanja; Haas, Roland; Laguna, Pablo; Shoemaker, Deirdre

**Eprint:** http://arxiv.org/abs/1010.2496

**Keywords:** accretion discs; astro-ph.CO; EM counterparts; gr-qc; massive binaries of black holes; numerical relativity

**Abstract:** What are the properties of accretion flows in the vicinity of coalescing supermassive black holes (SBHs)? The answer to this question has direct implications for the feasibility of coincident detections of electromagnetic (EM) and gravitational wave (GW) signals from coalescences. Such detections are considered to be the next observational grand challenge that will enable testing general relativity in the strong, nonlinear regime and improve our understanding of evolution and growth of these massive compact objects. In this paper we review the properties of the environment of coalescing binaries in the context of the circumbinary disk and hot, radiatively inefficient accretion flow models and use them to mark the extent of the parameter space spanned by this problem. We report the results from an initial, general relativistic, hydrodynamical study of the inspiral and merger of equal-mass, spinning black holes, motivated by the latter scenario. We find that correlated EM+GW oscillations can arise during the inspiral phase followed by the gradual rise and subsequent drop-off in the light curve at the time of coalescence. While there are indications that the latter EM signature is a more robust one, a detection of either signal coincidentally with GWs would be a convincing evidence for an impending SBH binary coalescence. The observability of an EM counterpart in the hot accretion flow scenario depends on the details of a model. In the case of the most massive binaries observable by the Laser Interferometer Space Antenna, upper limits on luminosity imply that they may be identified by EM searches out to z~ 0.1-1. However, given the radiatively inefficient nature of the gas flow, we speculate that a majority of massive binaries may appear as low luminosity AGN in the local universe.

## Conservative corrections to the innermost stable circular orbit (ISCO) of a Kerr black hole: a new gauge-invariant post-Newtonian ISCO condition, and the ISCO shift due to test-particle spin and the gravitational self-force

**Authors:** Favata, Marc

**Abstract:** The innermost stable circular orbit (ISCO) delimits the transition from circular orbits to those that plunge into a black hole. In the test-mass limit, well-defined ISCO conditions exist for the Kerr and Schwarzschild spacetimes. In the finite-mass case, there are a large variety of ways to define an ISCO in a post-Newtonian (PN) context. Here I generalize the gauge-invariant ISCO condition of Blanchet & Iyer (2003) to the case of spinning (non-precessing) binaries. The Blanchet-Iyer

ISCO condition has two desirable and unexpected properties: (1) it exactly reproduces the Schwarzschild ISCO in the test-mass limit, and (2) it accurately approximates the recently-calculated shift in the Schwarzschild ISCO frequency due to the conservative-piece of the gravitational self-force [Barack & Sago (2009)]. The generalization of this ISCO condition to spinning binaries has the property that it also exactly reproduces the Kerr ISCO in the test-mass limit (up to the order at which PN spin corrections are currently known). The shift in the ISCO due to the spin of the test-particle is also calculated. Remarkably, the gauge-invariant PN ISCO condition exactly reproduces the ISCO shift predicted by the Papapetrou equations for a fully-relativistic spinning particle. It is surprising that an analysis of the stability of the standard PN equations of motion is able (without any form of "resummation") to accurately describe strong-field effects of the Kerr spacetime. The ISCO frequency shift due to the conservative self-force in Kerr is also calculated from this new ISCO condition, as well as from the effective-one-body Hamiltonian of Barausse & Buonanno (2010). These results serve as a useful point-of-comparison for future gravitational self-force calculations in the Kerr spacetime.

## Testing Modified Gravity with Gravitational Wave Astronomy

**Authors:** Sopuerta, Carlos F.; Yunes, Nicolas

**Abstract:** The emergent area of gravitational wave astronomy promises to provide revolutionary discoveries in the areas of astrophysics, cosmology, and fundamental physics. One of the most exciting possibilities is to use gravitational-wave observations to test alternative theories of gravity. In this contribution we describe how to use observations of extreme-mass-ratio inspirals by the future Laser Interferometer Space Antenna to test a particular class of theories: Chern-Simons modified gravity.

## The Effect of Massive Perturbers on Extreme Mass-Ratio Inspiral Waveforms

**Authors:** Yunes, Nicolas; Miller, M. Coleman; Thornburg, Jonathan

**Abstract:** Extreme mass ratio inspirals, in which a stellar-mass object merges with a supermassive black hole, are prime sources for space-based gravitational wave detectors because they will facilitate tests of strong gravity and probe the spacetime around rotating compact objects. In the last few years of such inspirals, the total phase is in the millions of radians and details of the waveforms are sensitive to small perturbations. We show that one potentially detectable perturbation is the presence of a second supermassive black hole within a few tenths of a parsec. The acceleration produced by the perturber on the extreme mass-ratio system produces a steady drift that causes the waveform to deviate systematically from that of an isolated system. If the perturber is a few tenths of a parsec from the extreme-mass ratio system (plausible in as many as a few percent of cases) higher derivatives of motion might also be detectable. In that case, the mass and distance of the perturber can be derived independently, which would allow a new probe of merger dynamics.

## Secular Stellar Dynamics near a Massive Black Hole

**Authors:** Madigan, Ann-Marie; Hopman, Clovis; Levin, Yuri

**Abstract:** The angular momentum evolution of stars close to massive black holes (MBHs) is driven by secular torques. In contrast to two-body relaxation, where interactions between stars are incoherent, the resulting resonant relaxation (RR) process is characterized by coherence times of hundreds of orbital periods. In this paper, we show that all the statistical properties of RR can be reproduced in an autoregressive moving average (ARMA) model. We use the ARMA model, calibrated with extensive N-body simulations, to analyze the long-term evolution of stellar systems around MBHs with Monte Carlo simulations. We show that for a single mass system in steady state, a depression is carved out near a MBH as a result of tidal disruptions. In our Galactic center, the size of the depression is about 0.2 pc, consistent with the size of the observed "hole" in the distribution of bright late-type stars. We also find that the velocity vectors of stars around a MBH are locally not isotropic. In a second application, we evolve the highly eccentric orbits that result from the tidal disruption of binary stars, which are considered to be plausible precursors of the "S-stars" in the Galactic center. We find that in this scenario more highly eccentric (e > 0.9) S-star orbits are produced than have been observed to date.

## Multiwavelength VLBI observations of Sagittarius A*

**Authors:** Lu, R. -S.; Krichbaum, T. P.; Eckart, A.; König, S.; Kunneriath, D.; Witzel, G.; Witzel, A.; Zensus, J. A.

**Abstract:** The compact radio source Sgr A*, associated with the super massive black hole at the center of the Galaxy, has been studied with VLBA observations at 3 frequencies (22, 43, 86 GHz) performed on 10 consecutive days in May 2007. The total VLBI flux density of Sgr A* varies from day to day. The variability is correlated at the 3 observing frequencies with higher variability amplitudes appearing at the higher frequencies. For the modulation indices, we find 8.4 % at 22 GHz, 9.3 % at 43 GHz, and 15.5 % at 86 GHz. The radio spectrum is inverted between 22 and 86 GHz, suggesting inhomogeneous synchrotron self-absorption with a turnover frequency at or above 86 GHz. The radio spectral index correlates with the flux density, which is harder (more inverted spectrum) when the source is brighter. The average source size does not appear to be variable over the 10-day observing interval. However, we see a tendency for the sizes of the minor axis to increase with increasing total flux, whereas the major axis remains constant. Towards higher frequencies, the position angle of the elliptical Gaussian increases, indicative of intrinsic structure, which begins to dominate the scatter broadening. At cm-wavelength, the source size varies with wavelength as $\lambda^{2.12\pm0.12}$, which is interpreted as the result of interstellar scatter broadening. After removal of this scatter broadening, the intrinsic source size varies as $\lambda^{1.4...1.5}$. The VLBI closure phases at 22, 43, and 86 GHz are zero within a few degrees, indicating a symmetric or point-like source structure. In the context of an expanding plasmon model, we obtain an upper limit of the expansion velocity of about 0.1 c from the non-variable VLBI structure. This agrees with the velocity range derived from the radiation transport modeling of the flares from the radio to NIR wavelengths.

## Variability of black-hole accretion discs: a theoretical study

**Authors:** Ferreira, Barbara T.

**Abstract:** This thesis investigates phenomena occurring in black-hole accretion discs which are likely to induce high-frequency quasi-periodic variability. Two classes of pseudo-relativistic theoretical models are studied. The first is based on the stability of transonic accretion flows and its connection to a disc instability that takes the form of propagating waves (viscous overstability). The second class of models looks at accretion-disc oscillations which are trapped due to the non-monotonic variation of the epicyclic frequency in relativistic flows. In particular, it focuses on inertial waves trapped below the maximum of the epicyclic frequency which are excited in deformed, warped or eccentric, discs. The influence of a transonic background on the propagation of these inertial modes is also investigated.

## The Final Merger of Black-Hole Binaries

**Authors:** Centrella, Joan M.; Baker, John G.; Kelly, Bernard J.; van Meter, James R.

**Abstract:** Recent breakthroughs in the field of numerical relativity have led to dramatic progress in understanding the predictions of General Relativity for the dynamical interactions of two black holes in the regime of very strong gravitational fields. Such black-hole binaries are important astrophysical systems and are a key target of current and developing gravitational-wave detectors. The waveform signature of strong gravitational radiation emitted as the black holes fall together and merge provides a clear observable record of the process. After decades of slow progress, these mergers and the gravitational-wave signals they generate can now be routinely calculated using the methods of numerical relativity. We review recent advances in understanding the predicted physics of events and the consequent radiation, and discuss some of the impacts this new knowledge is having in various areas of astrophysics.

## The Effect of Data Gaps on LISA Galactic Binary Parameter Estimation

**Authors:** Carré, Jérôme; Porter, Edward K.

**Abstract:** In the last few years there has been an enormous effort in parameter estimation studies for different sources with the space based gravitational wave detector, LISA. While these studies have investigated sources of differing complexity, the one thing they all have in common is they assume continuous data streams. In reality, the LISA data stream will contain gaps from such possible events such as re-pointing of the satellite antennae, to discharging static charge build up on the satellites, to disruptions due to micro-meteor strikes. In this work we conduct a large scale Monte Carlo parameter estimation simulation for galactic binaries assuming data streams containing gaps. As the expected duration and frequency of the gaps are currently unknown, we have decided to focus on gaps of approximately one hour, occurring either once per day or once per week. We also study the case where, as well as the expected periodic gaps, we have a data drop-out of one continuous week. Our results show that for for galactic binaries, a gap of once per week introduces a bias of between 0.5% and 1% in the estimation of parameters, for the most important parameters such as the sky position, amplitude and frequency. This number rises to between 3% and 7% for the case of one gap a day, and to between 4% and 9% when we have one gap a day and a spurious gap of a week. A future study will investigate the effect of data gaps on supermassive black hole binaries and extreme mass ratio inspirals.

## LISA Sensitivities to Gravitational Waves from Relativistic Metric Theories of Gravity

**Authors:** Tinto, Massimo; Alves, Márcio Eduardo da Silva

**Abstract:** The direct observation of gravitational waves will provide a unique tool for probing the dynamical properties of highly compact astrophysical objects, mapping ultra-relativistic regions of space-time, and testing Einstein's general theory of relativity. LISA (Laser Interferometer Space Antenna), a joint NASA-ESA mission to be launched in the next decade, will perform these scientific tasks by detecting and studying low-frequency cosmic gravitational waves through their influence on the phases of six modulated laser beams exchanged between three remote spacecraft. By directly measuring the polarization components of the waves LISA will detect, we will be able to test Einstein's theory of relativity with good sensitivity. Since a gravitational wave signal predicted by the most general relativistic metric theory of gravity accounts for *six* polarization modes (the usual two Einstein's tensor polarizations as well as two vector and two scalar wave components), we have derived the LISA Time-Delay Interferometric responses and estimated their sensitivities to vector- and scalar-type waves. We find that (i) at frequencies larger than roughly the

*Notes & News for GW science*

inverse of the one-way light time ($\approx 6 \times 10^{-2}$ Hz.) LISA is more than ten times sensitive to scalar-longitudinal and vector signals than to tensor and scalar-transverse waves, and (ii) in the low part of its frequency band is equally sensitive to tensor and vector waves and somewhat less sensitive to scalar signals.

## Constraining properties of the black hole population using LISA

**Authors:** Gair, Jonathan R; Sesana, Alberto; Berti, Emanuele; Volonteri, Marta

**Abstract:** LISA should detect gravitational waves from tens to hundreds of systems containing black holes with mass in the range from 10 thousand to 10 million solar masses. Black holes in this mass range are not well constrained by current electromagnetic observations, so LISA could significantly enhance our understanding of the astrophysics of such systems. In this paper, we describe a framework for combining LISA observations to make statements about massive black hole populations. We summarise the constraints that LISA observations of extreme-mass-ratio inspirals might be able to place on the mass function of black holes in the LISA range. We also describe how LISA observations can be used to choose between different models for the hierarchical growth of structure in the early Universe. We consider four models that differ in their prescription for the initial mass distribution of black hole seeds, and in the efficiency of accretion onto the black holes. We show that with as little as 3 months of LISA data we can clearly distinguish between these models, even under relatively pessimistic assumptions about the performance of the detector and our knowledge of the gravitational waveforms.

## An Efficient Time-Domain Method to Model Extreme-Mass-Ratio Inspirals

**Authors:** Canizares, Priscilla; Sopuerta, Carlos F.

**Abstract:** The gravitational-wave signals emitted by Extreme-Mass-Ratio Inspirals will be hidden in the instrumental LISA noise and the foreground noise produced by galactic binaries in the LISA band. Then, we need accurate gravitational-wave templates to extract these signals from the noise and obtain the relevant physical parameters. This means that in the modeling of these systems we have to take into account how the orbit of the stellar-mass compact object is modified by the action of its own gravitational field. This effect can be described as the action of a local force, the self-force. We present a time-domain technique to compute the self-force for geodesic eccentric orbits around a non-rotating massive black hole. To illustrate the method we have applied it to a testbed model consisting of scalar charged particle orbiting a non-dynamical black hole. A key feature of our method is that it does not introduce a small scale associated with the stellar-mass compact object. This is achieved by using a multidomain framework where the particle is located at the interface between two subdomains. In this way, we just have to evolve homogeneous wave-like equations with smooth solutions that have to be communicated across the subdomain boundaries using appropriate junction conditions. The numerical technique that we use to implement this scheme is the pseudospectral collocation method. We show the suitability of this technique for the modeling of Extreme-Mass-Ratio Inspirals and show that it can provide accurate results for the self-force.

## Extreme Mass-Ratio Inspirals in the Effective-One-Body Approach: Quasi-Circular, Equatorial Orbits around a Spinning Black Hole

**Authors:** Yunes, Nicolas; Buonanno, Alessandra; Hughes, Scott A.; Pan, Yi; Barausse, Enrico; Miller, M. Coleman; Throwe, William

**Abstract:** We construct effective-one-body waveform models suitable for data analysis with LISA for extreme-mass ratio inspirals in quasi-circular, equatorial orbits about a spinning supermassive black hole. The accuracy of our model is established through comparisons against frequency-domain, Teukolsky-based waveforms in the radiative approximation. The calibration of eight high-order post-Newtonian parameters in the energy flux suffices to obtain a phase and fractional amplitude agreement of better than 1 radian and 1 % respectively over a period between 2 and 6 months depending on the system considered. This agreement translates into matches higher than 97 % over a period between 4 and 9 months, depending on the system. Better agreements can be obtained if a larger number of calibration parameters are included. Higher-order mass ratio terms in the effective-one-body Hamiltonian and radiation-reaction introduce phase corrections of at most 30 radians in a one year evolution. These corrections are usually one order of magnitude larger than those introduced by the spin of the small object in a one year evolution. These results suggest that the effective-one-body approach for extreme mass ratio inspirals is a good compromise between accuracy and computational price for LISA data analysis purposes.

## Accuracy and effectualness of closed-form, frequency-domain waveforms for non-spinning black hole binaries

**Authors:** Damour, T.; Trias, M.; Nagar, A.

**Abstract:** The coalescences of binary black hole (BBH) systems, here taken to be non-spinning, are among the most promising sources for gravitational wave (GW) ground-based detectors, such as LIGO and Virgo. To detect the GW signals emitted by BBHs, and measure the parameters of the source, one needs to have in hand a bank of GW templates that are both effectual (for detection), and accurate (for measurement). We study the effectualness and the accuracy of the two types of parametrized banks of templates that are directly defined in the frequency-domain by means of closed-form expressions, namely 'post-Newtonian' (PN) and 'phenomenological' models. In absence of knowledge of the exact waveforms, our study assumes as fiducial, target waveforms the ones generated by the most accurate version of the effective one body (EOB) formalism. We find that, for initial GW detectors the use, at each point of parameter space, of the best closed-form template (among PN and phenomenological models) leads to an effectualness >97% over the entire mass range and >99% in an important fraction of parameter space; however, when considering advanced detectors, both of the closed-form frequency-domain models fail to be effectual enough in significant domains of the two-dimensional [total mass and mass ratio] parameter space. Moreover, we find that, both for initial and advanced detectors, the two closed-form frequency-domain models fail to satisfy the minimal required accuracy standard in a very large domain of the two-dimensional parameter space. In addition, a side result of our study is the determination, as a function of the mass ratio, of the maximum frequency at which a frequency-domain PN waveform can be 'joined' onto a NR-calibrated EOB waveform without undue loss of accuracy.

# On the relevance of gravitational self-force corrections on parameter estimation errors for extreme-mass-ratio inspirals

**Authors:** Huerta, E. A.; Gair, Jonathan R

**Abstract:** It is not currently clear how important it will be to include conservative self-force (SF) corrections in the models for extreme-mass-ratio inspiral (EMRI) waveforms that will be used to detect such signals in LISA (Laser Interferometer Space Antenna) data. These proceedings will address this issue for circular-equatorial inspirals using an approximate EMRI model that includes conservative corrections at leading post-Newtonian order. We will present estimates of the magnitude of the parameter estimation errors that would result from omitting conservative corrections, and compare these to the errors that will arise from noise fluctuations in the detector. We will also use this model to explore the relative importance of the second-order radiative piece of the SF, which is not presently known.

# Searching for an Intermediate Mass Black Hole in the Blue Compact Dwarf galaxy MRK 996

**Authors:** Georgakakis, A.; Tsamis, Y. G.; James, B. L.; Aloisi, A.

**Abstract:** The possibility is explored that accretion on an intermediate mass black hole contributes to the ionisation of the interstellar medium of the Compact Blue Dwarf galaxy MRK996. Chandra observations set tight upper limits (99.7 per cent confidence level) in both the X-ray luminosity of the posited AGN, Lx(2-10keV)<3e40erg/s, and the black hole mass, <1e4/$\lambda$ Msolar, where $\lambda$, is the Eddington ratio. The X-ray luminosity upper limit is insufficient to explain the high ionisation line [OIV]25.89$\mu$ m, which is observed in the mid-infrared spectrum of the MRK996 and is proposed as evidence for AGN activity. This indicates that shocks associated with supernovae explosions and winds of young stars must be responsible for this line. It is also found that the properties of the diffuse X-ray emission of MRK996 are consistent with this scenario, thereby providing direct evidence for shocks that heat the galaxy's interstellar medium and contribute to its ionisation.

## The LISA PathFinder DMU and Radiation Monitor

**Authors:** Canizares, Priscilla; Conchillo, Aleix; Diaz–Aguilo, Marc; Garcia-Berro, Enrique; Gesa, Lluis; Gibert, Ferran; Grimani, Catia; Lloro, Ivan; Lobo, Alberto; Mateos, Ignacio; Nofrarias, Miquel; Ramos-Castro, Juan; Sanjuan, Josep; Sopuerta, Carlos F

**Abstract:** The LISA PathFinder DMU (Data Management Unit) flight model was formally accepted by ESA and ASD on 11 February 2010, after all hardware and software tests had been successfully completed. The diagnostics items are scheduled to be delivered by the end of 2010. In this paper we review the requirements and performance of this instrumentation, specially focusing on the Radiation Monitor and the DMU, as well as the status of their programmed use during mission operations, on which work is ongoing at the time of writing.

## Recoiling Black Holes in Merging Galaxies: Relationship to AGN Lifetimes, Starbursts, and the M-sigma Relation

**Authors:** Blecha, Laura; Cox, Thomas J.; Loeb, Abraham; Hernquist, Lars

**Abstract:** Gravitational-wave (GW) recoil of merging supermassive black holes (SMBHs) may influence the co-evolution of SMBHs and their host galaxies. We examine this possibility using SPH/N-body simulations of gaseous galaxy mergers in which the merged BH receives a recoil kick. This enables us to follow recoiling BHs in self-consistent, evolving merger remnants. In contrast to recent studies on similar topics, we conduct a large parameter study, generating a suite of over 200 simulations with more than 60 merger models and a range of recoil velocities (vk). Our main results are as follows. (1) BHs kicked at nearly the central escape speed (vesc) may oscillate on large orbits for up to a Hubble time, but in gas-rich mergers, BHs kicked with up to ∼ 0.7 vesc may be confined to the central few kpc of the galaxy, owing to gas drag and steep central potentials. (2) vesc in gas-rich mergers may increase rapidly during final coalescence, in which case trajectories may depend on the timing of the BH merger relative to the formation of the potential well. (3) Recoil events generally reduce the lifetimes of bright active galactic nuclei (AGN), but may actually extend AGN lifetimes at lower luminosities. (4) Kinematically-offset AGN ($v > 800$ km s$^{-1}$) may be observable for up to ~ 10 Myr either immediately after the recoil or during pericentric passages through a gas-rich remnant. (5) Spatially-offset AGN ($R > 1$ kpc) generally have low luminosities and lifetimes of ~ 1 - 100 Myr. (6) Rapidly-recoiling BHs may be up to ~ 5 times less massive than their stationary counterparts. This lowers the normalization of the M-sigma relation and contributes to both intrinsic and overall scatter. (7) Finally, the displacement of AGN feedback after a recoil event enhances central star formation rates, thereby extending the star-burst phase of the merger and creating a denser stellar cusp. [Abridged.]

## One-zone models for spheroidal galaxies with a central supermassive black-hole. Self-regulated Bondi accretion

**Authors:** Lusso, E.; Ciotti, L.

**Abstract:** By means of a one-zone evolutionary model we study the co-evolution of supermassive black holes and their host galaxies, as a function of the accretion radiative efficiency, dark matter content and cosmological infall of gas. In particular, the radiation feedback is computed by using the self-regulated Bondi accretion. The models are characterized by strong oscillations when the galaxy is in the AGN state with a high accretion luminosity. We found that these one-zone models are able to reproduce two important phases of galaxy evolution, namely an obscured-cold phase when the bulk of star formation and black hole accretion occur, and the following quiescent hot phase in which accretion remains highly sub-Eddington. A Compton-thick phase is also found in almost all models, associated with the cold phase. An exploration of the parameter space reveals that the best agreement with the present day Magorrian relation is obtained, indipendently of the dark matter halo mass, for galaxies with low-mass seed black hole, and the accretion radiative efficiency ~ 0.1.

# Transient resonances in the inspirals of point particles into black holes

**Authors:** Flanagan, Eanna E.; Hinderer, Tanja

**Abstract:** We show that transient resonances occur in the two body problem in general relativity, in the highly relativistic, extreme mass-ratio regime for spinning black holes. These resonances occur when the ratio of polar and radial orbital frequencies, which is slowly evolving under the influence of gravitational radiation reaction, passes through a low order rational number. At such points, the adiabatic approximation to the orbital evolution breaks down, and there is a brief but order unity correction to the inspiral rate. Corrections to the gravitational wave signal's phase due to resonance effects scale as the square root of the inverse of mass of the small body, and thus become large in the extreme-mass-ratio limit, dominating over all other post-adiabatic effects. The resonances make orbits more sensitive to changes in initial data (though not quite chaotic), and are genuine non-perturbative effects that are not seen at any order in a standard post-Newtonian expansion. Our results apply to an important potential source of gravitational waves, the gradual inspiral of white dwarfs, neutron stars, or black holes into much more massive black holes. It is hoped to exploit observations of these sources to map the spacetime geometry of black holes. However, such mapping will require accurate models of binary dynamics, which is a computational challenge whose difficulty is significantly increased by resonance effects. We estimate that the resonance phase shifts will be of order a few tens of cycles for mass ratios $\sim 10^{-6}$, by numerically evolving fully relativistic orbital dynamics supplemented with an approximate, post-Newtonian self-force.

# Conservative, gravitational self-force for a particle in circular orbit around a Schwarzschild black hole in a Radiation Gauge

**Authors:** Shah, Abhay; Keidl, Tobias; Friedman, John; Kim, Dong-Hoon; Price, Larry

**Abstract:** This is the second of two companion papers on computing the self-force in a radiation gauge; more precisely, the method uses a radiation gauge for the radiative part of the metric perturbation, together with an arbitrarily chosen gauge for the parts of the perturbation associated with changes in black-hole mass and spin and with a shift in the center of mass. We compute the conservative part of the self-force for a particle in circular orbit around a Schwarzschild black hole. The gauge vector relating our radiation gauge to a Lorenz gauge is helically symmetric, implying that the quantity $h_{\alpha\beta} u^\alpha u^\beta (= h_{uu})$ must have the same value for our radiation gauge as for a Lorenz gauge; and we confirm this numerically to one part in $10^{13}$. As outlined in the first paper, the perturbed metric is constructed from a Hertz potential that is in term obtained algebraically from the the retarded perturbed spin-2 Weyl scalar, $\psi_0$ . We use a mode-sum renormalization and find the renormalization coefficients by matching a series in L = $\ell$ + 1/2 to the large-L behavior of the expression for the self-force in terms of the retarded field $h_{\alpha\beta}^{ret}$; we similarly find the leading renormalization coefficients of $h_{uu}$ and the related change in the angular velocity of the particle due to its self-force. We show numerically that the singular part of the self-force has the form $f_\alpha \propto$ , the part of $\nabla_\alpha \rho^{-1}$ that is axisymmetric about a radial line through the particle. This differs only by a constant from its form for a Lorenz gauge. It is because we do not use a radiation gauge to describe the change in black-hole mass that the singular part of the self-force has no singularity along a radial line through the particle and, at least in this example, is spherically symmetric to subleading order in $\rho$.

## Relativistic encounters of more than two black holes

**Authors:** Amaro-Seoane, Pau; Freitag, Marc Dewi

**Abstract:** Two coalescing black holes (BHs) represent a conspicuous source of gravitational waves (GWs). The merger involves 17 parameters in the general case of Kerr BHs, so that a successful identification and parameter extraction of the information encoded in the waves will provide us with a detailed description of the physics of BHs. A search based on matched-filtering for characterization and parameter extraction requires the development of some $10^{15}$ waveforms. If a third additional BH perturbed the system, the waveforms would not be applicable, and we would need to increase the number of templates required for a valid detection. In this letter, we calculate the probability that more than two BHs interact in the regime of strong relativity in a dense stellar cluster. We determine the physical properties necessary in a stellar system for three black holes to have a close encounter in this regime and also for an existing binary of two BHs to have a strong interaction with a third hole. In both cases the event rate is negligible. While dense stellar systems such as galactic nuclei, globular clusters and nuclear stellar clusters are the breeding grounds for the sources of gravitational waves that ground-based and space-borne detectors like Advanced LIGO and LISA will be exploring, the analysis of the waveforms in full general relativity needs only to evaluate the two-body problem. This reduces the number of templates of waveforms to create by orders of magnitude.

## Dynamical friction of massive objects in galactic centres

**Authors:** Just, A.; Khan, F. M.; Berczik, P.; Ernst, A.; Spurzem, R.

**Abstract:** Dynamical friction leads to an orbital decay of massive objects like young compact star clusters or Massive Black Holes in central regions of galaxies. The dynamical friction force can be well approximated by Chandrasekhar's standard formula, but recent investigations show, that corrections to the Coulomb logarithm are necessary. With a large set of N-body simulations we show that the improved formula for the Coulomb logarithm fits the orbital decay very well for circular and eccentric orbits. The local scale-length of the background density distribution serves as the maximum impact parameter for a wide range of power-law indices of -1 ... -5. For each type of code the numerical resolution must be compared to the effective minimum impact parameter in order to determine the Coulomb logarithm. We also quantify the correction factors by using self-consistent velocity distribution functions instead of the standard Maxwellian often used. These factors enter directly the decay timescale and cover a range of 0.5 ... 3 for typical orbits. The new Coulomb logarithm combined with self-consistent velocity distribution functions in the Chandrasekhar formula provides a significant improvement of orbital decay times with correction up to one order of magnitude compared to the standard case. We suggest the general use of the improved formula in parameter studies as well as in special applications.

## Inspiral of Generic Black Hole Binaries: Spin, Precession, and Eccentricity

**Authors:** Levin, Janna; Contreras, Hugo

Notes & News for GW science

Abstract: We compile the equations of motion describing the most general black hole binaries as computed by Will and collaborators. We use the equations converted to Hamiltonian variables to consider spinning and precessing and eccentric pairs. We find that while spin-spin coupling corrections can destroy constant radius orbits in principle, the effect is so small that orbits will reliably tend to quasi-spherical as angular momentum and energy are lost to gravitational radiation. Still, highly eccentric pairs can retain eccentricity by the time of plunge. We also show that three natural frequencies of an orbit demonstrating both spin precession and perihelion precession are the frequency of angular motion in the orbital plane, the frequency of the plane precession, and the frequency of radial oscillations. These three shape the waveform. The pattern of energy lost during the inspiral is also directly related to these same natural harmonics.

## Hierarchical Assembly of Supermassive Black Holes: Adaptive Optics Imaging of Double-Peaked [O III] Active Galactic Nuclei

Authors: Fu, Hai; Myers, Adam D.; Djorgovski, S. G.; Yan, Lin

Abstract: Hierarchical galaxy assembly models predict the ubiquity of binary supermassive black holes (SMBHs). Nevertheless, observational confirmations of binary SMBHs are rare. We have obtained high-resolution near-infrared images of 50 double-peaked [O III] active galactic nuclei (AGNs) with Keck II laser guide star adaptive optics. The sample is compiled from the literature and consists of 17 type-1 and 33 type-2 AGNs over $0.03 < z < 0.56$. Eight type-1 and eight type-2 sources are apparently undergoing mergers with multiple components of comparable luminosities, separated between 0.6 and 12 kpc. Disturbed morphologies are evident in most cases. The merger fractions of type-1s and type-2s differ because the fraction increases with redshift, $f_{merger} \propto (1 + z)^4$, which is consistent with the evolution of major merger fraction of L* galaxies at $z < 1$. We show that type-1 AGNs in compact merging systems are outliers of the $M_{BH} - \sigma$ relation since stellar velocity dispersions could be over-estimated because of relative component velocities. It is thus important to cull mergers from AGN samples before comparing the $M_{BH} - \sigma$ relations of AGNs and normal galaxies. The emission-line properties are indistinguishable for spatially resolved and unresolved sources, emphasizing that multiple mechanisms can produce similar double-peaked profiles. This large sample of kpc-scale binary AGNs, if confirmed, is invaluable for studying the hierarchical assembly of SMBHs.

# Constraining the Black Hole Mass Spectrum with LISA Observations II: Direct comparison of detailed models

**Authors:** Plowman, Joseph E.; Hellings, Ronald W.; Tsuruta, Sachiko

**Abstract:** A number of scenarios have been proposed for the origin of the supermassive black holes (SMBHs) that are found in the centres of most galaxies. Many such scenarios predict a high-redshift population of massive black holes (MBHs), with masses in the range 100 to 100000 times that of the Sun. When the Laser Interferometer Space Antenna (LISA) is finally operational, it is likely that it will detect on the order of 100 of these MBH binaries as they merge. The differences between proposed population models produce appreciable effects in the portion of the population which is detectable by LISA, so it is likely that the LISA observations will allow us to place constraints on them. However, gravitational wave detectors such as LISA will not be able to detect all such mergers nor assign precise black hole parameters to the merger, due to weak gravitational wave signal strengths. This paper explores LISA's ability to distinguish between several MBH population models. In this way, we go beyond predicting a LISA observed population and consider the extent to which LISA observations could inform astrophysical modellers. The errors in LISA parameter estimation are applied with a direct method which generates random sample parameters for each source in a population realisation. We consider how the distinguishability varies depending on the choice of source parameters (1 or 2 parameters chosen from masses, redshift or spins) used to characterise the model distributions, with confidence levels determined by 1 and 2-dimensional tests based on the Kolmogorov-Smirnov test.

# The Black Hole Mass in Brightest Cluster Galaxy NGC 6086

**Authors:** McConnell, Nicholas J.; Ma, Chung-Pei; Graham, James R.; Gebhardt, Karl; Lauer, Tod R.; Wright, Shelley A.; Richstone, Douglas O.

Abstract: We present the first direct measurement of the central black hole mass, $M_{BH}$, in NGC 6086, the Brightest Cluster Galaxy (BCG) in Abell 2162. Our investigation demonstrates for the first time that stellar dynamical measurements of $M_{BH}$ in BCGs are possible beyond the nearest few galaxy clusters. We observed NGC 6086 with laser guide star adaptive optics and the integral-field spectrograph (IFS) OSIRIS at the W.M. Keck Observatory, and with the seeing-limited IFS GMOS-N at Gemini Observatory North. We combined the two IFS data sets with existing major-axis kinematics, and used axisymmetric stellar orbit models with an assumed dark matter halo to determine $M_{BH}$ and the R-band stellar mass-to-light ratio, $M*/L_R$. The best-fit values of $M_{BH}$ and $M*/L_R$ strongly depend on the assumed dark matter halo mass, $M_{halo}$: more massive halos yield larger $M_{BH}$ and smaller $M*/L_R$. For the most massive halo allowed within the gravitational potential of the host cluster, we find $M_{BH} = 3.6(+1.7)(-1.1) \times 10^9 M_\odot$ and $M*/L_R = 4.6(+0.3)(-0.7) \, M_\odot/L_\odot$ (68% confidence). The correlation between $M_{BH}$ and $M_{halo}$ could extend to dynamical models of other galaxies with central stellar cores, and new measurements of $M_{BH}$ from models with dark matter could steepen the empirical scaling relationships between black holes and their host galaxies. Further observations with adaptive optics will measure $M_{BH}$ in a larger sample of BCGs, progressing toward a statistical understanding of black hole-bulge scaling relationships in the most massive galaxies.

## Collisional formation of very massive stars in dense clusters

Authors: Moeckel, Nickolas; Clarke, Cathie J.

Abstract: We investigate the contraction of accreting protoclusters using an extension of n-body techniques that incorporates the accretional growth of stars from the gaseous reservoir in which they are embedded. Following on from Monte Carlo studies by Davis et al., we target our experiments toward populous clusters likely to experience collisions as a result of accretion-driven contraction. We verify that in less extreme star forming environments, similar to Orion, the stellar density is low enough that collisions are unimportant, but that conditions suitable for stellar collisions are much more easily satisfied in large-n clusters, i.e. n ~ 30,000 (we argue, however, that the density of the Arches cluster is insufficient for us to expect stellar collisions to have occurred in the cluster's prior evolution). We find that the character of the collision process is not such that it is a route toward smoothly filling the top end of the mass spectrum. Instead, runaway growth of one or two extreme objects can occur within less than 1 Myr after accretion is shut off, resulting in a few objects with masses several times the maximum reached by accretion. The rapid formation of these objects is due to not just the post-formation dynamical evolution of the clusters, but an interplay of dynamics and the accretional growth of the stars. We find that accretion-driven cluster shrinkage results in a distribution of gas and stars that offsets the disruptive effect of gas expulsion, and we propose that the process can lead to massive binaries and early mass segregation in star clusters.

## Extreme-Mass-Ratio-Black-Hole-Binary Evolutions with Numerical Relativity

**Authors:** Lousto, Carlos O.; Zlochower, Yosef

**Eprint:** http://arxiv.org/abs/1009.0292

**Keywords:** astro-ph.CO; astro-ph.GA; astro-ph.HE; astro-ph.SR; EMRI; gr-qc; numerical relativity

**Abstract:** We perform the first fully nonlinear numerical simulations of black-hole binaries with mass ratios 100:1. Our technique for evolving such extreme mass ratios is based on the moving puncture approach with a new gauge condition and an optimal choice of the mesh refinement (plus large computational resources). We achieve a convergent set of results for simulations starting with a small nonspinning black hole just outside the ISCO that then performs over two orbits before plunging into the 100 times more massive black hole. We compute the gravitational energy and momenta radiated as well as the final remnant parameters and compare these quantities with the corresponding perturbative estimates. The results show a close agreement. We briefly discuss the relevance of this simulations for Advanced LIGO, third-generation ground based detectors, and LISA observations, and self-force computations.

## The central black hole mass of the high-sigma but low-bulge-luminosity lenticular galaxy NGC 1332

**Authors:** Rusli, Stephanie P.; Thomas, Jens; Erwin, Peter; Saglia, Roberto P.; Nowak, Nina; Bender, Ralf

**Eprint:** http://arxiv.org/abs/1009.0515

**Abstract:** The masses of the most massive supermassive black holes (SMBHs) predicted by the $M_{BH}$-sigma and $M_{BH}$-luminosity relations appear to be in conflict. Which of the two relations is the more fundamental one remains an open question. NGC 1332 is an excellent example that represents the regime of conflict. It is a massive lenticular galaxy which has a bulge with a high velocity dispersion sigma of ~ 320 km/s; bulge–disc decomposition suggests that only 44% of the total light comes from the bulge. The $M_{BH}$-sigma and the $M_{BH}$-luminosity predictions for the central black hole mass of NGC 1332 differ by almost an order of magnitude. We present a stellar dynamical measurement of the SMBH mass using an axisymmetric orbit superposition method. Our SINFONI integral-field unit (IFU) observations of NGC 1332 resolve the SMBH's sphere of influence which has a diameter of ~ 0.76 arcsec. The sigma inside 0.2 arcsec reaches ~ 400 km/s. The IFU data allow us to increase the statistical significance of our results by modelling each of the four quadrants separately. We measure a SMBH mass of $(1.45 \pm 0.20) \times 10^9 M_\odot$ with a bulge mass-to-light ratio of 7.08 ±0.39 in the R-band. With this mass, the SMBH of NGC 1332 is offset from the $M_{BH}$-luminosity relation by a full order of magnitude but is consistent with the $M_{BH}$-sigma relation.

# Constraining scalar fields with stellar kinematics and collisional dark matter

**Authors:** Amaro-Seoane, Pau; Barranco, Juan; Bernal, Argelia; Rezzolla, Luciano

**Abstract:** The existence and detection of scalar fields could provide solutions to long-standing puzzles about the nature of dark matter, the dark compact objects at the center of most galaxies, and other phenomena. Yet, self-interacting scalar fields are very poorly constrained by astronomical observations, leading to great uncertainties in estimates of the mass $m_\phi$ and the self-interacting coupling constant lambda of these fields. To counter this, we have systematically employed available astronomical observations to develop new constraints, considerably restricting this parameter space. In particular, by exploiting precise observations of stellar dynamics at the center of our Galaxy and assuming that these dynamics can be explained by a single boson star, we determine an upper limit for the boson star compactness and impose significant limits on the values of the properties of possible scalar fields. Requiring the scalar field particle to follow a collisional dark matter model further narrows these constraints. Most importantly, we find that if a scalar dark matter particle does exist, then it cannot account for both the dark-matter halos and the existence of dark compact objects in galactic nuclei

## Radiatively inefficient accretion flows induced by gravitational-wave emission before massive black hole coalescence

**Authors:** Hayasaki, Kimitake

**Abstract:** We study an accretion flow during the gravitational-wave driven evolution of binary massive black holes. After the binary orbit decays due to interacting with a massive circumbinary disk, the binary is decoupled from the circumbinary disk because the orbital-decay timescale due to emission of gravitational wave becomes shorter than the viscous timescale evaluated at the inner edge of circumbinary disk. During the subsequent evolution, the accretion disk, which is truncated at the tidal radius because of the tidal torque, also shrinks as the orbital decay. Assuming that the disk mass changed by this process is all accreted, the whole region of the disk completely becomes radiatively inefficient when the semi-major axis is several hundred Schwarzschild radii. The disk temperature can become comparable with the virial temperature there in spite of a low disk luminosity. The prompt high-energy emission is hence expected long before black hole coalescence as well as the gravitational wave signals. Binary massive black holes finally merge without accretion disks.

## Wandering Black Holes in Bright Disk Galaxy Halos

**Authors:** Bellovary, Jillian; Governato, Fabio; Quinn, Tom; Wadsley, James; Shen, Sijing; Volonteri, Marta

**Abstract:** We perform SPH+N-body cosmological simulations of massive disk galaxies, including a formalism for black hole seed formation and growth, and find that satellite galaxies containing supermassive black hole seeds are often stripped as they merge with the primary galaxy. These events naturally create a population of "wandering" black holes that are the remnants of stripped satellite cores; galaxies like the Milky Way may host 5 – 15 of these objects within their halos. The satellites that harbor black hole seeds are comparable to Local Group dwarf galaxies such as the Small and Large Magellanic Clouds; these galaxies are promising candidates to host nearby intermediate mass black holes. Provided that these wandering black holes retain a gaseous accretion disk from their host dwarf galaxy, they give a physical explanation for the origin and observed properties of some recently discovered off-nuclear ultraluminous X-ray sources such as HLX-1.

## Bayesian parameter estimation in the second LISA Pathfinder Mock Data Challenge

**Authors:** Nofrarias, M.; Röver, C.; Hewitson, M.; Monsky, A.; Heinzel, G.; Danzmann, K.; Ferraioli, L.; Hueller, M.; Vitale, S.

**Abstract:** A main scientific output of the LISA Pathfinder mission is to provide a noise model that can be extended to the future gravitational wave observatory, LISA. The success of the mission depends thus upon a deep understanding of the instrument, especially the ability to correctly determine the parameters of the underlying noise model. In this work we estimate the parameters of a simplified model of the LISA Technology Package (LTP) instrument. We describe the LTP by means of a closed-loop model that is used to generate the data, both injected signals and noise. Then, parameters are estimated using a Bayesian framework and it is shown that this method reaches the optimal attainable error, the Cramer-Rao bound. We also address an important issue for the mission: how to efficiently combine the results of different experiments to obtain a unique set of parameters describing the instrument.

## Tidal breakup of binary stars at the Galactic Center. II. Hydrodynamic simulations

**Authors:** Antonini, Fabio; Lombardi, James C.; Merritt, David

**Abstract:** In Paper I, we followed the evolution of binary stars as they orbited near the supermassive black hole (SMBH) at the Galactic center, noting the cases in which the two stars would come close enough together to collide. In this paper we replace the point-mass stars by fluid realizations, and use a smoothed-particle hydrodynamics (SPH) code to follow the close interactions. We model the binary components as main-sequence stars with initial masses of 1, 3 and 6 Solar masses, and with chemical composition profiles taken from stellar evolution codes. Outcomes of the close interactions include mergers, collisions that leave both stars intact, and ejection of one star at high velocity accompanied by capture of the other star into a tight orbit around the SMBH. For the first time, we follow the evolution of the collision products for many ($\gtrsim 100$) orbits around the SMBH. Stars that are initially too small to be tidally disrupted by the SMBH can be puffed up by close encounters or collisions, with the result that tidal stripping occurs in subsequent periapse passages. In these cases, mass loss occurs episodically, sometimes for hundreds of orbits before the star is completely disrupted. Repeated tidal flares, of either increasing or decreasing intensity, are a predicted consequence. In collisions involving a low-mass and a high-mass star, the merger product acquires a high core hydrogen abundance from the smaller star, effectively resetting the nuclear evolution "clock" to a younger age. Elements like Li, Be and B that can exist only in the outermost envelope of a star are severely depleted due to envelope ejection during collisions and due to tidal forces from the SMBH. In the absence of collisions, tidal spin-up of stars is only important in a narrow range of periapse distances, $r_t/2 \lesssim r_p er \lesssim r_t$ with $r_t$ the tidal disruption radius.

> ### *Intention and purpose of GW Notes*
>
> *A succinct explanation*

The electronic publishing service **arXiv** is a dynamic, well-respected source of news of recent work and is updated daily. But, perhaps due to the large volume of new work submitted, it is probable that a member of our community might easily overlook relevant material. This new e-journal and its blog, **The LISA Brownbag (http://www.lisa-science.org/brownbag)**, both produced by the AEI, propose to offer scientist of the Gravitational Wave community the opportunity to more easily follow advances in the three areas mentioned: Astrophysics, General Relativity and Data Analysis. We hope to achieve this by selecting the most significant e-prints and list them in abstract form with a link to the full paper in both a single e-journal (GW Newsletter) and a blog (The LISA Brownbag). Of course, *this also implies that the paper will have its impact increased, since it will reach a broader public*, so that we encourage you to not forget submitting your own work

In addition to the abstracts, in each PDF issue of GW Notes, we will offer you a previously unpublished article written by a senior researcher in one of these three domains, which addresses the interests of all readers.

Thus the aim of The LISA Brownbag and GW Notes is twofold:

- Whenever you see an interesting paper on GWs science and LISA, you can submit the **arXiv** number to our **submission page (http://brownbag.lisascience.org)**. This is straightforward: No registration is required (although recommended) to simply type in the number in the entry field of the page, indicate some keywords and that's it

- We will publish a new full article in each issue, if available. This "feature article" will be from the fields of Astrophysics, General Relativity or the Data Analysis of gravitational waves and LISA. We will prepare a more detailed guide for authors, but for now would like to simply remind submitters that they are writing for colleagues in closely related but not identical fields, and that cross-fertilization and collaboration is an important goal of our concept

Subscribers get the issue distributed in PDF form. Additionally, they will be able to submit special announcements, such as meetings, workshops and jobs openings, to the list of registered people. For this, please register at the **registration page (http://lists.aei.mpg.de/cgi-bin/mailman/listinfo/lisa_brownbag)** by filling in your e-mail address and choosing a password.

> ### *The Astro-GR meetings*
>
> *Past, present and future*

Sixty two scientists attended the **Astro-GR@AEI** meeting, which took place September 18-22 2006 at the **Max-Planck Institut für Gravitationsphysik (Albert Einstein-Institut)** in Golm, Germany. The meeting was the brainchild of an AEI postdoc, who had the vision of bringing together Astrophysicists and experts in General Relativity and gravitational-wave Data Analysis to discuss sources for **LISA**, the planned Laser Interferometer Space Antenna. More specifically, the main topics were EMRIs and IMRIs (Extreme and Intermediate Mass-Ratio Inspiral events), i.e. captures of stellar-mass compact objects by supermassive black holes and coalescence of intermediate-mass black holes with supermassive black holes.

The general consensus was that the meeting was both interesting and quite stimulating. It was generally agreed that someone should step up and host a second round of this meeting. Monica Colpi kindly did so and this led to **Astro-GR@Como**, which was very similar in its informal format, though with a focus on all sources, meant to trigger new ideas, as a kind of brainstroming meeting.

Also, in the same year, in the two first weeks of September, we had another workshop in the Astro-GR series with a new "flavour", namely, the **Two Weeks At The AEI (2W@AEI)**, in which the interaction between the attendees was be even higher than what was reached in the previous meetings. To this end, we reduced the number of talks, allowing participants more opportunity to collaborate. Moreover, participants got office facilities and we combined the regular talks with the so-called "powerpointless" seminar, which will were totally informal and open-ended, on a blackboard. The next one was held in Barcelona in 2009 at the beginning of September, **Astro-GR@BCN** and last September 2010 it was the turn of Paris, at the APC, **LISA Astro-GR@Paris (Paris, APC Monday 13th to Friday September 17th 2010)**.

The upcoming meeting Astro-GR@Mallorca will be soon announced and the site is being prepared as these lines are written.

If you are interested in hosting in the future an Astro-GR meeting, please contact us. We are open to new formats, as long as the *Five Golden Rules* are respected.

A proper Astro-GR meeting **MUST** closely follow the *Five Golden Rules*:

I.    Bring together Astrophysicists, Cosmologists, Relativists and Data Analysts

II.   Motivate new collaborations and projects

III.  Be run in the style of Aspen, ITP, Newton Institute and Modest meetings, with plenty of time for discussions

|||| Grant access to the slides in a cross-platform format, such as PDF and, within reason, to the recorded movies of the talks in a free format which everybody can play like **Theora**, for those who could not attend, following the good principles of **Open Access**

||||| Keep It Simple and... Spontaneous